\documentclass[twocolumn,twocolappendix]{aastex631} 
\usepackage{amsmath}

\newcommand{\mni}{M_{\mathrm{Ni}}}
\newcommand{\ssd}{\hat{\sigma}}
\newcommand{\peak}{\mathrm{peak}}
\newcommand{\tail}{\mathrm{tail}}
\newcommand{\ztf}{\mathrm{ZTF}}

\shortauthors{Rodr\'iguez, Maoz, and Nakar}
\graphicspath{{./}{figures/}}

\begin{document}

\title{The Iron Yield of Core-collapse Supernovae}

\author[0000-0001-8651-8772]{\'Osmar Rodr\'iguez}
\affiliation{School of Physics and Astronomy, Tel Aviv University, Tel Aviv 69978, Israel}
\correspondingauthor{\'Osmar Rodr\'iguez}
\email{olrodrig@gmail.com}

\author{Dan Maoz}
\affiliation{School of Physics and Astronomy, Tel Aviv University, Tel Aviv 69978, Israel}

\author[0000-0002-4534-7089]{Ehud Nakar}
\affiliation{School of Physics and Astronomy, Tel Aviv University, Tel Aviv 69978, Israel}

\begin{abstract}
We present a systematic analysis of 191 stripped-envelope supernovae (SE~SNe), aimed to compute their $^{56}$Ni masses from the luminosity in their radioactive tails ($M_\mathrm{Ni}^\mathrm{tail}$) and/or in their maximum light, and the mean $^{56}$Ni and iron yields of SE~SNe and core-collapse SNe. Our sample consists of SNe~IIb, Ib, and Ic from the literature and from the Zwicky Transient Facility Bright Transient Survey. To calculate luminosities from optical photometry, we compute bolometric corrections using 49~SE~SNe with optical and near-IR photometry, and develop corrections to account for the unobserved UV and IR flux. We find that the equation of Khatami \& Kasen for radioactive $^{56}$Ni-powered transients with a single free parameter does not fit the observed peak time-luminosity relation of SE~SNe. Instead, we find a correlation between $M_\mathrm{Ni}^\mathrm{tail}$, peak time, peak luminosity, and decline rate, which allows measuring individual $^{56}$Ni masses to a precision of 14\%. Applying this method to the whole sample, we find, for SNe~IIb, Ib, and Ic, mean $^{56}$Ni masses of $0.066\pm0.006$, $0.082\pm0.009$, and $0.132\pm0.011$\,M$_{\sun}$, respectively. After accounting for their relative rates, for SE~SNe as a whole we compute mean $^{56}$Ni and iron yields of $0.090\pm0.005$ and $0.097\pm0.007$\,M$_{\sun}$, respectively. Combining these results with the recent Type~II~SN mean $^{56}$Ni mass derived by Rodr\'iguez et al., core-collapse SNe, as a whole, have mean $^{56}$Ni and iron yields of $0.055\pm0.006$ and $0.058\pm0.007$\,M$_{\sun}$, respectively. We also find that radioactive $^{56}$Ni-powered models typically underestimate the peak luminosity of SE~SNe by 60--70\%, suggesting the presence of an additional power source contributing to the luminosity at peak.
\end{abstract}

\keywords{transients: supernovae --- nuclear reactions, nucleosynthesis, abundances}


\section{Introduction}\label{sec:introduction}
Core-collapse (CC) supernovae (SNe) are the explosions of massive stars ($M_\mathrm{ZAMS}\gtrsim 8$\,M$_{\sun}$), triggered by the gravitational collapse of their iron cores (see \citealt{2021Natur.589...29B} for a current review of the explosion mechanism). Like all SNe, CC SNe are important in multiple astrophysical roles, including as element and dust factories, as kinetic energy sources affecting star formation and galaxy evolution, as the sites of cosmic-ray acceleration, as the progenitors of neutron stars and possibly black holes, and as distance indicators.

CC~SNe are spectroscopically separated into two classes: H-rich SNe (SNe~II; \citealt{1941PASP...53..224M}) and stripped-envelope (SE) SNe \citep{1997ApJ...491..375C}. The latter class includes the H-poor Type IIb \citep{1988AJ.....96.1941F}, the H-free Type Ib \citep{1985ApJ...294L..17W,1985ApJ...296..379E}, and the H-free/He-poor Type Ic \citep{1986ASIC..180...45W}. Within the latter group, those SNe having line widths $>15000$\,km\,s$^{-1}$ around maximum light are referred to as broad-line Ic (Ic-BL) SNe, while those associated with gamma-ray bursts (GRBs) are labelled GRB-SNe \citep[see][]{2016ApJ...832..108M}. Among SE~SNe there are some cases showing narrow emission lines of He \citep[SNe~Ibn;][]{2008MNRAS.389..113P} and C and/or O \citep[SNe~Icn;][]{2021TNSAN..76....1G,2022Natur.601..201G} in the spectra, indicative of interaction of the ejecta with circumstellar material (CSM). In addition, some SNe~Ib and Ic display peculiar double-peaked light curves, such as SNe~2005bf \citep{2005ApJ...631L.125A,2005ApJ...633L..97T,2006ApJ...641.1039F}, PTF11mnb \citep{2018AA...609A.106T}, 2019cad \citep{2021MNRAS.504.4907G}, and 2019stc \citep{2021ApJ...913..143G}. For those SNe, a double-peaked $^{56}$Ni distribution in radius within the ejecta and/or the presence of a magnetar are invoked to explain the morphology of their light curves. SNe belonging to the SN~Ibn and Icn subgroups, and those with peculiar double-peaked light curves are not included in the present analysis of SE~SNe.

The light curves of SE~SNe at early epochs are characterized by a peak in optical/UV bands followed by a rapid decline. The decline feature, thought to be produced by the cooling of the SN progenitor envelope after the shock breakout, depends on the properties of the progenitor system \citep[e.g.][]{2014ApJ...788..193N}. This so called cooling phase has been observed in some SNe with early-time data such as 
SN~IIb~1993J \citep[e.g.][]{1994AJ....107.1022R}, SN~Ib~2008D \citep[e.g.][]{2009ApJ...702..226M}, and SN~Ic~2006aj \citep[e.g.][]{2006Natur.442.1008C}, among others. After the cooling phase, the $\gamma$-rays and positrons produced by the decay of radioactive materials synthesized in the explosion power the light curves, which rise to the characteristic peak of SE~SNe. This peak, thought to be powered by the radioactive $^{56}$Ni decay chain ${{^{56}\mathrm{Ni}}\to{^{56}\mathrm{Co}}\to{^{56}\mathrm{Fe}}}$, is characterized by its maximum luminosity so-called peak luminosity ($L_\peak$) and the peak time ($t_L^\peak$), defined as the date of the peak luminosity minus the explosion epoch. The light curves then decline and, at epochs $\gtrsim60$\,d after explosion, the ejecta becomes optically thin and the luminosity begins to decrease exponentially with time. The latter phase, called the radioactive tail, is also powered by the radioactive $^{56}$Ni decay chain in an ejecta with transparency to $\gamma$-rays increasing over time \citep{1969ApJ...157..623C}. The dependence of the bolometric light curve of SE~SNe on the amount of $^{56}$Ni synthesized in the explosion enables an estimate of the $^{56}$Ni mass yield ($\mni$) of these events.

Measuring the $^{56}$Ni mass of CC~SNe is important for testing the various progenitor scenarios and explosion mechanisms that have been proposed for different CC~SNe, as the $\mni$ yield depends sensitively on the explosion properties and on the core structure of the progenitor \citep[e.g.][]{2019MNRAS.483.3607S}. Indeed, analyses of $^{56}$Ni mass distributions have shown that SE~SNe produce, on average, more $^{56}$Ni than SNe~II \citep[e.g.][]{2015arXiv150602655K,2019AA...628A...7A,2020AA...641A.177M,2020MNRAS.496.4517S,2021MNRAS.505.1742R,2021ApJ...918...89A}, which may suggest significant differences in the progenitor structures and/or explosion properties between both SN types. Equally important, the empirical estimate of the mean $^{56}$Ni mass and the iron yield of CC~SNe is a critical ingredient for studies of cosmic and Galactic chemical enrichment \citep[e.g.][]{2017ApJ...848...25M,2017ApJ...837..183W}.

The most accurate measurement of the $^{56}$Ni mass is obtained by observing the luminosity during the radioactive tail \citep[e.g.][]{2020MNRAS.496.4517S,2021ApJ...918...89A}. Given that luminosities in this phase are available only for a small fraction of SE~SNe, alternatives methods have been used to measure $^{56}$Ni masses from the early part of the light curve. In particular, the $^{56}$Ni masses of SE~SNe have been typically estimated through the analytical light-curve model of \citet{1982ApJ...253..785A} \citep[e.g.][]{2013MNRAS.434.1098C,2016MNRAS.457..328L,2015AA...574A..60T,2018AA...609A.136T,2019AA...621A..71T,2019MNRAS.485.1559P,2021AA...651A..81B} and through Arnett's rule \citep[e.g.][]{2016MNRAS.458.2973P}, which is a prediction of the Arnett model in which the peak luminosity is equal to the heating rate from the decay of $^{56}$Ni and $^{56}$Co at that time \citep[e.g.][]{2005AA...431..423S}. However, the model of \citet{1982ApJ...253..785A} was developed for SNe~Ia \citep{2016MNRAS.458.1618D,2021ApJ...913..145W}, so there is no reason to expect that this model and Arnett's rule provide $^{56}$Ni masses of SE~SNe much better than order of magnitude estimates. Indeed, numerical models have shown that the Arnett light-curve model and Arnett's rule overestimate the $^{56}$Ni masses of SE~SNe \citep[e.g.][]{2015MNRAS.453.2189D,2016MNRAS.458.1618D,2019ApJ...878...56K,2021ApJ...913..145W}. This was confirmed empirically by \citet{2021ApJ...918...89A}, who found that the $^{56}$Ni masses of 27~SE~SNe inferred through Arnett's rule are, on average, greater than those computed from the luminosity in their radioactive tail by a factor of $\sim2$ \citep[see also][]{2020MNRAS.496.4517S}. The reason why Arnett's rule is not accurate for SE~SNe is that the peak of the luminosity is seen roughly at the time that the time of diffusion of radiation through most of the ejecta towards the observer becomes comparable to the dynamical time. At that time, the observed luminosity includes two components: a significant fraction (but not necessary all) of the instantaneous radioactive heating; and energy that was deposited at earlier times and suffered some adiabatic losses during the ejecta expansion, before it was able to diffuse to the observer. There is no reason to expect that the combination of these two components will match exactly the total instantaneous radioactive energy deposition.

The $^{56}$Ni masses of SE~SNe have also been computed through hydrodynamical modeling of light curves \citep[e.g.][]{1996AA...306..219U,2001ApJ...550..991N,2002ApJ...572L..61M,2006MNRAS.369.1939S,2009ApJ...692.1131T,2009PZ.....29....2T,2014ApJ...792....7F,2015ApJ...811..147F,2012ApJ...757...31B,2014AJ....148...68B,2018Natur.554..497B,2015AA...580A.142E,2015MNRAS.454...95M,2014MNRAS.439.1807B,2016AA...593A..68F,2016AA...592A..89T,2018AA...609A.136T,2021MNRAS.501.5797B}. In particular, \citet{2018AA...609A.136T} found that the $^{56}$Ni masses inferred with the Arnett model are consistent with those computed with the hydrodynamical models generated with the code of \citet{2011ApJ...729...61B}. This means that if the Arnett model and Arnett's rule provide similar $^{56}$Ni masses, then hydrodynamical models could also overestimate the $^{56}$Ni masses of SE~SNe.

Based on the energy conservation equation of \citet{2013arXiv1301.6766K}, \citet{2019ApJ...878...56K} proposed a new relation that, assuming that peak luminosity is powered only by radioactive $^{56}$Ni heating, allows to estimate $\mni$ 
as a function of $L_\peak$, $t_L^\peak$, and a dimensionless parameter called $\beta$ that depends on the spatial distribution of $\mni$ and on the ejecta opacity, among others parameters. This method was employed by \citet{2020AA...641A.177M} for estimating lower limits for $^{56}$Ni masses of 37~SE~SNe, using the $\beta$ parameters suggested in \citet{2019ApJ...878...56K}. On the other hand, \citet{2021ApJ...918...89A} calculated $\beta$ values for 27~SE~SNe with $^{56}$Ni masses computed from the luminosity in the radioactive tail. They found that these $^{56}$Ni masses and those computed with the \citet{2019ApJ...878...56K} relation and empirical median $\beta$ values for each SN subtypes are, on average, consistent within 17\%. They also found that their empirical $\beta$ values are systematically lower than those based on the numerical light-curve models of \citet{2016MNRAS.458.1618D} and \citet{2020ApJ...890...51E}, among others. This is primarily because the observed sample of \citet{2021ApJ...918...89A} has significantly higher peak luminosities for a given $^{56}$Ni mass than the models mentioned above. Indeed, \citet{2020ApJ...890...51E} and \citet{2021ApJ...913..145W} (who recomputed the light curves of \citealt{2020ApJ...890...51E} with a better treatment of the radiation transport) reported that a substantial fraction of observed SE~SNe is more luminous than their brightest models, while \citet{2022AA...657A..64S} found that 36\% of the SNe~Ib/Ic in their sample are brighter than the maximum $r$-band brightness predicted by models of \citet{2021ApJ...913..145W}. Given that numerical models seem to underestimate the peak luminosities of SE~SNe, the use of the \citet{2019ApJ...878...56K} relation and the mean $\beta$ values computed with these models could, on average, overestimate the $^{56}$Ni masses of SE~SNe.

The $^{56}$Ni mass distribution of SNe~IIb, Ib, and Ic, along with their mean $^{56}$Ni masses have been widely analyzed in the literature \citep[e.g.][]{2011ApJ...741...97D,2013MNRAS.434.1098C,2015arXiv150602655K,2015AA...574A..60T,2018AA...609A.136T,2019AA...621A..71T,2016MNRAS.457..328L,2016MNRAS.458.2973P,2019MNRAS.485.1559P,2019AA...628A...7A,2020AA...641A.177M,2020MNRAS.496.4517S,2021ApJ...918...89A,2021AA...651A..81B,2021ApJ...922..141O}. In these works, however, we identify a number of shortcomings affecting the inferred $^{56}$Ni masses and the mean $\mni$ values:

\begin{enumerate}
\item The $^{56}$Ni masses are mostly computed with the Arnett model or Arnett rule, so those values and the reported mean $^{56}$Ni masses are overestimated. Among the works mentioned above, only \citet{2020MNRAS.496.4517S} and \citet{2021ApJ...918...89A} reported accurate $^{56}$Ni masses measured from the luminosity in the radioactive tail. However, their samples contain only 11 and 27~SE~SNe, respectively, so the reported mean $^{56}$Ni masses for each SE~SN subtype may not be representative of the SN~IIb, Ib, and Ic populations.

\item The host galaxy reddenings ($E_{B-V}$) are mostly inferred from the equivalent width of the host galaxy \ion{Na}{1}\,D absorption line\ ($\mathrm{EW_{NaID}}$, e.g. \citealt{2003fthp.conf..200T,2012MNRAS.426.1465P}) and from color curves \citep{2018AA...609A.135S}. The methods of \citet{2003fthp.conf..200T} and \citet{2012MNRAS.426.1465P} are based on the observed correlation between $\mathrm{EW_{NaID}}$ and reddening for SNe~Ia and the Milky Way (MW), respectively, so there is no reason to expect such correlations to hold for regions where SE~SNe explode. The color method of \citet{2018AA...609A.135S} assumes that SNe in the SN~IIb, Ib, and Ic groups have the same intrinsic color curves between zero and 20\,d since the time of $V$-band maximum light. SE~SNe are photometrically and spectroscopically diverse, so it is not expected that the color method provide precise $E_{B-V}$ values, while the accuracy of the methodology strongly depends on the completeness of the SN sample used to calibrate the method. \citet{2018AA...609A.135S} used only three SNe per SN subtype for calibrating the color method, therefore their $E_{B-V}$ estimates could be, on average, potentially under- or overestimated.

\item The bolometric light curves, necessary to estimate $\mni$, can be computed by integrating the spectral energy distribution (SED) constructed with UV, optical, and IR photometry \citep[e.g.][]{2018AA...609A.136T,2020MNRAS.496.4517S}. However, UV and IR photometry is usually available only for bright SNe, so the bolometric correction (BC) technique is used as an alternative method to compute luminosities from optical photometry. \citet{2014MNRAS.437.3848L,2016MNRAS.457..328L} presented BCs for SE~SNe for the various bands and colors, which are typically used to infer luminosities \citep[e.g.][]{2016MNRAS.457..328L,2019AA...621A..71T,2021ApJ...918...89A,2021AA...651A..81B}.
These BCs, however, are based on six or fewer SNe per subtype, and were computed for all SE~SNe as a whole rather than separately for each SN subtype. In addition, the BCs for Sloan bands are based on synthetic, rather than observed, photometry.
\end{enumerate}

Thus, accurate and statistically robust estimations of the mean $^{56}$Ni masses of SNe~IIb, Ib, and Ic has not been possible so far due to (1) the low number of SNe used to compute BCs, calibrate the color method, and calculate the mean $^{56}$Ni masses, and (2) shortcomings in the methodology to compute BCs, $E_{B-V}$ from $\mathrm{EW_{NaID}}$, and $^{56}$Ni masses from the early part of the light curves. The number of SE~SNe with optical photometry and the fraction with additional UV/IR photometry have increased dramatically in the last ten years. Therefore, it is now feasible to improve the BC determination, the color method calibration, and the $^{56}$Ni mass distribution with the newly available data. On the other hand, to solve the shortcomings mentioned above, it is necessary to carry out a new analysis to improve the methods used to compute BC, $E_{B-V}$, and $\mni$. In particular, to further increase the number of SE~SNe with $^{56}$Ni estimates, it is necessary to investigate alternative methods to accurately infer $^{56}$Ni masses from peak times and peak luminosities, such as the \citet{2019ApJ...878...56K} relation and empirical correlations.

The goal of this work is to accurately estimate the mean $^{56}$Ni masses of SNe~IIb, Ib, and Ic which, combined with the recent mean $^{56}$Ni mass for SNe~II reported by \citet{2021MNRAS.505.1742R} and the relative SN rates of \citet{2017PASP..129e4201S}, allow an evaluation of the mean $^{56}$Ni and iron yields of CC~SNe as a whole. To this end, we collect and analyze data from a variety of sources for almost 200 SE~SNe having photometry near maximum light.

The paper is organized as follows. In Section~\ref{sec:data_set} we present our sample of SNe and their basic properties. In Section~\ref{sec:methodology} we describe the methods we use to measure host galaxy reddenings, bolometric corrections, bolometric light curves, and $^{56}$Ni masses. In Section~\ref{sec:results} we present the $\mni$ distribution and the mean $^{56}$Ni mass for each SN subtype, and evaluate the mean $^{56}$Ni and iron yield of SE~SNe and CC~SNe as a whole. In particular, in Section~\ref{sec:LtDM} we show a new correlation between $^{56}$Ni mass, peak time, peak luminosity, and decline rate. Comparison to previous work and discussion of systematics appear in Section~\ref{sec:discussion}. Our conclusions are summarised in Section~\ref{sec:conclusions}.

\section{Data Set}\label{sec:data_set}
For this work, we select SNe~IIb, Ib, and Ic (including SNe~Ic-BL) from the literature and from the Zwicky Transient Facility (ZTF; \citealt{2019PASP..131g8001G,2019PASP..131a8002B,2019PASP..131a8003M}) Bright Transient Survey\footnote{\url{https://sites.astro.caltech.edu/ztf/bts/explorer.php}} (BTS; \citealt{2020ApJ...895...32F,2020ApJ...904...35P}). From among these SNe we select those (1) in galaxies with known redshift\footnote{Available in the NASA/IPAC Extragalactic Database; \url{http://ned.ipac.caltech.edu}.} or with redshifts measured from \ion{H}{2} region narrow emission lines in the spectra, and (2) having photometry at maximum light in at least two optical bands in at least one of the following filter systems: Johnson--Kron--Cousins $BV\!RI$, Sloan $gri$, and/or ZTF $gr$. For the selected SNe, we collect UV, optical, and IR photometry, along with optical and IR spectroscopy. We include the UV photometry taken by the Swift's Ultraviolet/Optical Telescope (UVOT), which is available at the Swift's Optical/Ultraviolet Supernova Archive\footnote{\url{http://swift.gsfc.nasa.gov/docs/swift/sne/swift\_sn.html}} (SOUSA; \citealt{2014ApSS.354...89B}). When necessary, we convert Sloan $gr$ to ZTF $gr$, and vice versa, using the transformations provided in Appendix~\ref{sec:mag_conv}. Among the photometry we collect, that obtained by the Carnegie Supernova Project I (CSP-I; \citealt{2018AA...609A.134S}) is in natural CSP-I photometric systems. We transform the latter photometry to the standard system using the conversions provided in \citet{2017AJ....154..211K}.

Our sample of 191~SNe is listed in Appendix Table~\ref{table:SN_sample}. This includes the SN name (Column~1) and spectral subtype (Column~2), the host galaxy name (Column~3), the MW reddening $E_{B-V}^\mathrm{MW}$ (Column~4), the heliocentric SN redshift $z$ (Column~5), the bands of the photometry we use (Column~6), and the references for the data (Column~7). MW reddenings are taken from \citet{2011ApJ...737..103S}, which have an associated random error of 16\% \citep{1998ApJ...500..525S}. Throughout this work, for the MW we assume the extinction curve given by \citet{1999PASP..111...63F} with a ratio of total to selective extinction ${R_V=A_V/E_{B-V}}$ of $3.1$. In addition, all phases are divided by $(1+z)$ to account for time-dilation.

\section{Methodology}\label{sec:methodology}
\subsection{$^{56}$Ni Mass}\label{sec:Ni_mass}
\subsubsection{$^{56}$Ni mass from the radioactive tail luminosity}\label{sec:MNi_tail}
As mentioned earlier, the most accurate measurement of the $^{56}$Ni mass is obtained by observing the bolometric luminosity during the radioactive tail, when the radiation diffusion time $t_\mathrm{diff}$ is much shorter than the time since the explosion $t$. At that point in time, all of the deposited radioactive heat that thermalizes in the ejecta escapes almost immediately in the form of IR, optical, and UV radiation, so
\begin{equation}\label{eq:tail}
L(t)=Q(t).
\end{equation}
Here, $L$ is the bolometric luminosity and $Q$ is the instantaneous energy deposition rate that thermalizes within the ejecta. Since the peak time is observed when $t_\mathrm{diff} \approx t$ and as $t_\mathrm{diff}$ drops roughly as $t^{-2}$, the $^{56}$Ni mass can be accurately determined from the bolometric light curve at $t \gtrsim 3\,t_L^\peak$.

Under the assumption that the radioactive decay chain ${{^{56}\mathrm{Ni}}\to{^{56}\mathrm{Co}}\to{^{56}\mathrm{Fe}}}$ is the unique heating source, 
\begin{equation}\label{eq:QMq}
Q(t)=\mni\,q(t,t_\mathrm{esc}),
\end{equation}
where $\mni$ is in units of M$_{\sun}$, and
\begin{equation}\label{eq:qt}
q(t,t_\mathrm{esc}) = q_\gamma(t) f_\mathrm{dep}(t,t_\mathrm{esc})+q_\mathrm{pos}(t).
\end{equation}
Here, $q_\gamma(t)$ and $q_\mathrm{pos}(t)$ are the total energy release rates of $\gamma$-rays and positron kinetic energy per unit $\mni$, respectively, while $f_\mathrm{dep}(t,t_\mathrm{esc})$ is the $\gamma$-ray deposition function, which describes the fraction of the generated $\gamma$-ray energy deposited in the ejecta. The terms $q_\gamma(t)$ and $q_\mathrm{pos}(t)$ are given by
\begin{equation}
q_\gamma(t)=\left[6.45\,e^{-\frac{t}{\tau_\mathrm{Ni}}}+1.38\,e^{-\frac{t}{\tau_\mathrm{Co}}}\right]\times 10^{43}\,\mathrm{erg}\,\mathrm{s}^{-1},
\end{equation}
and
\begin{equation}\label{eq:q_pos}
q_\mathrm{pos}(t)=0.046\left[-e^{-\frac{t}{\tau_\mathrm{Ni}}}+e^{-\frac{t}{\tau_\mathrm{Co}}}\right]\times 10^{43}\,\mathrm{erg}\,\mathrm{s}^{-1}
\end{equation}
\citep{2019MNRAS.484.3941W}, where $\tau_\mathrm{Ni}=8.76$\,d and $\tau_\mathrm{Co}=111.4$\,d. To represent $f_\mathrm{dep}(t,t_\mathrm{esc})$ we adopt
\begin{equation}\label{eq:fdep}
f_\mathrm{dep}(t,t_\mathrm{esc})=1-\exp{[-(t_\mathrm{esc}/t)^2]}
\end{equation}
\citep{1997ApJ...491..375C,1999astro.ph..7015J}, where $t_\mathrm{esc}$ is a characteristic time-scale that represents the $\gamma$-ray escape time.

To estimate $\mni$ and $t_\mathrm{esc}$ for a given observed bolometric light curve in its radioactive-tail phase, we rewrite equation~(\ref{eq:tail}) as
\begin{equation}\label{eq:MNi_eq}
\log L(t) = \log\mni + \log q(t,t_\mathrm{esc}).
\end{equation}
Given that typically $t_L^\peak \sim 20$\,d, and following \citet{2021ApJ...918...89A}, we consider luminosities at $t>60$\,d as corresponding to the radioactive tail.

Let $\bar{y}(\theta,x)$ be the model that describes the correlation between the observables $x$ and $y$, where $\theta$ is a vector containing the free parameters of the model. Given $n$ measurements of $x$, $y$, and their $1\,\sigma$ errors ($\sigma_x,\sigma_y$), we compute $\theta$ maximizing the posterior probability
\begin{equation}\label{eq:posterior_prob}
p(\theta|x,y,\sigma_x,\sigma_y)=p(\theta)\mathcal{L}(\theta|x,y,\sigma_x,\sigma_y).
\end{equation}
Here, $p(\theta)$ is the prior function (assumed to be uninformative in this work), and $\mathcal{L}(\theta|x,y,\sigma_x,\sigma_y)$ is the likelihood function given by
\begin{equation}\label{eq:lnL}
\ln\mathcal{L}(\theta|x,y,\sigma_x,\sigma_y)=-\frac{1}{2}\sum_{j=1}^n \left[\ln V_j+\frac{(y_j-\bar{y}(\theta,x_j))^2}{V_j}\right],
\end{equation}
where $V_j=\sigma_{y_j}^2+(\partial\bar{y}/\partial x)^2 \sigma_{x_j}^2+\sigma_0^2$ is the variance, and $\sigma_0$ is the error not accounted for in the errors in $x_j$ and $y_j$. To evaluate whether it is necessary to include $\sigma_0$ in the variance, we use the Bayesian information criterion \citep[BIC;][]{1978AnSta...6..461S}. We compute the maximum likelihood $\ln\mathcal{L}_\mathrm{max}$ for two models, one with and one without $\sigma_0$. If the inclusion of $\sigma_0$ reduces the value of $\mathrm{BIC}=-2\ln\mathcal{L}_\mathrm{max} + k\ln n$, where $k$ is the number of free parameters, then we include $\sigma_0$ as a parameter to compute. We maximize the posterior probability in equation~(\ref{eq:posterior_prob}) by means of a Markov Chain Monte Carlo process using the python package \textsc{emcee} \citep{2013PASP..125..306F}, which also provides the marginalized distributions of the parameters. We adopt the sample standard deviation ($\ssd$) of those distributions as the $1\,\sigma$ error of the free parameters.

In our case, $x=t$, $y=\log L$, and $\theta=\{\log\mni,t_\mathrm{esc},\sigma_0\}$, while $\bar{y}$ is given by equation~(\ref{eq:MNi_eq}). As priors we adopt uniform distributions between $-4.0$ and 0.3\,dex for $\log\mni$, between 1 and 500\,d for $t_\mathrm{esc}$, and between $-5.0$ and $-0.7$ for $\log(\sigma_0)$. Since there are three free parameters, we need at least four $\log L$ measurements at ${t>60}$\,d. In addition, we require a minimum time coverage of 20\,d in order to obtain a reliable $t_\mathrm{esc}$ value.

\subsubsection{$^{56}$Ni mass from the peak luminosity}
Only the best sampled SNe have a detailed bolometric light curve of the radioactive tail that allows an accurate determination of the $^{56}$Ni mass. For all other SNe, the $^{56}$Ni mass is often determined based on the luminosity and the time of the peak. Since Arnett's rule is not accurate enough, in this paper we use two different approaches. The first is purely empirical, where we use the sample of SNe with $^{56}$Ni mass estimation that is based on the radioactive tail to search for the best relation between peak time, peak luminosity, and $^{56}$Ni mass. We also extend the search looking for other properties of the light curve near the peak that can improve the accuracy of the $^{56}$Ni mass estimate.

The second approach is motivated by theory. \cite{2013arXiv1301.6766K} have shown that the integral
\begin{equation}\label{eq:EQL}
E(t)=\frac{1}{t}\int_0^t t'(Q(t')-L(t'))dt',
\end{equation}
where $E$ is the internal energy of the ejecta, offers an energy conservation law that takes the adiabatic losses of the trapped radiation into account.\footnote{The equation in this form assumes that the initial heat in the ejecta, deposited by the SN shock, is negligible compared to the energy deposited by radioactive decay.} Based on this equation, \citet{2019ApJ...878...56K} found that the relation between $L_\peak$ and $t_L^\peak$ can be approximated by
\begin{equation}\label{eq:KK19}
L_\peak = \frac{2}{(\beta t_L^\peak)^2}\int_0^{\beta t_L^\peak} Q(t')t'\,dt'.
\end{equation}
Here, $\beta$ is a dimensionless parameter of order unity that depends mainly on the spatial distribution of the heating source and on the opacity of the ejecta. Thus, $\beta$ can vary from one SN to another (as we later show is indeed the case). Under the assumption that SE~SNe at peak are powered {\em only} by the radioactive $^{56}$Ni decay chain, the heating source is given by equation~(\ref{eq:QMq}), so equation~(\ref{eq:KK19}) can be written as
\begin{equation}\label{eq:KK19_MNi}
\begin{split}
\log (L_\peak/\mni)&= \log\left(\frac{12.808}{x_\mathrm{Ni}^2}\left[1-\frac{1+x_\mathrm{Ni}}{e^{x_\mathrm{Ni}}}\right]\right.\\
       &+ \left.\frac{2.852}{x_\mathrm{Co}^2}\left[1-\frac{1+x_\mathrm{Co}}{e^{x_\mathrm{Co}}}\right]\right),
\end{split}
\end{equation}
where $L_\peak$ is in units of $10^{43}$\,erg\,s$^{-1}$, $x_\mathrm{Ni}=\beta t_L^\peak/\tau_\mathrm{Ni}$, and $x_\mathrm{Co}=\beta t_L^\peak/\tau_\mathrm{Co}$. The average value of $\beta$ and its variation between different SNe can be inferred from a sample of SNe with known $\mni$, $L_\peak$, and $t_L^\peak$ values \citep[e.g.][]{2021ApJ...918...89A}. The hope is that once calibrated, the peak time-luminosity relation (equation~\ref{eq:KK19_MNi}) allows, in principle, to estimate the $^{56}$Ni mass of SE~SNe based solely on measurements of $t_L^\peak$ and $\log L_\peak$.

\subsection{Light curve fits}\label{sec:LC_fits}
\begin{figure}
\includegraphics[width=1.0\columnwidth]{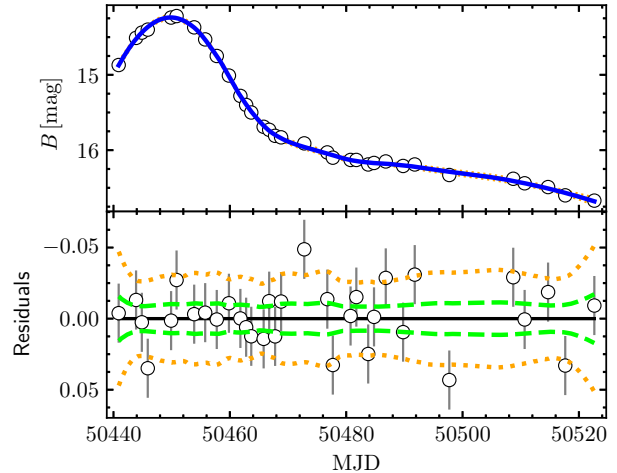} 
\caption{Top panel: $B$-band light curve of SN~1996cb, where the blue line is the \textsc{alr} fit. Bottom panel: fit residuals, where dashed and dotted lines are the $\pm1\,\sigma$ and $\pm3\,\sigma$ limits, respectively. Error bars are $1\,\sigma$ errors.}
\label{fig:ALR_fit}
\end{figure}

The derivation of $^{56}$Ni masses from observed photometric light curves, as outlined above, requires their interpolation in time. To interpolate the light curves of SE~SNe, some authors have used analytical functions developed for SNe~Ia \citep[e.g.][]{2018AA...609A.136T,2019AA...621A..71T,2021AA...651A..81B}. In this work, in order not to assume heuristic models, we perform \textsc{loess} non-parametric regression \citep{Cleveland_etal1992}. For this purpose we use the Automated Loess Regression (\textsc{alr}) code\footnote{\url{https://github.com/olrodrig/ALR}} \citep{2019MNRAS.483.5459R}, which performs \textsc{loess} regressions to the input photometry, taking into account observed and intrinsic errors, along with the presence of possible outliers. Figure~\ref{fig:ALR_fit} shows, as an example, the \textsc{alr} fit applied to the $B$-band photometry of SN~IIb~1996cb \citep{1999AJ....117..736Q}.

\subsection{Host galaxy distance moduli}\label{sec:mu}
Translating the observed light curve fluxes to luminosities naturally requires host-galaxy distances. We estimate host-galaxy distance moduli ($\mu$) proceeding in the same way as in \citet{2021MNRAS.505.1742R}. Here we briefly summarize the general procedure.

We collect Cepheids period-luminosity distances ($\mu_\mathrm{CPL}$) and Tip of the Red Giant Branch distances ($\mu_\mathrm{TRGB}$) from the literature, along with Tully-Fisher distances ($\mu_\mathrm{TF}$) from the Extragalactic Distance Database\footnote{\url{http://edd.ifa.hawaii.edu/}\label{edd_web}} \citep[EDD,][]{2009AJ....138..323T}. Then, we adopt the weighted average of $\mu_\mathrm{CPL}$, $\mu_\mathrm{TRGB}$, and $\mu_\mathrm{TF}$ as $\mu$ and the weighted average error as the distance uncertainty $\sigma_\mu$. If $\mu_\mathrm{CPL}$ and $\mu_\mathrm{TRGB}$ are not available, then we include distances calculated from recessional redshifts ($\mu_z$) assuming a local Hubble-Lema\^{i}tre constant ($H_0$) of ${74.03\pm1.42}$\,km\,s$^{-1}$\,Mpc$^{-1}$ \citep{2019ApJ...876...85R}, ${\Omega_\mathrm{m}=0.27}$, ${\Omega_\Lambda=0.73}$, and a velocity dispersion of 382\,km\,s$^{-1}$ \citep{2006ApJ...641...50W} to account for the effect of peculiar velocities over $\mu_z$. We also include distances computed with distance-velocity calculators based on smoothed velocity fields ($\mu_\mathrm{SVF}$) given by \citet{2017ApJ...850..207S} and \citet{2019MNRAS.488.5438G}.\footnote{These calculators are available on the EDD website and described in \citet{2020AJ....159...67K}.} We then adopt the weighted average of $\mu_\mathrm{TF}$, $\mu_z$, and $\mu_\mathrm{SVF}$ as $\mu$ and the weighted average error as $\sigma_\mu$. In the case of NGC~3938 (the host of SN~2017ein) we include the distance modulus reported in \citet{2019MNRAS.483.5459R} for SN~II~2005ay, which also exploded in NGC~3938.

All of the collected and final adopted distances are reported in Appendix Table~\ref{table:mu_values}. The mean $\mu$ uncertainty is 0.13\,mag.

\subsection{Explosion epochs}\label{sec:texp}
The estimation of $^{56}$Ni masses requires the knowledge of the explosion epochs $t_\mathrm{expl}$. The value of $t_\mathrm{expl}$ can be estimated as the midpoint time ($t_\mathrm{midpoint}$) between the last non-detection ($t_\mathrm{non-det}$) and the first SN detection ($t_\mathrm{detect}$) epochs (e.g. \citealt{2015AA...574A..60T}). This method is useful when $t_\mathrm{non-det}$ is a few days before $t_\mathrm{detect}$, otherwise other methods are necessary to improve the estimation of $t_\mathrm{expl}$. An alternative technique to calculate $t_\mathrm{expl}$ is by using the date of the peak of the $x$-band light curve, $t_x^\peak$, where $x$ represents any photometric filter.\footnote{If the peak related to the shock cooling is observed, then $t_x^\peak$ is the date of the second peak.}
In this method, the quantity $t_x^\peak-t_\mathrm{expl}$ for each SN subtype is assumed to be a constant, whose value is determine with a sample of SNe with well-constrained explosion epochs \citep[e.g.][]{2018AA...609A.136T}. However, for SNe~Ic, \citet{2019AA...621A..71T} and \citet{2021AA...651A..81B} found a correlation between the increase in magnitude during the 10\,d before the peak and the decrease in magnitude 15\,d after the peak ($\Delta m_{15}(x)$, the decline rate). We therefore expect that, at least for SNe~Ic, $t_x^\peak-t_\mathrm{expl}$ is not constant but varies with $\Delta m_{15}(x)$. Indeed, as we show in this section, for SNe~Ib and Ic there is a correlation between both quantities for the $V$-band. This correlation provides a method to estimate explosion epochs relying not only on $t_V^\peak$ but also on $\Delta m_{15}(V)$.

We use our \textsc{alr} light-curve fits (Section~\ref{sec:LC_fits}) to compute $t_x^\peak$ and $\Delta m_{15}(x)$ for the various optical bands, which are listed in Table~\ref{table:tmax} and Table~\ref{table:Dm15}, respectively. The $t_\mathrm{non-det}$ and $t_\mathrm{detect}$ values are collected in Table~\ref{table:t_values}.

To estimate $t_\mathrm{expl}$, we use the $V$-band as a fiducial filter and proceed as follows. First, we calibrate the correlation between $t_V^\peak-t_\mathrm{expl}$ and $\Delta m_{15}(V)$ using SNe with $t_\mathrm{detect}-t_\mathrm{non-det}$ lower than 8\,d and adopting $t_\mathrm{midpoint}$ as a first approximation for $t_\mathrm{expl}$. Next, we use these correlations along with $t_V^\peak$ and $\Delta m_{15}(V)$ to estimate a second approximation for explosion epochs, $t_\mathrm{expl}^\peak$. Then, to include the constraint that the explosion epoch cannot be longer (shorter) than $t_\mathrm{detect}$ ($t_\mathrm{non-det}$), for each SN we randomly generate $10^6$ values from the normal distribution $\mathcal{N}(t_\mathrm{expl}^\peak, \sigma_{t_\mathrm{expl}^\peak})$ and select those values between $t_\mathrm{non-det}$ and $t_\mathrm{detect}$. Finally, we adopt the average of the selected values as our best estimate of $t_\mathrm{expl}$. The detailed process used to compute our best estimates for $t_V^\peak$, $\Delta m_{15}(V)$, and $t_\mathrm{expl}$ is presented below.

\subsubsection{Peak epochs}
Figure~\ref{fig:tmax_comparison} shows the cumulative distribution functions (CDFs) for the $t_x^\peak-t_V^\peak$ values, while Table~\ref{table:xi_x} summarises their average ($\xi_x$), $\sigma_{0,x}$, and $\ssd$ values. We see that $t_x^\peak$ tends to increase as the $x$-band effective wavelength increases, which was previously reported by \citet{2015AA...574A..60T,2018AA...609A.136T}. The reported $\xi_x$ values are such that $t_x^\peak=t_V^\peak+\xi_x$, so each $t_x^\peak$ measurement provides an independent estimate of $t_V^\peak$. For each SN, we combine its $t_V^\peak$ estimates into the single observable
\begin{equation}
\langle t_V^\peak\rangle = \langle t_x^\peak-\xi_x\rangle,
\end{equation}
where angle brackets denote a weighted average with weights $w_x=1/(\sigma_{t_x^\peak}^2+\sigma_{0,x}^2)$. The $\langle t_V^\peak\rangle$ values and their weighted-average errors are listed in Column~11 of Table~\ref{table:tmax}. Given that $\langle t_V^\peak\rangle$ is our best estimate of $t_V^\peak$, we adopt it as the final $t_V^\peak$.

\begin{figure}
\includegraphics[width=1.0\columnwidth]{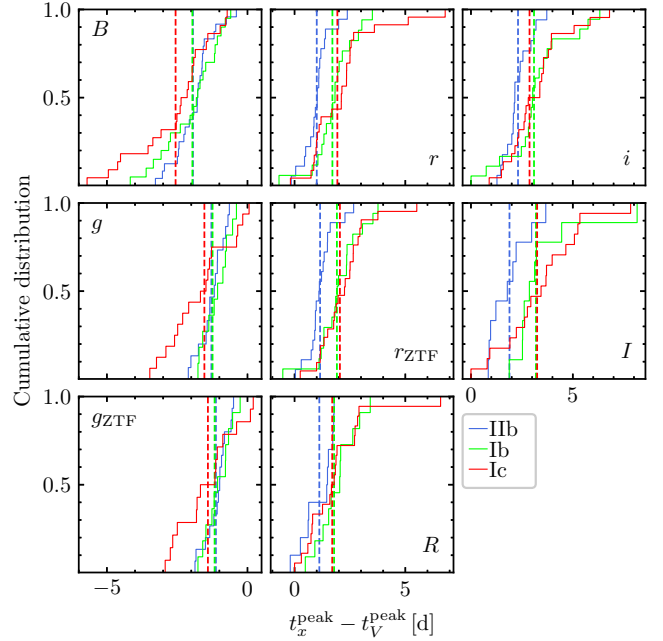}
\caption{Cumulative distributions for the $t_x^\peak-t_V^\peak$ values. Dashed lines indicate mean values.}
\label{fig:tmax_comparison}
\end{figure}

\begin{deluxetable*}{CCCCCCCCCCCCCCC}
\tablecaption{Average $t_x^\peak-t_V^\peak$ values\label{table:xi_x}}
\tablehead{
         & \multicolumn{4}{c}{SNe~IIb} & & \multicolumn{4}{c}{SNe~Ib} & & \multicolumn{4}{c}{SNe~Ic} \\
         \cline{2-5} \cline{7-10} \cline{12-15}
\colhead{$x$}  & \colhead{$\xi_x$ (d)} & \colhead{$\sigma_{0,x}$ (d)} & \colhead{$\ssd$ (d)} & \colhead{$N$} & & \colhead{$\xi_x$ (d)} & \colhead{$\sigma_{0,x}$ (d)} & \colhead{$\ssd$ (d)} &\colhead{$N$} &  & \colhead{$\xi_x$ (d)} & \colhead{$\sigma_{0,x}$ (d)} & \colhead{$\ssd$ (d)} & \colhead{$N$}  
}
\startdata
 B      & -1.9 & 0.6 & 0.7 & 24 & & -2.0 & 0.8 & 1.0 & 20 & & -2.6 & 1.2 & 1.4 & 22 \\
 g      & -1.3 & 0.0 & 0.5 & 15 & & -1.2 & 0.4 & 0.5 & 11 & & -1.5 & 0.9 & 1.1 & 16 \\
 g_\ztf & -1.1 & 0.0 & 0.4 & 15 & & -1.2 & 0.3 & 0.5 & 11 & & -1.4 & 0.8 & 1.0 & 14 \\
 r      &  1.0 & 0.0 & 0.6 & 18 & &  1.7 & 0.7 & 1.0 & 17 & &  1.9 & 1.0 & 1.5 & 23 \\
 r_\ztf &  1.1 & 0.0 & 0.6 & 18 & &  1.9 & 0.8 & 1.0 & 17 & &  2.0 & 0.6 & 1.1 & 21 \\
 R      &  1.1 & 0.6 & 0.7 & 10 & &  1.8 & 0.7 & 0.9 & 11 & &  1.7 & 0.9 & 1.5 & 18 \\
 i      &  2.3 & 0.6 & 0.7 & 17 & &  3.1 & 1.4 & 1.6 & 18 & &  2.9 & 0.8 & 1.4 & 22 \\
 I      &  1.9 & 0.9 & 1.0 &  9 & &  3.2 & 0.0 & 1.9 &  9 & &  3.2 & 1.6 & 2.0 & 17 \\
\enddata
\tablecomments{For $x=V$, $\xi_x=\sigma_{0,x}=0$.}
\end{deluxetable*}

\begin{figure}
\includegraphics[width=1.0\columnwidth]{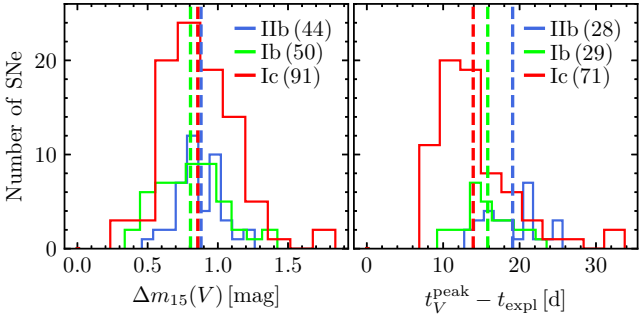}
\caption{Histograms of $\Delta m_{15}(V)$ and $t_V^\peak-t_\mathrm{expl}$. Dashed lines indicate mean values.}
\label{fig:Dm15_tmax_cdf}
\end{figure}

We estimate $t_V^\peak-t_\mathrm{expl}$ for SNe with $t_\mathrm{non-det}$ up to eight days before $t_\mathrm{detect}$, where we adopt $t_\mathrm{expl}=t_\mathrm{midpoint}$. The right-hand panel of Figure~\ref{fig:Dm15_tmax_cdf} shows the histograms for the $t_V^\peak-t_\mathrm{expl}$ values of SNe~IIb, Ib, and Ic. Those distributions have mean ($\ssd$) values of $19.1\,(3.5)$, $15.8\,(3.5)$, and $13.9\,(5.3)$\,d, respectively. This is consistent with the picture that the peak time of SNe~Ic is on average earlier than that of SNe~IIb and Ib \citep[e.g.][]{2011MNRAS.416.3138V,2015AA...574A..60T,2016MNRAS.458.2973P}.

\subsubsection{Decline rates}
\begin{figure*}
\includegraphics[width=0.32\textwidth]{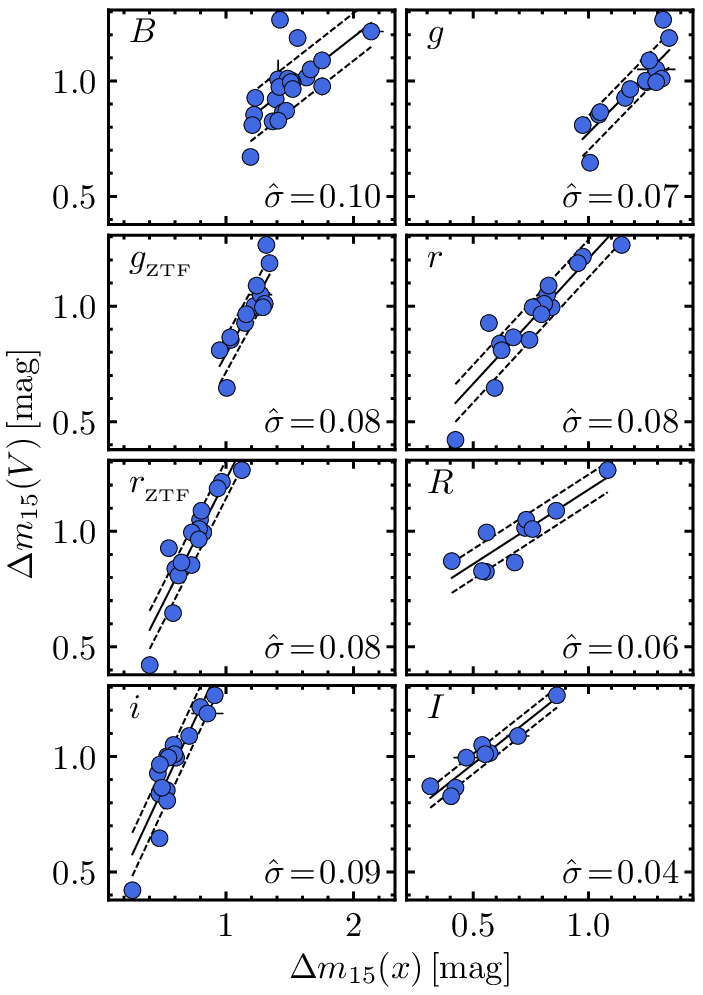}
\includegraphics[width=0.32\textwidth]{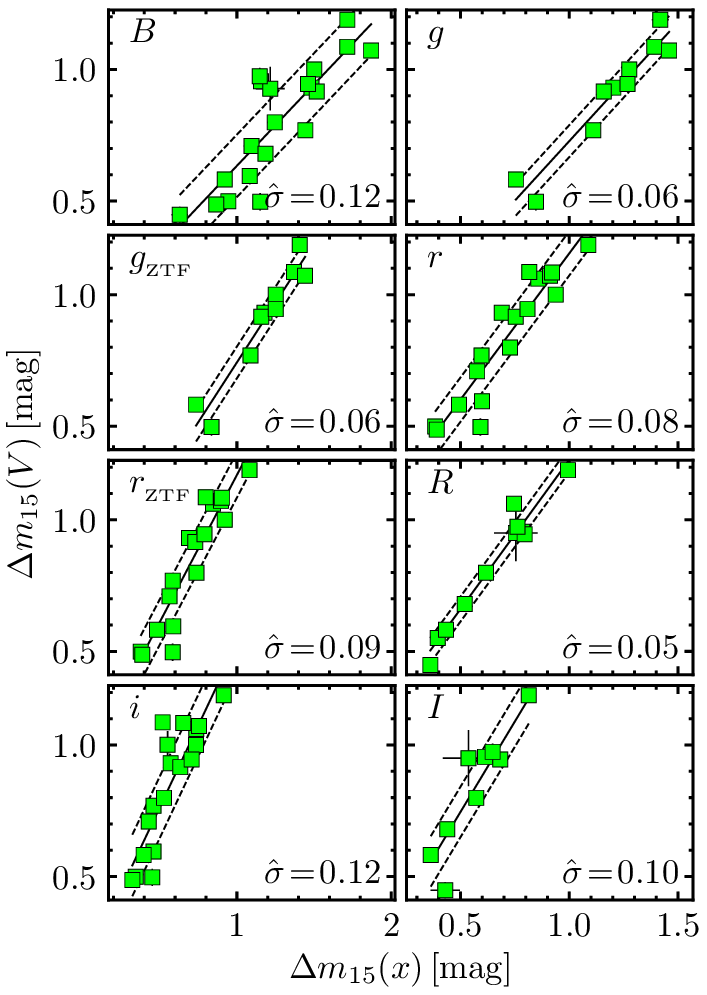}
\includegraphics[width=0.32\textwidth]{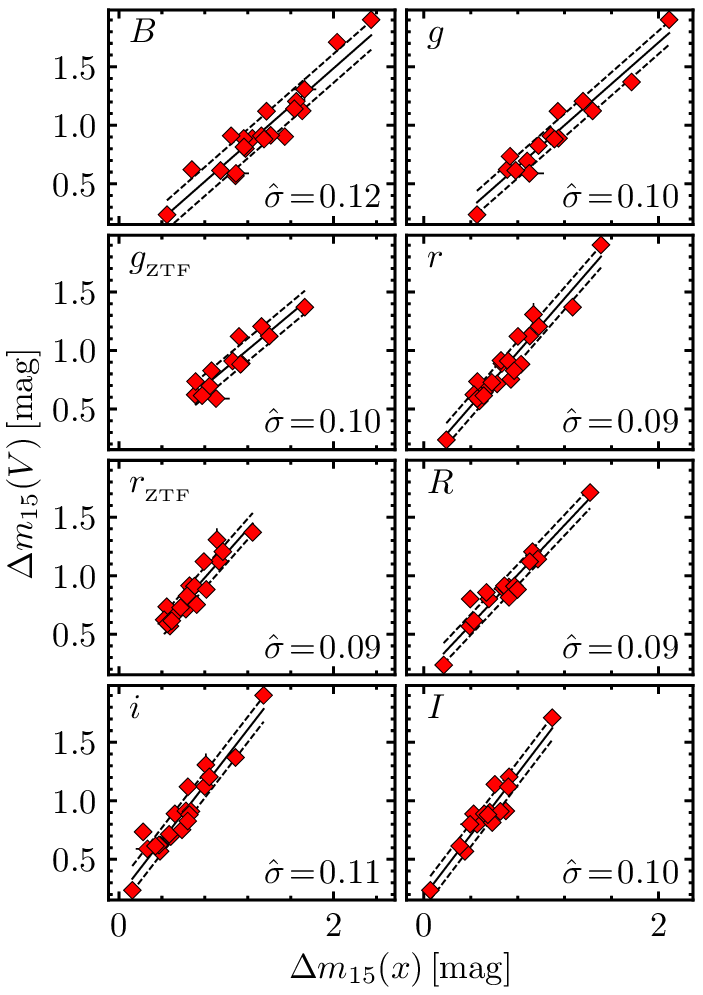}
\caption{$\Delta m_{15}(V)$ as a function of $\Delta m_{15}(x)$ for SNe~IIb (blue circles), Ib (green squares), and Ic (red diamonds). Solid lines are linear fits to the data, and dashed lines are the $\pm1\,\ssd$ limits around the fits. Error bars indicate $\pm1\,\sigma$ errors.}
\label{fig:Dm15_comparison}
\end{figure*}

Figure~\ref{fig:Dm15_comparison} shows $\Delta m_{15}(V)$ against $\Delta m_{15}(x)$ for optical bands. To describe the correlation between both quantities, we adopt a linear fit
\begin{equation}\label{eq:DV15_Dm15}
\Delta m_{15}(V)=a_x+b_x\Delta m_{15}(x),
\end{equation}
where $a_x$, $b_x$, $\sigma_{0,x}$ (calculated with equation~\ref{eq:lnL}) and $\ssd$ values are collected in Table~\ref{table:DV15_vs_Dm15}. Similar to $\langle t_V^\peak\rangle$, for each SN we compute the weighted average of the $\Delta m_{15}(V)$ values obtained from different bands, i.e.,
\begin{equation}
\langle\Delta m_{15}(V)\rangle = \langle a_x+b_x\Delta m_{15}(x)\rangle,
\end{equation} with weights $w_x=1/(b_x^2\sigma_{\Delta m_{15}(x)}^2+\sigma_{0,x}^2)$. The $\langle\Delta m_{15}(V)\rangle$ values and their weighted-mean errors are summarized in Column~11 of Table~\ref{table:Dm15}. Similar to $\langle t_V^\peak\rangle$, we adopt $\langle\Delta m_{15}(V)\rangle$ as the final $\Delta m_{15}(V)$.

\begin{deluxetable*}{cccccccccccccccccc}
\tablecaption{Parameters of the correlation between $\Delta m_{15}(V)$ and $\Delta m_{15}(x)$\label{table:DV15_vs_Dm15}}
\tablehead{
         & \multicolumn{5}{c}{SNe~IIb} & & \multicolumn{5}{c}{SNe~Ib} & & \multicolumn{5}{c}{SNe~Ic} \\
         \cline{2-6} \cline{8-12} \cline{14-18}
\colhead{$x$} & \colhead{$a_x$} & \colhead{$b_x$} & \colhead{$\sigma_{0,x}$} & \colhead{$\ssd$} & \colhead{$N$} & \nocolhead{} &\colhead{$a_x$} & \colhead{$b_x$} & \colhead{$\sigma_{0,x}$} & \colhead{$\ssd$} & \colhead{$N$} & \nocolhead{} & \colhead{$a_x$} & \colhead{$b_x$} & \colhead{$\sigma_{0,x}$} & \colhead{$\ssd$} & \colhead{$N$} 
}
\startdata   
 $B     $ &  $ 0.33$ & $0.43$ & $0.10$ & $0.10$ & $23$ &  & $ 0.01$ & $0.62$ & $0.11$ & $0.12$ & $20$ &  & $-0.12$ & $0.81$ & $0.12$ & $0.12$ & $21$ \\
 $g     $ &  $-0.25$ & $1.02$ & $0.07$ & $0.07$ & $14$ &  & $-0.18$ & $0.91$ & $0.05$ & $0.06$ & $10$ &  & $-0.06$ & $0.88$ & $0.07$ & $0.10$ & $15$ \\
 $g_\ztf$ &  $-0.21$ & $1.00$ & $0.06$ & $0.08$ & $14$ &  & $-0.17$ & $0.91$ & $0.04$ & $0.06$ & $10$ &  & $ 0.09$ & $0.77$ & $0.07$ & $0.10$ & $13$ \\
 $r     $ &  $ 0.12$ & $1.08$ & $0.07$ & $0.08$ & $17$ &  & $ 0.04$ & $1.11$ & $0.08$ & $0.08$ & $17$ &  & $ 0.06$ & $1.15$ & $0.08$ & $0.10$ & $22$ \\
 $r_\ztf$ &  $ 0.14$ & $1.08$ & $0.07$ & $0.08$ & $17$ &  & $ 0.05$ & $1.12$ & $0.07$ & $0.09$ & $17$ &  & $ 0.15$ & $1.04$ & $0.06$ & $0.09$ & $20$ \\
 $R     $ &  $ 0.54$ & $0.64$ & $0.06$ & $0.07$ & $10$ &  & $ 0.09$ & $1.14$ & $0.04$ & $0.05$ & $11$ &  & $ 0.15$ & $1.06$ & $0.07$ & $0.09$ & $17$ \\
 $i     $ &  $ 0.26$ & $1.19$ & $0.08$ & $0.09$ & $17$ &  & $ 0.14$ & $1.25$ & $0.11$ & $0.12$ & $18$ &  & $ 0.19$ & $1.19$ & $0.09$ & $0.11$ & $21$ \\
 $I     $ &  $ 0.57$ & $0.79$ & $0.03$ & $0.04$ & $ 9$ &  & $ 0.06$ & $1.38$ & $0.04$ & $0.10$ & $ 9$ &  & $ 0.18$ & $1.31$ & $0.10$ & $0.10$ & $16$ \\
\enddata
\tablecomments{For $x=V$, $a_x=\sigma_{0,x}=0$ and $b_x=1.0$.}
\end{deluxetable*}

The left-hand panel of Figure~\ref{fig:Dm15_tmax_cdf} shows the histograms for the $\Delta m_{15}(V)$ estimates. The distributions for SNe~IIb, Ib, and Ic have mean ($\ssd$) values of $0.88\,(0.16)$, $0.81\,(0.24)$, and $0.86\,(0.27)$\,mag, respectively. Given the apparent similarity between the distributions, we use the $k$-sample Anderson-Darling (AD) test \citep[e.g.][]{Scholz_Stephen1987} to test whether the $\Delta m_{15}(V)$ samples of SNe~IIb, Ib, and Ic are drawn from a common unspecified distribution (the null hypothesis). We obtain a standardized test statistics ($T_\mathrm{AD}$) of $1.87$ with an AD $p$-value ($p_\mathrm{AD}$) of 0.05, meaning that the null hypothesis cannot be rejected at the 5\% significance level. This is consistent with the study by \citet{2015AA...574A..60T}, which found similar $\Delta m_{15}(r/R)$ and $\Delta m_{15}(g)$ distributions for SNe~Ib and Ic.

\subsubsection{Peak epoch versus decline rate}
\begin{figure}
\includegraphics[width=1.0\columnwidth]{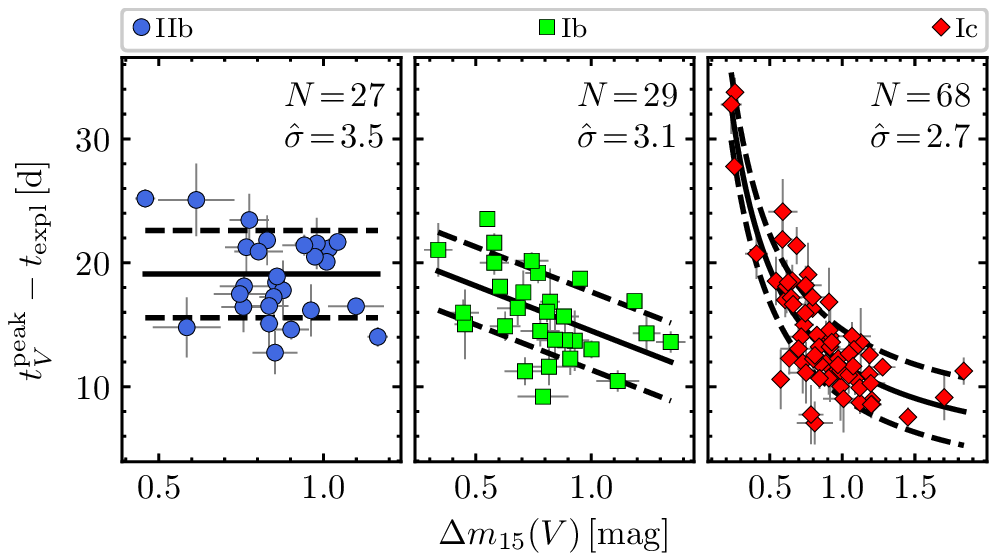}
\caption{$t_V^\peak-t_\mathrm{expl}$ versus $\Delta m_{15}(V)$. Solid lines are the best fits to the data, while dashed lines are the $\pm1\,\ssd$ limits. Error bars are 68\% errors.} 
\label{fig:tVmax}
\end{figure}

Figure~\ref{fig:tVmax} shows $t_V^\peak-t_\mathrm{expl}$ as a function of $\Delta m_{15}(V)$ for SNe with $t_\mathrm{detect}-t_\mathrm{non-det}<8$\,d, for which we adopt $t_\mathrm{expl}=t_\mathrm{midpoint}$. For SNe~IIb we find that $ t_V^\peak-t_\mathrm{expl}$ is consistent with a constant value, so we adopt $t_V^\peak-t_\mathrm{expl}=19.1\pm3.5$\,d. On the other hand, for SNe~Ib and Ic we find a dependence of $t_V^\peak-t_\mathrm{expl}$ on $\Delta m_{15}(V)$, which we express as
\begin{equation}\label{eq:tp_vs_Dm15_Ib}
t_V^\peak-t_\mathrm{expl}\,[\mathrm{d}] = 21.8-7.3\, \Delta m_{15}(V)
\end{equation}
($\ssd=3.1$\,d) for SNe~Ib, and
\begin{equation}\label{eq:tp_vs_Dm15_Ic}
t_V^\peak-t_\mathrm{expl}\,[\mathrm{d}] = 3.0+\frac{9.4}{ \Delta m_{15}(V)}-\frac{0.6}{ \Delta m_{15}(V)^{2}}
\end{equation}
($\ssd=2.7$\,d) for SNe~Ic. 

We use these $t_V^\peak-t_\mathrm{expl}$ calibrations to convert $t_V^\peak$ into $t_\mathrm{expl}^\peak$, which are listed along with their errors in Column~5 of Table~\ref{table:t_values}. For SNe~Ic~2006nx, 2013F, 2013dk, 2014ft, and 2021bm, which do not have $\Delta m_{15}(V)$ estimates, we adopt $t_V^\peak-t_\mathrm{expl}=13.9\pm5.3$\,d (the mean value of the distribution).

\begin{figure}
\includegraphics[width=1.0\columnwidth]{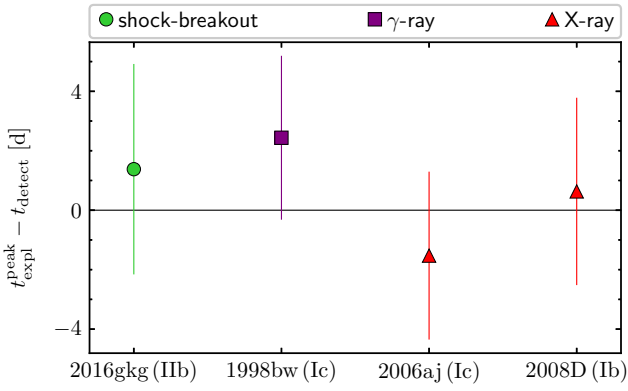}
\caption{Differences between $t_\mathrm{expl}^\peak$ and the first detection epoch for SNe~1998bw ($\gamma$-ray detection), 2006aj and 2008D (X-ray detection), and 2016gkg (shock-breakout detection). Error bars are $1\,\sigma_{t_\mathrm{expl}^\peak}$ errors.}
\label{fig:t_comparison}
\end{figure}

To evaluate whether the $t_V^\peak-t_\mathrm{expl}$ calibrations provide reasonable values for explosion epochs, in Figure~\ref{fig:t_comparison} we show differences between $t_\mathrm{expl}^\peak$ and $t_\mathrm{detect}$ for SNe~1998bw, 2006aj, 2008D, and 2016gkg. For these SNe, $t_\mathrm{detect}$ corresponds to the shock-breakout detection (for SN~2016gkg), the first $\gamma$-ray detection (for SN~1998bw), or the first X-ray detection (for SNe~2006aj and 2008D), thus $t_\mathrm{detect}$ is the best estimate for explosion epoch. In Figure~\ref{fig:t_comparison} we see that the $t_\mathrm{expl}^\peak-t_\mathrm{detect}$ value of each SN is consistent with zero within $1\,\sigma_{t_\mathrm{expl}^\peak}$. In addition, the average ($\ssd$) of the $t_\mathrm{expl}^\peak-t_\mathrm{detect}$ estimates is of 0.7\,d (1.7\,d), so $t_\mathrm{expl}^\peak$ is statistically consistent with $t_\mathrm{detect}$ to $1\,\ssd/\sqrt{N}$. Therefore, even we do not know the exact explosion epoch for almost all SNe in our sample, we at least do not find systematic differences between our $t_\mathrm{expl}^\peak$ estimates and the explosion epochs for those SNe with well-known $t_\mathrm{expl}$.

To include the constraint provided by $t_\mathrm{non-det}$ and $t_\mathrm{detect}$ on the explosion epoch, for each SN we randomly generate $10^6$ values from the normal distribution $\mathcal{N}(t_\mathrm{expl}^\peak, \sigma_{t_\mathrm{expl}^\peak})$. We then select the values between $t_\mathrm{non-det}$ and $t_\mathrm{detect}$, and compute the mean and $\ssd$ value, which we adopt as the final $t_\mathrm{expl}$ and its error, respectively. Those estimates are reported in Column~6 of Table~\ref{table:t_values}. The mean $t_\mathrm{expl}$ error for our SN sample is 1.4\,d.

\subsection{Host galaxy reddenings}\label{sec:EhBV}
To obtain luminosities, light curves must be corrected for the effects of dust extinction and reddening. In this work, we use color curves to compute $E_{B-V}$, and employ those estimates to calibrate the relation between $E_{B-V}$ and $\mathrm{EW_{NaID}}$ for SE~SNe.

\subsubsection{Color curve shapes}
For a given SN, we model its $c=x-y$ (where $x$ and $y$ represent any two photometric bands) color curve (corrected for MW reddening and $K$-correction, see Appendix~\ref{sec:AK_correction}) as
\begin{equation}
c(t) = \delta_c+\Psi_c(t- t_y^\peak),
\end{equation}
where $\delta_c$ is the vertical intercept of the color curve, and $\Psi_c$ is a polynomial representing the dependence of $c$ on $t- t_y^\peak$ (i.e. the shape of the color curve). Under the assumption that all SNe of a given subtype have the same intrinsic color curve given by
\begin{equation}\label{eq:c0}
c_0(t) = \delta_{0,c}+\Psi_{0,c}(t- t_y^\peak),
\end{equation}
then $\Psi_c=\Psi_{0,c}$ and
\begin{equation}\label{eq:delta_c}
\delta_c=\delta_{0,c}+E_{B-V} R_c.
\end{equation}
Here, $R_c=E_{x-y}/E_{B-V}$, which depends on the $R_V$ value characterising a host galaxy's extinction along the SN line of sight. The $\delta_c$ value of each color is therefore an indicator of $E_{B-V}$. In this work we use the independent colors $\bv$, $g-i$, $V\!-r$, $V\!-\!R$, $V\!-i$, $V\!-I$, and $(g-r)_\ztf$.

To measure $\delta_c$ for each SN in a sample of size $N$ along with the parameters of $\Psi_{0,c}$, we minimize
\begin{equation}\label{eq:s2}
s^2=\sum_j^N\sum_k^{M_j>1}\left[Y_{j,k}-\delta_{Y,j}-\Psi_{0,Y}(X_{j,k})\right]^2.
\end{equation}
Here, $Y=c$, $X=t- t_y^\peak$, $M_j$ is the number of observations for the $j$-th SN, while the polynomial order of $\Psi_{0,Y}$ is determined with the BIC. To estimate $\delta_c$ we use data in the time range between $t_y^\peak$ and $t_y^\peak+t_c^\mathrm{max}$, where $t_c^\mathrm{max}$ is the time where the color curves reach their maximum value. We choose this time interval because color curves are found to be similar at such epochs \citep[e.g.][]{2018AA...609A.135S}, and because $\Psi_{0,c}$ can then be represented by a low-order polynomial. For a given SN, the error on $\delta_c$ is
\begin{equation}
\sigma_{\delta_c}=\sqrt{\ssd_c^2+[\langle\dot{\Psi}_{0,c}\rangle\sigma_{t_y^\peak}]^2},
\end{equation}
where $\ssd_c$ is the $\ssd$ dispersion of the $c-\delta_c$ values around $\Psi_{0,c}$, while $\langle\dot{\Psi}_{0,c}\rangle$ is the average of the time derivative of $\Psi_{0,c}$.

\begin{figure}
\includegraphics[width=1.0\columnwidth]{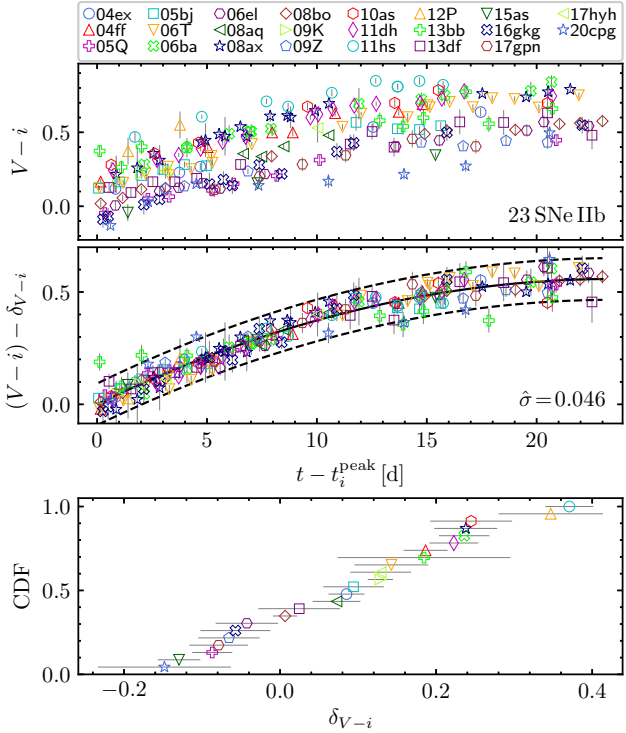}
\caption{$V\!-i$ color curves of SNe~IIb corrected for MW reddening and $K$-correction (top panel), and shifted by $\delta_{V\!-i}$ (middle panel). The solid line is a quadratic fit, corresponding to $\Psi_{0,V\!-i}$, while dashed lines indicate the $\pm2\,\ssd$ limits. Bottom panel: cumulative distribution for the $\delta_{V\!-i}$ values. Error bars are $1\,\sigma$ errors.} 
\label{fig:BV_IIb}
\end{figure}

The top panel of Figure~\ref{fig:BV_IIb} shows the $V\!-i$ color curves of 23~SNe~IIb in our sample, with $t-t_i^\peak$ between zero and $t_{V\!-i}^\mathrm{max}=23.3$\,d. The observed spread is due to the host galaxy reddenings of each SN and the intrinsic diversity of the $V\!-i$ color curves. The $\Psi_{0,V\!-i}$ curve and the $\delta_{V\!-i}$ values, obtained from the minimization of equation~(\ref{eq:s2}), are shown in the middle and bottom panel of Figure~\ref{fig:BV_IIb}, respectively. We find that $\Psi_{0,V\!-i}$ is quadratic in $t-t_i^\peak$ (solid line), where the $\ssd$ value of 0.046\,mag corresponds to the minimum error on $V\!-i$ induced by random errors and intrinsic differences. 

\begin{figure}
\includegraphics[width=1.0\columnwidth]{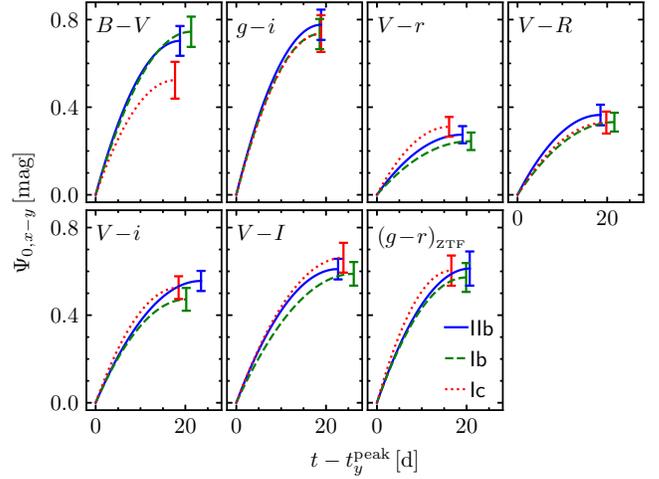}
\caption{$\Psi_{0,c}$ curves for different colors of SNe~IIb (blue solid lines), Ib (green dashed lines), and Ic (red dotted lines). Error bars indicate the $\pm1\,\ssd$ limits around the curves.} 
\label{fig:Psi_c}
\end{figure}

For all the independent colors we find a quadratic dependence of $\Psi_{0,c}$ on $t-t_y^\peak$, given by
\begin{equation}\label{eq:Psi_0}
\Psi_{0,c}=a_c\left(\frac{t-t_y^\peak}{10\,\mathrm{d}}\right)+b_c\left(\frac{t-t_y^\peak}{10\,\mathrm{d}}\right)^2.
\end{equation}
The $a_c$, $b_c$, and $t_c^\mathrm{max}$ values for different colors and SN subtypes are listed in Table~\ref{table:Psi_c}, while Figure~\ref{fig:Psi_c} shows the $\Psi_{0,c}$ curves. The $\delta_c$ estimates are collected in \ref{table:delta}. 

\begin{deluxetable*}{ccccccccccccccccccc}
\tablecaption{Color curve shape parameters\label{table:Psi_c}}
\tablehead{
 & \multicolumn{5}{c}{SNe~IIb} & & \multicolumn{5}{c}{SNe~Ib} & & \multicolumn{5}{c}{SNe~Ic} \\
 \cline{2-6} \cline{8-12} \cline{14-18}
\colhead{$c$} & \colhead{$a_c$} & \colhead{$b_c$}   & \colhead{$\ssd_c$} & \colhead{${t_c^\mathrm{max}}$} & \colhead{$N$} & & \colhead{$a_c$} & \colhead{$b_c$}   & \colhead{$\ssd_c$} & \colhead{${t_c^\mathrm{max}}$} & \colhead{$N$} & & \colhead{$a_c$} & \colhead{$b_c$}   & \colhead{$\ssd_c$} & \colhead{${t_c^\mathrm{max}}$} & \colhead{$N$}
   }
\startdata
 $B\!-\!V$        & $0.748$ & $-0.199$ & $0.068$ & $18.8$ & $30$ & & $0.701$ & $-0.165$ & $0.069$ & $21.3$ & $25$ & & $0.598$ & $-0.170$ & $0.083$ & $17.6$ & $39$ \\
 $g\!-\!i$        & $0.821$ & $-0.217$ & $0.069$ & $18.9$ & $21$ & & $0.789$ & $-0.212$ & $0.069$ & $18.6$ & $19$ & & $0.786$ & $-0.219$ & $0.083$ & $18.0$ & $28$ \\
 $V\!-\!r$        & $0.287$ & $-0.075$ & $0.039$ & $19.1$ & $23$ & & $0.234$ & $-0.056$ & $0.040$ & $21.0$ & $19$ & & $0.385$ & $-0.121$ & $0.046$ & $15.9$ & $30$ \\
 $V\!-\!R$        & $0.395$ & $-0.107$ & $0.047$ & $18.5$ & $11$ & & $0.307$ & $-0.071$ & $0.043$ & $21.6$ & $12$ & & $0.332$ & $-0.085$ & $0.050$ & $19.6$ & $20$ \\
 $V\!-\!i$        & $0.474$ & $-0.101$ & $0.046$ & $23.5$ & $23$ & & $0.466$ & $-0.115$ & $0.052$ & $20.2$ & $19$ & & $0.558$ & $-0.151$ & $0.055$ & $18.5$ & $30$ \\
 $V\!-\!I$        & $0.539$ & $-0.119$ & $0.047$ & $22.7$ & $10$ & & $0.450$ & $-0.086$ & $0.054$ & $26.2$ & $10$ & & $0.547$ & $-0.115$ & $0.070$ & $23.7$ & $19$ \\
 $(g\!-\!r)_\ztf$ & $0.590$ & $-0.142$ & $0.078$ & $20.7$ & $36$ & & $0.576$ & $-0.145$ & $0.066$ & $19.9$ & $37$ & & $0.711$ & $-0.211$ & $0.070$ & $16.9$ & $75$ \\
\enddata
\end{deluxetable*}

\subsubsection{$E_{B-V}$ and representative $R_V$ values}
For each SN subtype, using $N$ SNe with $\delta_c$ values in more than one color, we can compute the $\delta_{0,c}$ values, the $E_{B-V}$ estimates for each SN, and the representative host galaxy $R_V$ maximizing equation~(\ref{eq:posterior_prob}), where the log-likelihood of equation~(\ref{eq:delta_c}) is given by
\begin{eqnarray}\label{eq:lnL_delta}
\ln\mathcal{L}=&-&\frac{1}{2}\sum_{j=1}^N \sum_c^{\{c\}_j} \left[\ln(\sigma_{\delta_{c,j}}^2+\sigma_{0,c}^2)\right.\nonumber\\
               &+& \left.\frac{[\delta_{c,j}-\tilde{\delta}_{0,c}-\tilde{E}_{{B-V},j} R_c(R_V)]^2}{\sigma_{\delta_{c,j}}^2+\sigma_{0,c}^2}\right].
\end{eqnarray}
Here, $\{c\}_j$ are the available colors for the $j$-th SN, and $\sigma_{0,c}$ is the uncertainty not accounted for in the error on $\delta_{c,j}$. In the last equation we use $\tilde{E}_{B-V}$ and $\tilde{\delta}_{0,c}$ instead of $E_{B-V}$ and $\delta_{0,c}$ because there is a degeneracy between both parameters. Indeed, we can express $E_{B-V}$ and $\delta_{0,c}$ as
\begin{equation}\label{eq:EhBV}
E_{B-V}=\tilde{E}_{B-V}+\mathrm{ZP}_{E_{B-V}}
\end{equation}
and
\begin{equation}\label{eq:delta_0}
\delta_{0,c}=\tilde{\delta}_{0,c}-R_c\,\mathrm{ZP}_{E_{B-V}}
\end{equation}
such that $\delta_{0,c}+E_{B-V} R_c=\tilde{\delta}_{0,c}+\tilde{E}_{B-V} R_c$. The constant $\mathrm{ZP}_{E_{B-V}}$ corresponds to the zero-point for the reddening scale.

Table~\ref{table:delta_0} collects $\tilde{\delta}_{0,c}$, $\sigma_{0,c}$, and the representative $R_V$ values obtained from the maximization of equation~(\ref{eq:posterior_prob}). The $\tilde{E}_{B-V}$ values and their weighted mean errors, given by
\begin{equation}
\sigma_{\tilde{E}_{B-V}}=\left[\sum_c R_c^2/(\sigma_{\delta_c}^2+\sigma_{0,c}^2)\right]^{-1/2},
\end{equation}
are listed in Column~4 of Table~\ref{table:EhBV}. For those SNe with only one $\delta_c$ measurement, we use $\tilde{E}_{B-V}=(\delta_c-\tilde{\delta}_{0,c})/R_c$.

\begin{deluxetable*}{cccccccccccc}
\tablecaption{Results of the posterior probability maximization for $\delta_c$\label{table:delta_0}}
\tablehead{
 \nocolhead{}       & \multicolumn{3}{c}{SNe IIb ($N=27$)} & \nocolhead{} & \multicolumn{3}{c}{SNe Ib ($N=30$)} & \nocolhead{} & \multicolumn{3}{c}{SNe Ic ($N=44$)} \\
 \nocolhead{}       & \multicolumn{3}{c}{$R_V=2.6\pm0.4$} &  \nocolhead{} & \multicolumn{3}{c}{$R_V=2.7\pm0.5$} &  \nocolhead{} & \multicolumn{3}{c}{$R_V=3.8\pm0.4$} \\
 \cline{2-4} \cline{6-8} \cline{10-12}
\colhead{$c$} & \colhead{$\tilde{\delta}_{0,c}$} & \colhead{$\sigma_{0,c}$} & \colhead{$R_c(R_V)$} & \nocolhead{} & \colhead{$\tilde{\delta}_{0,c}$} & \colhead{$\sigma_{0,c}$}  & \colhead{$R_c(R_V)$} & \nocolhead{} & \colhead{$\tilde{\delta}_{0,c}$} & \colhead{$\sigma_{0,c}$} & \colhead{$R_c(R_V)$} 
}
\startdata  
$B\!-\!V$        & $ 0.0  $ & $0.032$ & $1.00$ & & $ 0.0  $ & $0.115$ & $1.00$ & & $ 0.0  $ & $0.177$ & $1.00$ \\
$g\!-\!i$        & $-0.525$ & $0.0  $ & $1.69$ & & $-0.447$ & $0.0  $ & $1.71$ & & $-1.162$ & $0.0  $ & $1.93$ \\
$V\!-\!r$        & $-0.132$ & $0.020$ & $0.46$ & & $-0.130$ & $0.047$ & $0.46$ & & $-0.298$ & $0.048$ & $0.47$ \\
$V\!-\!R$        & $-0.077$ & $0.042$ & $0.61$ & & $-0.064$ & $0.027$ & $0.61$ & & $-0.299$ & $0.035$ & $0.66$ \\
$V\!-\!i$        & $-0.466$ & $0.0  $ & $1.00$ & & $-0.420$ & $0.024$ & $1.02$ & & $-0.985$ & $0.0  $ & $1.24$ \\
$V\!-\!I$        & $-0.215$ & $0.050$ & $1.18$ & & $-0.136$ & $0.0  $ & $1.21$ & & $-0.741$ & $0.0  $ & $1.54$ \\
$(g\!-\!r)_\ztf$ & $-0.226$ & $0.049$ & $1.17$ & & $-0.236$ & $0.037$ & $1.17$ & & $-0.604$ & $0.087$ & $1.20$ \\
\enddata
\end{deluxetable*}

For SNe~IIb, Ib, and Ic we find $R_V$ ($\pm1\,\sigma$ error) values of ${2.6\pm0.4}$, ${2.7\pm0.5}$, and ${3.8\pm0.4}$, respectively. Although SNe~Ic seem to have a larger $R_V$ value compared to SNe~IIb and Ib (as previously found by \citealt{2018AA...609A.135S}), this is only at the $\pm2.1\,\sigma$ level, and we cannot exclude on this basis a single $R_V$ for all three types of SNe.

To estimate $\mathrm{ZP}_{E_{B-V}}$ in equation~(\ref{eq:EhBV}), we compare $\tilde{E}_{B-V}$ against $\mathrm{EW_{NaID}}$, which we use as a proxy for reddening. Column~3 of Table~\ref{table:EhBV} lists the $\mathrm{EW_{NaID}}$ values for the SNe in our sample, which we compute using the procedure presented in Appendix~\ref{sec:pEW}. We also include $\mathrm{EW_{NaID}}$ values calculated in the literature from high-resolution spectra. We focus our analysis on SNe with $\mathrm{EW_{NaID}}<1$\,\AA\, since this observable becomes a poor tracer  of reddening for larger values \citep[e.g.][]{2013ApJ...779...38P}, and with $\sigma_\mathrm{EW_{NaID}}<0.2$\,\AA\, to minimize the induced scatter.

\begin{figure}
\includegraphics[width=1.0\columnwidth]{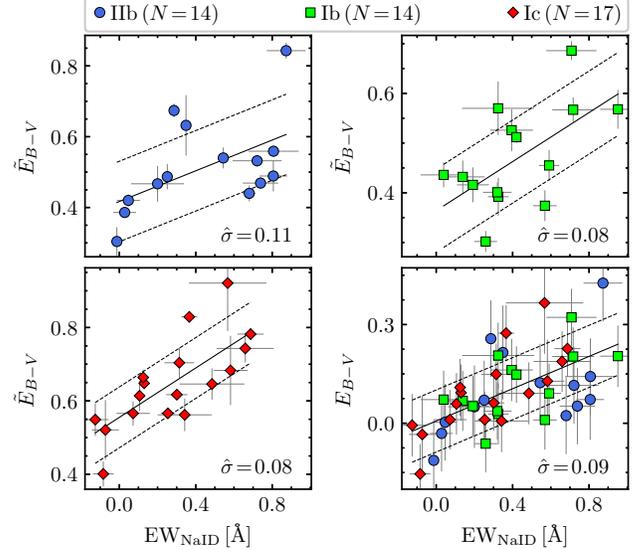}
\caption{$\tilde{E}_{B-V}$ versus $\mathrm{EW_{NaID}}$ for SNe~IIb (upper left-hand panel), Ib (upper right-hand panel), and Ic (bottom left-hand panel). Bottom right-hand panel: host galaxy reddening versus $\mathrm{EW_{NaID}}$. Solid and dashed lines are linear fits and $1\,\ssd$ limits, respectively. Error bars indicate $\pm1\,\sigma$ errors.}
\label{fig:ErBV}
\end{figure}

Figure~\ref{fig:ErBV} shows $\tilde{E}_{B-V}$ against $\mathrm{EW_{NaID}}$ for each SN subtype. We express the dependence of $\tilde{E}_{B-V}$ on $\mathrm{EW_{NaID}}$ as
\begin{equation}\label{eq:Er_vs_NaID}
\tilde{E}_{B-V} = a + b\,\mathrm{EW_{NaID}}[\mathring{\mathrm{A}}],
\end{equation}
where the values of $a$, $b$, $\sigma_0$ (computed maximizing equation~\ref{eq:posterior_prob}), and of  $\ssd$ are reported in Table~\ref{table:zp_E}. To evaluate the linear correlations, we calculate the Pearson correlation coefficient ($r_\mathrm{P}$) and the corresponding null-hypothesis probability $p$-value ($p_\mathrm{P}$). For SNe~IIb, Ib, and Ic we obtain moderate/strong correlations, with $r_\mathrm{P}$ ($p_\mathrm{P}$) values of 0.52 (0.05), 0.57 (0.03), and 0.75 ($<0.01$), respectively. If we assume that $E_{B-V}$ is zero for $\mathrm{EW_{NaID}}=0$, then from equations~(\ref{eq:EhBV}) and (\ref{eq:Er_vs_NaID}) we obtain $\mathrm{ZP}_{E_{B-V}}=-a$. Therefore we adopt $-a$ and $\ssd$ as $\mathrm{ZP}_{E_{B-V}}$ and its error, respectively. The $E_{B-V}$ values computed with equation~(\ref{eq:EhBV}), which we call $E_{B-V}^\mathrm{CC}$ (CC is for color curves), and their errors given by
\begin{equation}
\sigma_{E_{B-V}^\mathrm{CC}}=\sqrt{\sigma_{\tilde{E}_{B-V}}^2+\sigma_{\mathrm{ZP}_{E_{B-V}}}^2}
\end{equation}
are reported in Column~5 of Table~\ref{table:EhBV}.

\begin{deluxetable}{lcccccc}
\tablecaption{$\tilde{E}_{B-V}$ versus $\mathrm{EW_{NaID}}$ parameters\label{table:zp_E}}
\tablehead{
 \colhead{Type} & \colhead{$N$} & \colhead{$a$} & \colhead{$b$} & \colhead{$\sigma_0$} & \colhead{$\ssd$} & \colhead{$b$\tablenotemark{a}}
}
\startdata
IIb & $14$ & $ 0.417(59)$ & $0.217(108)$ & $0.107$ & $0.114$ & $0.206$ \\
Ib  & $14$ & $ 0.364(51)$ & $0.247(101)$ & $0.074$ & $0.084$ & $0.229$ \\
Ic  & $17$ & $ 0.555(30)$ & $0.336( 92)$ & $0.057$ & $0.083$ & $0.376$ \\
\enddata
\tablecomments{Numbers in parentheses are $1\,\sigma$ errors in units of 0.001.}
\tablenotetext{a}{Assuming $R_V=3.1$.}
\end{deluxetable}

\subsubsection{Relation between $E_{B-V}$ and Na\,ID EW}
We note that although the $b$ estimates differ somewhat among the three SE~SN types, they are consistent within the errors, i.e. the relation between host-galaxy reddening and $\mathrm{EW_{NaID}}$ does not appear to depend on SN subtype. Based on this, we use the $E_{B-V}^\mathrm{CC}$ and $\mathrm{EW_{NaID}}$ values of SNe~IIb, Ib, and Ic (bottom right-hand panel of Figure~\ref{fig:ErBV}) to infer the relation between host galaxy reddening and $\mathrm{EW_{NaID}}$. We find a linear fit
\begin{equation}\label{eq:Eh_NaID}
E_{B-V}=0.007(\pm0.024)+0.246(\pm0.054)\,\mathrm{EW_{NaID}}[\mathring{\mathrm{A}}],
\end{equation}
with $\ssd=0.094$\,mag, $r_\mathrm{P}=0.60$, and $p_\mathrm{P}<0.01$.

Using equation~(\ref{eq:Eh_NaID}) we compute host galaxy reddenings for the SNe with $\mathrm{EW_{NaID}}$ measurements ($E_{B-V}^\mathrm{NaID}$), which are listed in Column~6 of Table~\ref{table:EhBV}. We adopt the weighted average of $E_{B-V}^\mathrm{CC}$ and $E_{B-V}^\mathrm{NaID}$ as the final $E_{B-V}$, which are collected in Column~7 of Table~\ref{table:EhBV}. The typical $E_{B-V}$ error is 0.108\,mag.

\subsubsection{Accuracy of the host galaxy reddening scale}\label{sec:EBV_accuracy}
As expected, our $E_{B-V}$ estimates are not very precise due to the photometric and spectroscopic diversity of SE~SNe. On the other hand, the accuracy of the reddening scale depends on the similarity between the SN sample that we use to calculate $\mathrm{ZP}_{E_{B-V}}$ and a complete sample. To evaluate this accuracy, we compare the SN~IIb, Ib, and Ic samples used to compute $\mathrm{ZP}_{E_{B-V}}$ with volume-limited (VL) samples, which we use as approximations for complete samples. For this comparison, we use absolute $r$-band magnitudes at peak ($M_r^\peak$). For 22 (6) SNe without photometry in the $r$ ($r$ and $R$) band, we measure $M_r^\peak$ using $R$ ($V$) band photometry and the mean $r-R$ ($V-r$) color at peak of $0.13\pm0.06$ ($0.04\pm0.07$). The $M_r^\peak$ values are listed in Table~\ref{table:Mrpeak}.

\begin{figure}
\includegraphics[width=1.0\columnwidth]{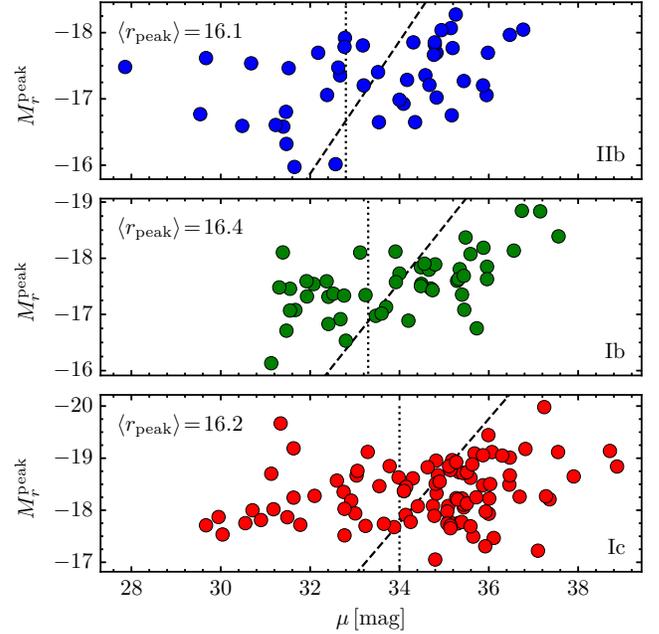}
\caption{$M_r^\peak$ against $\mu$ for SNe~IIb, Ib, and Ic. Dashed lines correspond to $r_\peak=\langle r_\peak\rangle$, and dotted lines are the limits for the volume-limited samples.} 
\label{fig:Miller}
\end{figure}

Figure~\ref{fig:Miller} shows $M_r^\peak$ against $\mu$, where the dashed lines indicate the average apparent $r$-band magnitude at peak, $\langle r_\peak\rangle$. For SNe~IIb and Ib it is evident that there is a reduction in the number of objects as we move to greater $r_\peak$ values (right-hand side of the dashed line), which is a consequence of the Malmquist selection bias. To minimize this bias, we select VL samples with $\mu\leq\mu_\mathrm{VL}$, with $\mu_\mathrm{VL}$ values of 32.8 for SNe~IIb, 33.3 for Ib, and 34.0 for SNe~Ic. By using these $\mu_\mathrm{VL}$ values, the selection bias affecting the VL samples becomes relevant only in the small region with $r_\peak>\langle r_\peak\rangle$. 

\begin{figure}
\includegraphics[width=1.0\columnwidth]{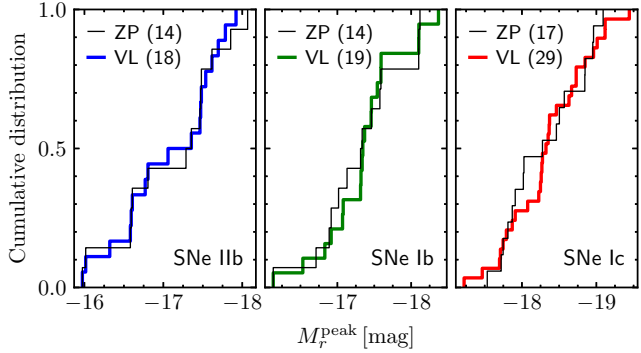}
\caption{Cumulative distributions for the $M_r^\peak$ values in the volume-limited samples (thick lines) and the samples used to compute the zero-points of the reddening scales (thin lines). Numbers in parentheses are the sample sizes.} 
\label{fig:VL_and_calib}
\end{figure}

Figure~\ref{fig:VL_and_calib} shows the cumulative distributions for the $M_r^\peak$ values in the VL samples and those of the SNe used to compute $\mathrm{ZP}_{E_{B-V}}$. The similarity of the distributions for each SN subtype is evident, with $p_\mathrm{AD}>0.92$. Therefore we can expect that our $E_{B-V}$ estimates are not, on average, under- or overestimated.

\subsubsection{Host galaxy reddening distribution}
\begin{figure}
\includegraphics[width=1.0\columnwidth]{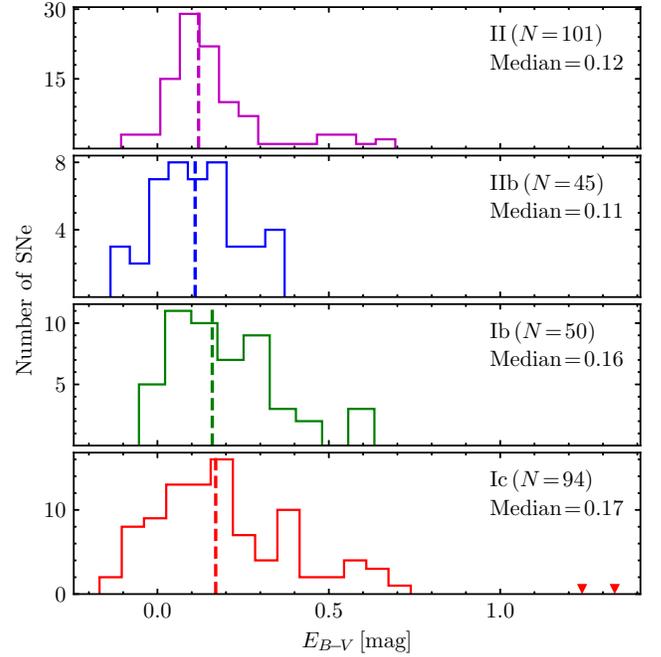}
\caption{Histograms for the $E_{B-V}$ values of the SNe~IIb, Ib, and Ic in our sample, and of the SNe~II in the sample of \citet{2021MNRAS.505.1742R}. Dashed lines indicate median values, while red triangles mark the locations of the outliers.} 
\label{fig:EhBV_hist}
\end{figure}

Figure~\ref{fig:EhBV_hist} shows the histograms of $E_{B-V}$ for SNe~IIb, Ib, and Ic, where for comparison we also include the SNe~II from the sample of \citet{2021MNRAS.505.1742R}. For 26~SE~SNe we obtain negative $E_{B-V}$ values, which range between $-0.005$ and $-0.169$\,mag with a mean of $-0.06$\,mag, and are consistent with zero to within $-1.4\,\sigma_{E_{B-V}}$. Although negative reddenings have no physical meaning, such values reflect the uncertainty of our methodology to measure $E_{B-V}$, which is due mainly to the diversity of SE~SNe. Indeed, negative $E_{B-V}$ values are inferred for SNe in our sample with the bluest intrinsic colors. Similarly, our methodology overestimates $E_{B-V}$ for the SNe with the reddest intrinsic colors. For SNe with negative reddenings, a $E_{B-V}$ of zero is a more appropriate value. However, correcting only the lower boundary of the $E_{B-V}$ distribution by replacing negative reddenings by zero would bias the sample, producing an overestimation of the mean $E_{B-V}$, and therefore an overestimation of the mean $^{56}$Ni masses of SE~SNe. Given that we aim to infer accurate mean $^{56}$Ni and iron yields of SE~SNe, we perform the host galaxy reddening correction keeping the negative $E_{B-V}$ values. For studies of single events, our reported luminosities and $^{56}$Ni masses for SNe with negative $E_{B-V}$ should be considered lower limits.

To identify extreme values in the distributions, we use the \citet{1863mspa.book.....C} criterion. For SNe~Ic we find that SNe~2005kl and 2013F have $E_{B-V}$ values greater than the Chauvenet upper rejection limit ($E_{B-V}=0.93$\,mag), and hence we consider them outliers. The mean, $\ssd$, skewness, and median values of the $E_{B-V}$ distributions for SNe~II, IIb, Ib, and Ic (after removing SNe~2005kl and 2013F) are listed in Table~\ref{table:Eh_stats}. Since the distributions are asymmetric (with positive skewness), we use the median rather than the mean as the representative value.  We see that the median $E_{B-V}$ tends to grow in the sequence of II--IIb--Ib--Ic subtypes.

\begin{deluxetable}{lcccccc}
\tablecaption{Statistics of the $E_{B-V}$ distributions\label{table:Eh_stats}}
\tablehead{
 \colhead{Type} & \colhead{$N$}  & \colhead{Mean}  & \colhead{$\ssd$} & \colhead{Skewness} & \colhead{Median}
}
\startdata
II  & $101$ & $0.16$ & $0.15$ & $1.53$ & $0.12$ \\
IIb & $ 45$ & $0.12$ & $0.13$ & $0.14$ & $0.11$ \\
Ib  & $ 50$ & $0.19$ & $0.16$ & $0.96$ & $0.16$ \\
Ic  & $ 94$ & $0.20$ & $0.20$ & $0.65$ & $0.17$ \\
\enddata
\end{deluxetable}

Using the $k$-sample AD test to compare the $E_{B-V}$ distributions of the \{II,IIb\}, \{II,Ib\}, \{II,Ic\}, \{IIb,Ib\}, \{IIb,Ic\}, and \{Ib,Ic\} samples, we obtain $p_\mathrm{AD}$ values of 0.07, 0.16, 0.02, 0.05, 0.03, and 0.36, respectively. In other words, the $E_{B-V}$ distributions for SNe~II and IIb are statistically similar, the distribution for SNe~Ib is similar to those of SNe~II, IIb, and Ic, while the distribution for SNe~Ic and those for SNe~II and SNe~IIb are different (null hypothesis rejected at the 2--3\% significance level). These findings remain unchanged if we adopt $R_V=3.1$ for SE~SNe. The difference between the $E_{B-V}$ distributions for SNe~IIb and Ic has been previously reported by \citet{2018AA...609A.135S}. For the $E_{B-V}$ distributions of the \{II,IIb,Ib\} sample we find a $p_\mathrm{AD}=0.04$, which increases to 0.05 if we use $R_V=3.1$. Therefore, it is still not clear-cut whether or not the $E_{B-V}$ values for SNe~II, IIb, and Ib are drawn from a common distribution.

\subsubsection{Intrinsic color curves}
\begin{figure}
\includegraphics[width=1.0\columnwidth]{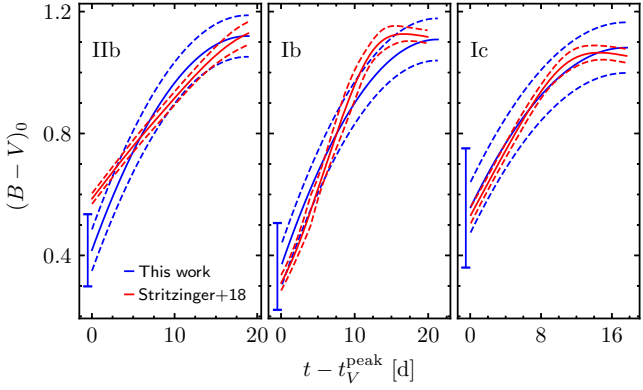}
\caption{Intrinsic $\bv$ color curves (solid lines) estimated in this work (blue) and in \citet{2018AA...609A.135S} (red). Dashed lines are $\pm1\,\ssd$ limits, while error bars represent the error on $(\bv)_0$ due to the intrinsic uncertainty and $\mathrm{ZP}_{E_{B-V}}$ error.} 
\label{fig:BV_Vi}
\end{figure}

The intrinsic color curves (given by equation~\ref{eq:c0}) can be constructed using equations~(\ref{eq:Psi_0}) and (\ref{eq:delta_0}), along with the derived $a_c$, $b_c$, $\tilde{\delta}_{0,c}$, $R_c$, and $\mathrm{ZP}_{E_{B-V}}$ values. Figure~\ref{fig:BV_Vi} shows the intrinsic $\bv$ color curves for SE~SNe, along with those presented in \citet{2018AA...609A.135S}. We see that despite the evident differences in shape, they are consistent within their $1\,\ssd$ limits. The $\ssd$ scatter around the color curves of \citet{2018AA...609A.135S} is 2--3 times lower than in our templates. We notice that the intrinsic color curves of \citet{2018AA...609A.135S} were constructed with only 3~SNe per subtype, as opposed to the 25--39 SNe per subtype that we use to determine the shape of the $(\bv)_0$ curves. The figure also shows the error on $(\bv)_0$ due to the intrinsic uncertainty and $\mathrm{ZP}_{E_{B-V}}$ error, which is $\sim$2 times greater than the $\ssd$ value. The latter means that the intrinsic $\bv$ color curves of SE~SNe are more diverse than apparent from  the templates of \citet{2018AA...609A.135S}.

\subsection{Bolometric flux}\label{sec:BC}
Let $f_\lambda$ be the SED and $f_{x_1\leq\lambda\leq x_n}$ the integral of $f_\lambda$ between $\bar{\lambda}_{x_1}$ and $\bar{\lambda}_{x_n}$. The bolometric flux
can be expressed as
\begin{equation}\label{eq:fbol}
f_\mathrm{bol}=f_{\lambda<B}+f_{B\leq\lambda\leq x_\mathrm{IR}}+f_{\lambda>x_\mathrm{IR}}.
\end{equation}
Here, $x_\mathrm{IR}$ is the reddest available photometric IR band, while $f_{\lambda<B}$ and $f_{\lambda>x_\mathrm{IR}}$ are the flux at wavelengths below $\bar{\lambda}_B$ and beyond $\bar{\lambda}_{x_\mathrm{IR}}$, respectively.

\subsubsection{Observed flux}
Given measured photometry in filters $S=\{x_1,\ldots,x_n\}$, a proxy for $f_{x_1\leq\lambda\leq x_n}$ is the quantity
\begin{equation}\label{eq:fS}
f_S=\frac{1}{2}\sum_{j=1}^{n-1}(\bar{\lambda}_{x_{j+1}}-\bar{\lambda}_{x_j})(\bar{f}_{x_{j+1}}+\bar{f}_{x_j})
\end{equation}
(e.g. \citealt{2017PASP..129d4202L}). Here,
\begin{equation}\label{eq:f_eff}
\bar{f}_x=10^{-0.4(m_{x,0}-\mathrm{ZP}_{\mathrm{flux},x})}
\end{equation}
is the monochromatic flux (in erg\,s$^{-1}$\,cm$^{-2}$\,\AA$^{-1}$) associated with the intrinsic $x$-band magnitude $m_{x,0}$, and $\mathrm{ZP}_{\mathrm{flux},x}$ is a constant providing the conversion from magnitudes to monochromatic fluxes (listed in Table~\ref{table:zp_and_leff}). To quantify the relative difference between $f_{x_1\leq\lambda\leq x_n}$ and $f_S$, we define
\begin{equation}\label{eq:alpha_S}
\alpha_S = f_{x_1\leq\lambda\leq x_n}/f_S.
\end{equation}

\begin{figure}
\includegraphics[width=1.0\columnwidth]{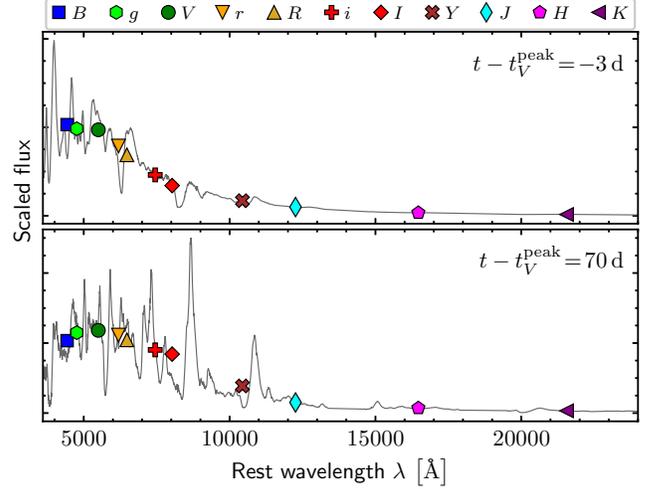}
\caption{Optical/IR spectra of SN~2011dh at $-3$ (top panel) and 70\,d (bottom panel) from $t_V^\peak$. Colored symbols are the monochromatic fluxes estimated with the magnitudes computed from the spectra.}
\label{fig:SED_11dh}
\end{figure}

To estimate $\alpha_S$ we assemble 34 observed optical/IR spectra of eight SE~SNe, covering the $\lambda$ range between 0.36 and 2.4\,\micron. From those spectra, we compute synthetic magnitudes for optical $BgV\!r\!RiI$ and IR $Y\!\!J\!H\!K$ bands (see Appendix~\ref{sec:syn_mag}), which we convert to $\bar{f}_x$ using equation~(\ref{eq:f_eff}). Figure~\ref{fig:SED_11dh} shows the optical/IR spectra of SN~2011dh at two different epochs and the $(\bar{\lambda}_x,\bar{f}_x)$ values for the aforementioned filters. We then calculate $f_S$ (equation~\ref{eq:fS}) and  $f_{x_1\leq\lambda\leq x_n}$ for different filter combinations. 

\begin{figure}
\includegraphics[width=1.0\columnwidth]{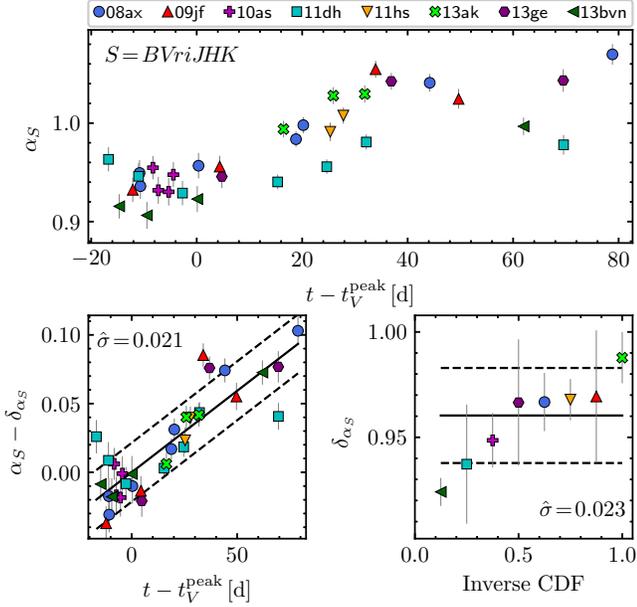}
\caption{Top panel: $\alpha_S$ against $t-t_V^\peak$ for $S=BV\!riJ\!H\!K$. Bottom left panel: $\alpha_S$ values shifted by $\delta \alpha_S$, where the solid line is a straight line fit corresponding to $\Psi_{0,\alpha_S}$. Bottom right panel: inverse CDF for the $\delta_{\alpha_S}$ values, where the solid line is the mean value. Dashed lines are $\pm1\,\ssd$ limits, and error bars are $1\,\sigma$ errors.} 
\label{fig:alpha_x}
\end{figure}

The top panel of Figure~\ref{fig:alpha_x} shows the $\alpha_S$ values for ${S=BV\!riJ\!H\!K}$. We see a dependence of $\alpha_S$ on $t-t_V^\peak$, which we express as
\begin{equation}
\alpha_S = \delta_{0,\alpha_S} + \Psi_{0,\alpha_S}(t- t_V^\peak),
\end{equation}
with
\begin{equation}
\Psi_{0,\alpha_S}(t- t_V^\peak) = b_S\frac{t- t_V^\peak}{100\,\mathrm{d}}.
\end{equation}
Minimising equation~(\ref{eq:s2}) with $X=t-t_V^\peak$ and $Y=\alpha_S$, we infer the $\delta_{\alpha_S}$ estimates for each SN (bottom right panel) and $\Psi_{0,\alpha_S}$ (bottom left panel). We adopt the mean and $\ssd$ of the $\delta_{\alpha_S}$ estimates as $\delta_{0,\alpha_S}$ and its error, respectively.

Table~\ref{table:alpha_parameters} summarizes the parameters of the $\alpha_S$ calibrations for different sets of optical/IR filters, which are valid in the range $-17\,\mathrm{d}<t- t_V^\peak<79\,\mathrm{d}$. To avoid relying on extrapolations, for $t - t_V^\peak$ greater than $79$\,d we adopt the $\alpha_S$ value evaluated at $79$\,d.

\begin{deluxetable}{c c c c}
\tablecaption{Parameters of the $\alpha_S$ calibrations\label{table:alpha_parameters}}
\tablehead{
 \colhead{$S$} & \colhead{$b_S$} & \colhead{$\delta_{0,\alpha_S}$} & \colhead{$t-t_V^\peak$ range}\\
 \nocolhead{}  & \nocolhead{}    &  \nocolhead{}                   & \colhead{ (d)}
}
\startdata 
 $BV\!riJ\!H$           &  $0.124$ & $0.960\pm0.023$ & $-17,79$  \\
 $BV\!riJ\!H\!K$        &  $0.119$ & $0.960\pm0.023$ & $-17,79$  \\
 $BgV\!riJ\!H\!K$       &  $0.112$ & $0.963\pm0.023$ & $-17,79$  \\
 $BgV\!riY\!\!J\!H$     &  $0.058$ & $0.988\pm0.021$ & $-17,79$  \\
 $BgV\!r\!RiI\!J\!H$    &  $0.021$ & $0.989\pm0.016$ & $-17,79$  \\
 $BV\!RI\!J\!H\!K$      &   $0.0$  & $0.990\pm0.019$ & $-17,79$  \\
 $BgV\!r\!RiIY\!\!J\!H$ &   $0.0$  & $1.004\pm0.015$ & $-17,79$  \\
 $BV\!r\!RiI\!J\!H\!K$  &   $0.0$  & $0.992\pm0.018$ & $-17,79$  \\
 $BgV\!r\!RiI\!J\!H\!K$ &   $0.0$  & $0.993\pm0.016$ & $-17,79$  \\
 $u\!B$                 & $-0.484$ & $1.247\pm0.120$ & $-17,47$  \\
 $U\!B$                 & $-0.021$ & $1.174\pm0.118$ & $-17,271$ \\
\enddata
\end{deluxetable}

\subsubsection{Unobserved flux at $\lambda<\bar{\lambda}_B$}
To estimate $f_{\lambda<B}$, we use the intrinsic $B$-band magnitude ($B_0$) and the UV correction (UVC; \citealt{2014ApJ...787..157P}), defined as
\begin{equation}\label{eq:UVC}
\mathrm{UVC}=-2.5\log(f_{\lambda<B})-B_0.
\end{equation} 
To calibrate UVC, we first compute $f_{\lambda<B}$ for 21~SE~SNe with $BV$ and UVOT $u$, $w1$, and $w2$ photometry. For those SNe, we express $f_{\lambda<B}$ as
\begin{equation}
f_{\lambda<B} = f_{\lambda<u}+\alpha_{u\!B}f_{u\!B},
\end{equation}
where $\alpha_{u\!B}$ is the factor to convert $f_{u\!B}$ to $f_{u\leq\lambda\leq B}$ (i.e. equation~\ref{eq:alpha_S} with $S=uB$). To calculate $\alpha_{u\!B}$, we use the spectra of 15~SE~SNe (scaled to match the $BV$ photometry) to estimate $f_{\lambda_u\leq\lambda\leq\lambda_B}$, and their $uB$ photometry to compute $f_{u\!B}$. The calibration of $\alpha_{u\!B}$ as a function of $t-t_V^\peak$ is provided in Table~\ref{table:alpha_parameters}.

\begin{figure}
\includegraphics[width=1.0\columnwidth]{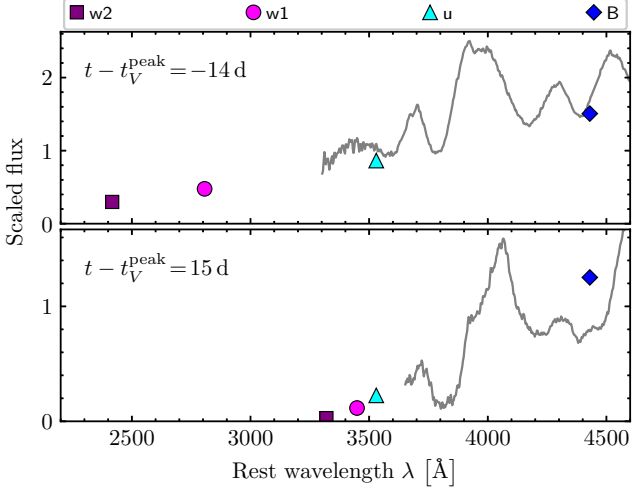}
\caption{Spectra of SN~2011dh at $-14$ (top panel) and 15\,d (bottom panel) since $t_V^\peak$. Colored symbols are monochromatic fluxes computed from observed photometry.}
\label{fig:UV_SED_11dh}
\end{figure}

Figure~\ref{fig:UV_SED_11dh} shows spectra of SN~2011dh at $-14$\,d and $15$\,d since $t_V^\peak$, along with the monochromatic fluxes estimated from the observed $B$ and UVOT photometry. The effective wavelength of the $w2$ and $w1$ bands is not constant because it depends strongly on the SED shape, being larger for redder SEDs (see Appendix~\ref{sec:syn_mag}). To estimate the unobserved UV flux (in our case, $f_{\lambda<w2}$), some authors extrapolate the bluest $(\bar{\lambda}_x,\bar{f}_x)$ point linearly to zero flux at 200\,nm \citep[e.g.][]{2014MNRAS.437.3848L,2018AA...609A.136T,2020MNRAS.496.4517S}. This has been justified by the argument that the contribution of the flux  to the bolometric flux is negligible below this wavelength (e.g. see the UV spectra of \citealt{2012ApJ...760L..33B,2015ApJ...803...40B}, and \citealt{2022ApJ...937...40K}). Applying such an extrapolation, we find that the contribution of $f_{\lambda<w2}$ to $f_{\lambda<B}$ is lower than 10\% for 95\% of the available epochs, with a typical contribution value around 3\%. Therefore, we choose to ignore this flux contribution. In addition, given the scarcity of UV spectra in the UVOT filter wavelength ranges, we adopt the flux computed with equation~(\ref{eq:fS}) and $S=\{w2,w1,u\}$ as $f_{\lambda<u}$.

\begin{figure}
\includegraphics[width=1.0\columnwidth]{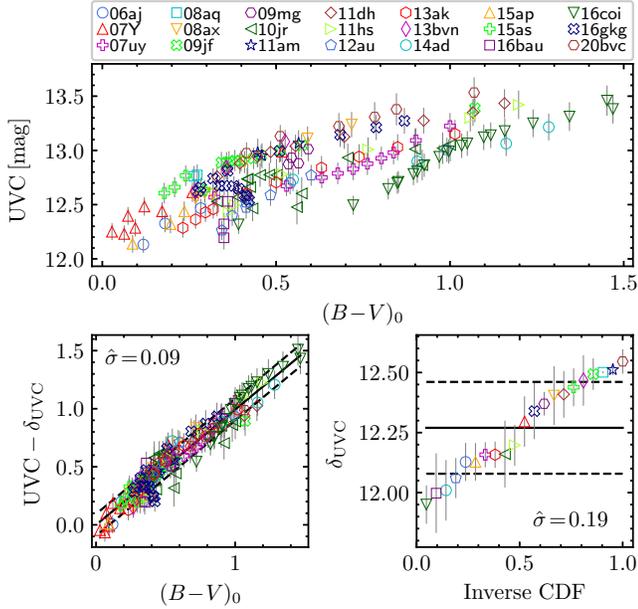}
\caption{Top panel: UV corrections against the intrinsic $\bv$ color for $t-t_\mathrm{expl}<40$\,d. Bottom left panel: UVC values shifted by $\delta \alpha_\mathrm{UVC}$, where the solid line is a straight line fit corresponding to $\Psi_{0,\mathrm{UVC}}$. Bottom right panel: inverse CDF for the $\delta_\mathrm{UVC}$ values, where the solid line is the mean value. Dashed lines are $\pm1\,\ssd$ limits, and error bars are $1\,\sigma$ errors.} 
\label{fig:q}
\end{figure}

The top panel of Figure~\ref{fig:q} shows the UVC values as a function of the intrinsic $\bv$ color for epochs between $3$ and $40$\,d since explosion. We express the observed correlation between UVC and $(\bv)_0$ as
\begin{equation}
\mathrm{UVC}=\delta_{0,\mathrm{UVC}}+\Psi_{0,\mathrm{UVC}}((\bv)_0).
\end{equation}
Minimising equation~(\ref{eq:s2}) with $X=(\bv)_0$ and $Y=\mathrm{UVC}$, we obtain the $\delta_{\mathrm{UVC}}$ values for each SN (bottom right panel) and $\Psi_{0,\mathrm{UVC}}$ (bottom left panel). We adopt the mean and $\ssd$ value of the $\delta_{\mathrm{UVC}}$ estimates as $\delta_{0,\mathrm{UVC}}$ and its error, respectively. The expression for UVC is then
\begin{equation}
\mathrm{UVC} = 12.27+0.99(\bv)_0\pm0.19,
\end{equation}
which is valid for $t-t_\mathrm{expl}<40$\,d. The UVC $\ssd$ value we obtain is much lower than that previously reported by \citet{2014ApJ...787..157P} ($\ssd=0.63$\,mag). The SN sample of that work is more heterogeneous, as it includes SNe~II and IIn, which may partly explain their higher $\ssd$ value.

In our sample, only SNe~2011dh, 2012au, and 2016coi have UVOT photometry at $t-t_\mathrm{expl}>40$\,d. For those SNe, we obtain roughly constant UVC estimates, with mean values of $13.44\pm0.04$, $12.93\pm0.05$ and $13.56\pm0.07$\,mag, respectively. To check the constancy of UVC for such epochs, we use the $U\!BV$ photometry of 21~SE~SNe and assume $f_{\lambda<B} =\alpha_{U\!B}f_{U\!B}$. The calibration of $\alpha_{U\!B}$, computed in the same way as $\alpha_{u\!B}$, is reported in Table~\ref{table:alpha_parameters}.

\begin{figure}
\includegraphics[width=1.0\columnwidth]{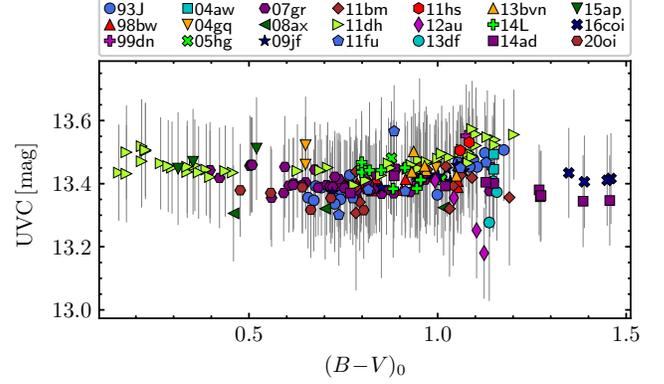}
\caption{UV corrections against the intrinsic $\bv$ color for $t-t_\mathrm{expl}>40$\,d.} 
\label{fig:UV_late}
\end{figure}

Figure~\ref{fig:UV_late} shows UVC against $(\bv)_0$ for $t-t_\mathrm{expl}>40$\,d, where we see that the UVC estimates remain nearly constant. For SNe~2011dh, 2012au, and 2016coi we obtain mean $\mathrm{UVC}$ values of $13.49\pm0.04$, $13.31\pm0.10$, and $13.43\pm0.07$\,mag, which are consistent with our previous finding within their errors. For the 21~SNe in the figure we compute a mean UVC of $13.42\pm0.06$. Hereafter for $t-t_\mathrm{expl}>40$\,d we adopt $\mathrm{UVC} = 13.42\pm0.19$, where we assume an error of 0.19\,mag in order not to underestimate the UVC uncertainty with respect to the error obtained for $t-t_\mathrm{expl}<40$\,d.

\subsubsection{Unobserved IR flux}
We model the flux at $\lambda>\bar{\lambda}_{x_\mathrm{IR}}$ as a power law, given by
\begin{equation}\label{eq:k}
f_\lambda=C\cdot(\lambda/\mathrm{\AA})^k.
\end{equation}
To measure the power law index $k$, we fit equation~(\ref{eq:k}) to the $(\bar{\lambda}_x, \bar{f}_x)$ points computed with $J\!H\!K$ or $Y\!J\!H$ photometry.

\begin{figure}
\includegraphics[width=1.0\columnwidth]{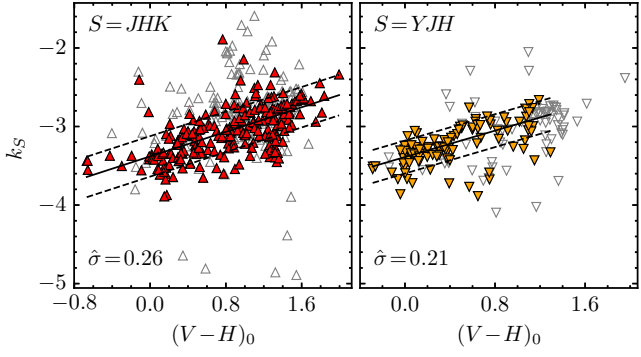}
\caption{Power-law indices for the $J\!H\!K$ (left-hand panel) and $Y\!J\!H$ (right-hand panel) photometry against the intrinsic ${V\!-\!H}$. Filled symbols are $k$ values with relative errors lower than 10\%. Solid lines are straight line fits to the filled symbols, dashed lines are $\pm1\,\ssd$ limits. Error bars are omitted for clarity.} 
\label{fig:k_JHK}
\end{figure}

Figure~\ref{fig:k_JHK} shows the $k$ values obtained with the $J\!H\!K$ ($k_{J\!H\!K}$) and $Y\!J\!H$ ($k_{Y\!J\!H}$) photometry as a function of the intrinsic $V\!-H$ color. We can see that the $k$ estimates are almost all larger than the power law index for the Rayleigh-Jeans tail of a Planck function ($k=-4$), which is typically assumed to model the IR flux of SE~SNe \citep[e.g.][]{2014MNRAS.437.3848L,2018AA...609A.136T}. Furthermore, $k$ grows with redder color.

A linear fit only to the data points having relative $k$ errors lower than 10\% gives
\begin{equation}\label{eq:k_JHK}
k=-3.38[\pm0.03]+0.39[\pm0.03](V\!-\!H)_0
\end{equation}
for $J\!H\!K$, and
\begin{equation}\label{eq:k_YJH}
k=-3.39[\pm0.03]+0.42[\pm0.06](V\!-\!H)_0
\end{equation}
for $Y\!J\!H$. The parameters of both fits are consistent with each other. As the $k_{J\!H\!K}$ and $k_{Y\!J\!H}$ values come from independent SN samples, equations~(\ref{eq:k_JHK}) and (\ref{eq:k_YJH}) are two independent estimates of the dependence of $k$ on $(V\!-\!H)_0$. Combining both samples, we obtain the linear fit
\begin{equation}
k=-3.38[\pm0.02]+0.40[\pm0.02](V\!-\!H)_0,
\end{equation}
with $\sigma_0=0.12$ and $\ssd=0.24$. We use this expression to indirectly estimate $k$ from the observed $(V\!-\!H)_0$ color, and then we adopt the weighted average of this value and $k_{J\!H\!K}$ (or $k_{Y\!J\!H}$) as the final $k$.

Once we determine $k$ for each SN, we compute the corresponding $C$ value fitting equation~(\ref{eq:k}) to the IR photometry. We then calculate $f_{\lambda>x_\mathrm{IR}}$ by integrating the power law from $\bar{\lambda}_{x_\mathrm{IR}}$ to infinity, i.e.
\begin{equation}\label{eq:fIR}
f_{\lambda>x_\mathrm{IR}}=-C\cdot(\bar{\lambda}_{x_\mathrm{IR}}/\mathring{\mathrm{A}})^{k+1}/(k+1).
\end{equation}

\subsubsection{Fractional flux evolution}
Using equations~(\ref{eq:fbol}), (\ref{eq:alpha_S}), (\ref{eq:UVC}), and (\ref{eq:fIR}) we next compute bolometric fluxes for 15~SNe~IIb, 15~SNe~Ib, and 19~SNe~Ic having optical/IR photometry in bands between $B$ and $H/K$.

\begin{figure}
\includegraphics[width=1.0\columnwidth]{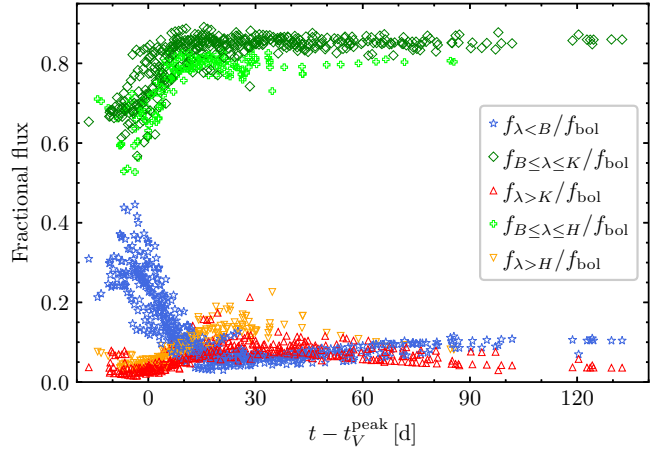}
\caption{Fractional flux in a given wavelength range as a function of the time since the $V$-band maximum light.}
\label{fig:flux_ratios}
\end{figure}

Figure~\ref{fig:flux_ratios} shows the temporal evolution of the fractional flux in different wavelength ranges. Before $t_V^\peak$, the contribution of $f_{\lambda<B}$ to $f_\mathrm{bol}$ is comparable to that of $f_{B\leq\lambda\leq x_\mathrm{IR}}$, while $f_{\lambda>x_\mathrm{IR}}/f_\mathrm{bol}$ is $\lesssim 5$\%. After $t_V^\peak$, $f_{\lambda<B}$ decreases drastically, accounting for $<10$\% of the bolometric flux at $t- t_V^\peak\sim20$\,d. At that time, the contribution of $f_{\lambda>x_\mathrm{IR}}$ to $f_\mathrm{bol}$ is about 5--15\%, while $f_{B\leq\lambda\leq x_\mathrm{IR}}$ accounts for $>80$\% of the bolometric flux. For $t- t_V^\peak>60$\,d, the contribution of $f_{B\leq\lambda\leq K}$, $f_{\lambda<B}$, and $f_{\lambda>K}$ to $f_\mathrm{bol}$ remains at about 85, 10, and 5\%, respectively. 

\subsection{Bolometric corrections}
We define the bolometric correction as
\begin{equation}
\mathrm{BC}_x=-2.5\log{(f_\mathrm{bol})}-m_{x,0},
\end{equation}
which can be calibrated as a function of the color \citep[e.g.][]{2014MNRAS.437.3848L}. However, given that the color curves of SE~SNe are non-monotonic \citep[e.g.][]{2018AA...609A.135S}, the same color can correspond to different epochs with different SEDs and therefore with different BC values.

\begin{figure}
\includegraphics[width=1.0\columnwidth]{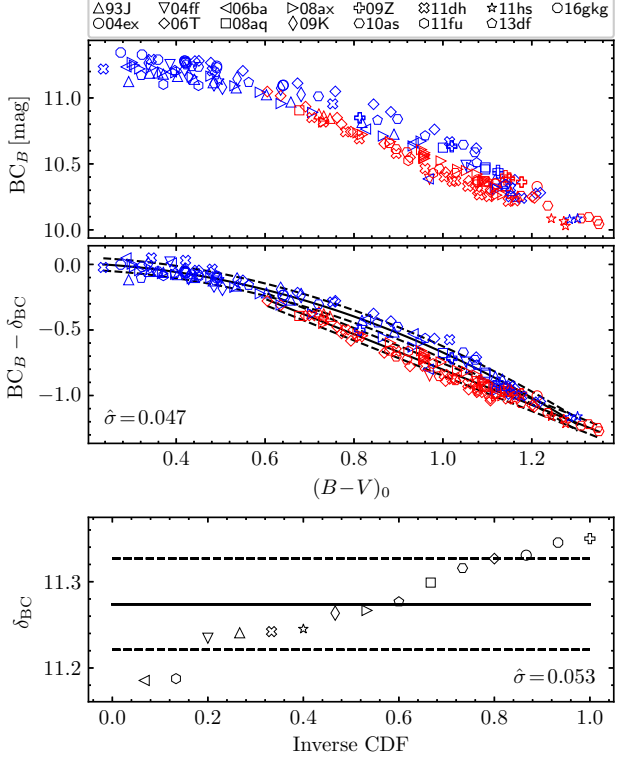}
\caption{Top panel: $\mathrm{BC}_B$ values for SNe~IIb versus the intrinsic $\bv$ color, where blue (red) symbols correspond to epochs before (after) the maximum of the $\bv$ color. Middle panel: $\mathrm{BC}_B$ values shifted by $\delta_{\mathrm{BC}}$, where solid lines are polynomial fits. Bottom panel: Inverse CDF for the $\delta_{\mathrm{BC}}$ estimates, where the solid line indicates the mean value. Dashed lines are the $\pm1\,\ssd$ limits.} 
\label{fig:BC_IIb}
\end{figure}

The top panel of Figure~\ref{fig:BC_IIb} shows the $\mathrm{BC}_B$ estimates of SNe~IIb against the intrinsic $\bv$ color, where blue (red) symbols correspond to $t-t_V^\peak$ epochs before (after) the time of maximum $\bv$ color, $t_{B-V}^\mathrm{max}$. We see that the $\mathrm{BC}_B$ values for $t-t_V^\peak>t_{B-V}^\mathrm{max}$ tend to be lower than those for $t-t_V^\peak<t_{B-V}^\mathrm{max}$ within the same color range. To disentangle this degeneracy, we parametrize the dependence of $\mathrm{BC}_x$ on a given color, $c=x-y$, as
\begin{eqnarray}
\mathrm{BC}_{x,c} = \left\{
\begin{array}{lr}
 \mathrm{ZP}_{x,c} + a_1c+a_2c^2, & \,\,t-t_y^\peak \leq t_c^\mathrm{max}\\
 \mathrm{ZP}_{x,c} + a_3 +a_4c,   & t-t_y^\peak > t_c^\mathrm{max}
\end{array}\right. .
\end{eqnarray}
Here, $\mathrm{ZP}_{x,c}$ is the zero-point for the BC calibration, while the $a_i$ parameters are obtained by minimizing 
\begin{eqnarray}\label{eq:s2_BC}
s^2&=&\sum_j^N \left[\sum_k^{\leq t_c^\mathrm{max}}\left(\mathrm{BC}_{x,j,k}-\delta_{\mathrm{BC},j}-a_1c_{j,k}-a_2c_{j,k}^2\right)^2\right. \nonumber\\
   &+&\left.\sum_k^{> t_c^\mathrm{max}}\left(\mathrm{BC}_{x,j,k}-\delta_{\mathrm{BC},j}-a_3-a_4c_{j,k}\right)^2\right].
\end{eqnarray}
We adopt the mean and sample standard deviation of the $\delta_{\mathrm{BC},j}$ values as $\mathrm{ZP}_{x,c}$ and its error, respectively.

The middle and bottom panels of Figure~\ref{fig:BC_IIb} show the result of the minimisation of equation~(\ref{eq:s2_BC}), while Table~\ref{table:BC_cal} summarizes the BC parameters for different combinations of bands and colors. We also compare the $M_r^\peak$ distributions of the SNe~IIb, Ib, and Ic used to calibrate BCs for different band-color combinations with the $M_r^\peak$ distributions of VL samples (see Section~\ref{sec:EBV_accuracy}), finding $p_\mathrm{AD}$ values greater than 0.12, 0.18, and 0.35 for SNe~IIb, Ib, and Ic, respectively. Based on this, we can expect that the luminosities computed with the BC technique are not, on average, over- or underestimated.

\begin{figure}
\includegraphics[width=1.0\columnwidth]{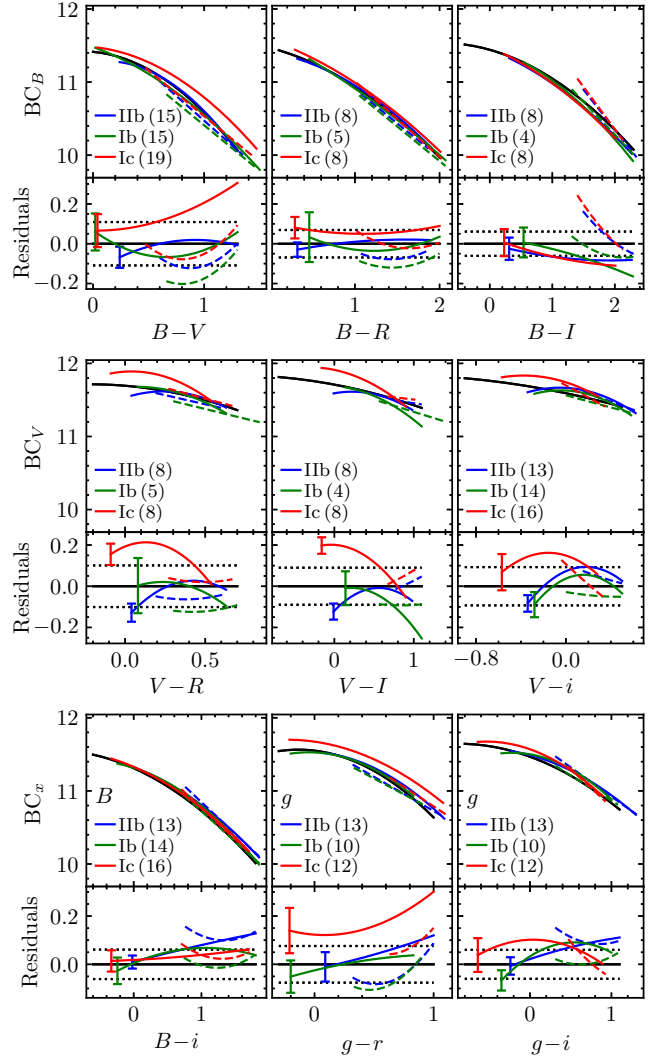}
\caption{BC calibrations along with the residuals with respect to the \citet{2014MNRAS.437.3848L,2016MNRAS.457..328L} calibrations (black lines). Colored solid (dashed) lines are our BC calibrations for times before (after) the epoch of maximum color. The number of SNe used to calibrate the BC is indicated in parentheses. Error bars indicates the $1\,\sigma$ error around our calibrations, while dotted lines indicate the $\pm1\,\ssd$ limits of the \citet{2014MNRAS.437.3848L,2016MNRAS.457..328L} calibrations.} 
\label{fig:BC_templates}
\end{figure}

In Figure~\ref{fig:BC_templates} we compare our BC calibrations with those of \citet{2014MNRAS.437.3848L,2016MNRAS.457..328L}, which were computed with 15~SE~SNe (3~SNe~IIb, 6~SNe~Ib, and 6~SNe~Ic). From the latter calibrations we subtract the zero-point used by the authors to define the bolometric magnitude scale (equivalent to $-11.497$\,mag). In the residuals panels we see that our BC calibrations are consistent with those of \citet{2014MNRAS.437.3848L,2016MNRAS.457..328L} to within their errors. 

\subsection{Bolometric light curves}
\begin{figure}
\includegraphics[width=1.0\columnwidth]{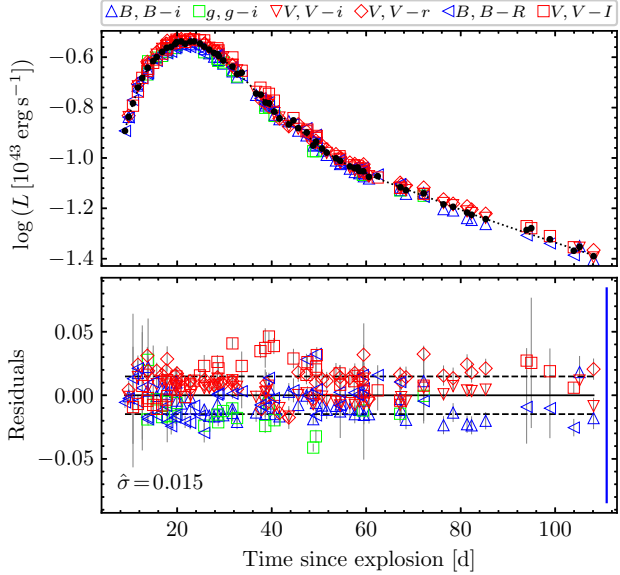}
\caption{Top panel: bolometric light curve of SN~2009jf computed with different band-color combinations (empty symbols). The dotted line is the \textsc{alr} fit to the weighted average light curve (black circles). Bottom panel: residuals around the \textsc{alr} fit. Error bars are $1\,\sigma$ errors due to uncertainties on photometry, while the blue vertical bar depicts the $1\,\sigma$ error at peak due to the error on $\mu$, reddening, and BC. Dashed lines are the $\pm1\,\ssd$ limits.} 
\label{fig:Lbol}
\end{figure}

\begin{figure}
\includegraphics[width=1.0\columnwidth]{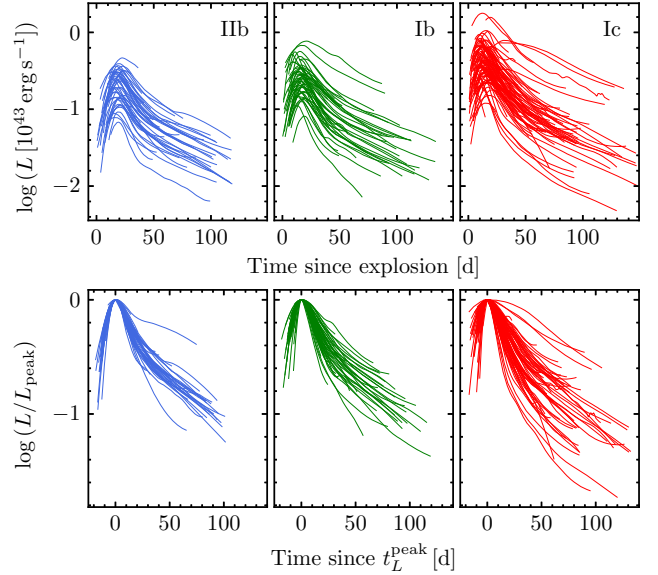}
\caption{Bolometric SN light curves interpolated with the \textsc{alr} code (top panels) and normalized to the peak (bottom panels). } 
\label{fig:logL_vs_t}
\end{figure}

We now apply the results of the previous sections to the data in order to derive bolometric light curves for the SN sample. For a SN with photometry in $n$ optical bands, we compute luminosities for each of the $n-1$ independent band-color combinations using
\begin{equation}
\log L_{x,c}=(\mu-\mathrm{BC}_{x,c}-m_{x,0})/2.5 -2.922
\end{equation}
\citep[e.g.][]{2021MNRAS.505.1742R}. Then, we adopt the weighted mean of the $\log L_{x,c}$ estimates in windows of 0.25\,d as the final $\log L$ values, with weights  
\begin{equation}\label{eq:w_phot}
w_{x,c}=\left[\left[1+\frac{d\,\mathrm{BC}_{x,c}}{d c}\right]^2\sigma_{m_{x}}^2+\left[\frac{d\,\mathrm{BC}_{x,c}}{d c}\right]^2\sigma_{m_{y}}^2\right]^{-1}.
\end{equation}
As an example, Figure~\ref{fig:Lbol} shows the bolometric light curve of SN~2009jf computed with $BgV\!r\!RiI$ photometry.  Figure~\ref{fig:logL_vs_t} shows the final bolometric light curves for the sample, interpolated with the \textsc{alr} code.

The error on $\log L$ is given by
\begin{equation}
\sigma_{\log L} = 0.4\sqrt{\sigma_\mathrm{phot}^2+\sigma_{\mathrm{BC}}^2 + \sigma_\mu^2+\sigma_E^2},
\end{equation}
where $\sigma_\mathrm{phot}$, $\sigma_{\mathrm{BC}}$, and $\sigma_E$ are the errors due to uncertainties on photometry, BC ZP, and reddening, respectively. The $\sigma_\mathrm{phot}$ and $\sigma_{\mathrm{BC}}$ errors are the weighted mean errors computed with equation~(\ref{eq:w_phot}) and with $w_{x,c}=1/\sigma_{\mathrm{ZP}_{x,c}}^2$, respectively, while
\begin{equation}
\sigma_E=\sqrt{\zeta_\mathrm{MW}^2\sigma_{E_{B-V}^\mathrm{MW}}^2+\zeta_\mathrm{h}^2\sigma_{E_{B-V}}^2}.
\end{equation}
Here,
\begin{equation}\label{eq:zeta}
\zeta_s = \frac{1}{n-1}\sum_{x,c}\left[R_x^s + (R_x^s-R_y^s) \frac{d\,\mathrm{BC}_{x,c}}{d c}\right],
\end{equation}
where $s=\{\mathrm{h},\mathrm{MW}\}$ and $R_x^s=A_x^s/E_{B-V}^s$ (see Appendix~\ref{sec:AK_correction}).

\subsection{Explosion parameters}\label{sec:explosion_parameters}
\begin{figure}
\includegraphics[width=1.0\columnwidth]{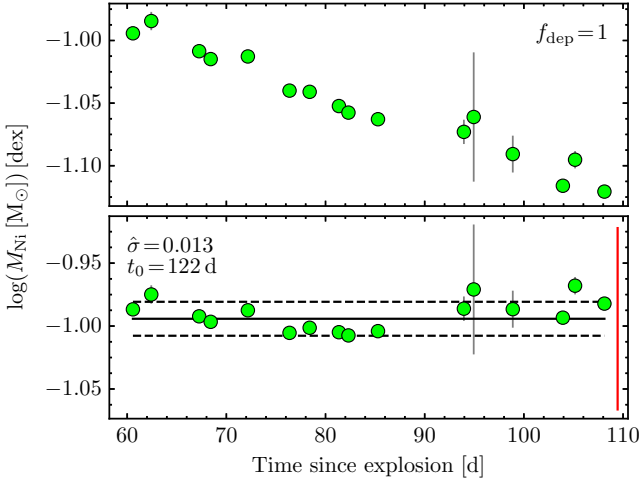}
\caption{$\log\mni$ estimates for SN~2009jf versus the time since explosion, before (top panel) and after (bottom panel) correcting for the $\gamma$-ray deposition function. The solid horizontal lines corresponds to the $\log\mni$ value that maximizes the posterior probability, while dashed lines are the $\pm1\,\ssd$ limits. Error bars are $1\,\sigma$ errors due to uncertainties on photometry, while the red vertical bar depicts the $1\,\sigma$ error due to the uncertainty on $\mu$, reddening, explosion epoch, and BC.} 
\label{fig:2009jf_MNi}
\end{figure}

From the bolometric light curves derived above we now measure $\log L_\peak$, $t_L^\peak$, and the decline rate $\Delta m_{15}(\mathrm{bol})=2.5\log(L_\peak/L(t_L^\peak+15\,\mathrm{d}))$, which are listed in Table~\ref{table:MNi}. We then compute $\log\mni$ and $t_\mathrm{esc}$ using the procedure described in Section~\ref{sec:MNi_tail}. Figure~\ref{fig:2009jf_MNi} shows, for example, the $\log\mni$ estimates of SN~2009jf as a function of the time since explosion before (top panel) and after (bottom panel) correcting for the $\gamma$-ray deposition function. 

\begin{figure}
\includegraphics[width=1.0\columnwidth]{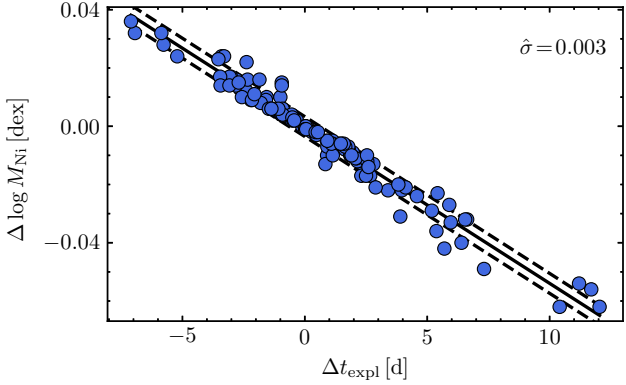}
\caption{Change in $\log\mni$ against change in explosion time. Negative (positive) $\Delta t_\mathrm{expl}$ values are quantities computed using $t_\mathrm{non-det}$ ($t_\mathrm{detect}$) as explosion time. The solid line is a straight line fit and dashed lines are $\pm1\ssd$ limits.} 
\label{fig:Delta_lMNi_vs_Delta_t}
\end{figure}

The error on $\log\mni$ is given by
\begin{equation}
\sigma_{\log\mni} = \sqrt{\tilde{\sigma}_{\log\mni}^2+\sigma_t^2  + \frac{\sigma_\mathrm{BC}^2+\sigma_\mu^2+{\sigma_E^\tail}^2}{6.25}},
\end{equation}
where $\tilde{\sigma}_{\log\mni}$ is the uncertainty due to errors on photometry and $t_\mathrm{esc}$ (provided by \textsc{emcee}), $\sigma_t$ is the error in $\log\mni$ due to the uncertainty in $t_\mathrm{expl}$, while $\sigma_E^\tail$ is the $\sigma_E$ error in the radioactive tail. To estimate $\sigma_t$, we compute $\log\mni$ assuming $t_\mathrm{expl}=t_\mathrm{non-det}$ and $t_\mathrm{expl}=t_\mathrm{detect}$. These epochs correspond to the lowest and greatest possible value for $t_\mathrm{expl}$. Figure~\ref{fig:Delta_lMNi_vs_Delta_t} shows the change in $\log\mni$ ($\Delta \log\mni$) versus the change in $t_\mathrm{expl}$ ($\Delta t_\mathrm{expl}$). The distribution, which is well represented by the straight line $\Delta\log\mni=-0.0001-0.0054(\Delta t_\mathrm{expl}/\mathrm{d})$, indicates that $\log\mni$ decreases as $t_\mathrm{expl}$ advances. Based on this, we adopt $\sigma_t=0.0054\,\sigma_{t_\mathrm{expl}}\,\mathrm{d}^{-1}$. The $\log\mni$ and $t_\mathrm{esc}$ values are summarized in Columns~8 and 9 of Table~\ref{table:MNi}, respectively. The mean $\log\mni$ error for our SN sample is 0.120\,dex, of which 65, 31, 2, 1.5, and 0.5\% is induced by $\sigma_E^\tail$, $\sigma_\mu$, $\sigma_\mathrm{BC}$, $\tilde{\sigma}_{\log\mni}$, and $\sigma_t$, respectively.

We also compute $\log (L_\peak/\mni)$ for the 20~SNe~IIb, 21~SNe~Ib, and 34~SNe~Ic with $\log\mni$ estimates. The error on $\log(L_\peak/\mni)$ is given by
\begin{equation}\label{eq:eLambda}
\sigma_{\log (L_\peak/\mni)}=\sqrt{\tilde{\sigma}_{\log\mni}^2+\sigma_t^2+\frac{{\sigma_\mathrm{phot}^\peak}^2+{\sigma_E^\mathrm{pt}}^2}{6.25}},
\end{equation}
where $\sigma_\mathrm{phot}^\peak$ is the $\sigma_\mathrm{phot}$ error at peak time, and
\begin{equation}
\sigma_E^\mathrm{pt} = \sqrt{[\zeta_\mathrm{MW}^\peak-\zeta_\mathrm{MW}^\tail]^2\sigma_{E_{B-V}^\mathrm{MW}}^2+[\zeta_\mathrm{h}^\peak-\zeta_\mathrm{h}^\tail]^2\sigma_{E_{B-V}}^2},
\end{equation}
where the terms $\zeta_s^\peak$ and $\zeta_s^\tail$ for $s=\{\mathrm{MW},\mathrm{h}\}$ are given by equation~(\ref{eq:zeta}) evaluated at $t_L^\peak$ and in the radioactive tail, respectively. The $\log (L_\peak/\mni)$ estimates and their errors are listed in Column~10 of Table~\ref{table:MNi}. The typical $\log (L_\peak/\mni)$ error is 0.031\,dex, of which 72, 19, 7, and 2\% is induced by $\sigma_E^\mathrm{pt}$, $\tilde{\sigma}_{\log\mni}$, $\sigma_t$, and $\sigma_\mathrm{phot}^\peak$, respectively.

A rough estimate of the ejecta mass ($M_\mathrm{ej}$) and the kinetic energy ($E_\mathrm{k}$) can be computed with the relations
\begin{equation}\label{eq:Mej}
M_\mathrm{ej} = \frac{1}{2}\left(\frac{13.7c}{\kappa}\right) (t_L^\peak)^2 v_\peak
\end{equation}
\citep[e.g.][]{2019MNRAS.485.1559P} and 
\begin{equation}\label{eq:Ek}
E_\mathrm{k}=0.3M_\mathrm{ej} v_\peak^2
\end{equation}
\citep[e.g.][]{2016MNRAS.458.2973P}, where $v_\peak$ is the velocity of the ejecta at $t_L^\peak$, $c$ is the light speed, and $\kappa=0.07$\,cm$^2$\,g$^{-1}$ is the opacity of the ejecta\footnote{The relations for $M_\mathrm{ej}$ and $E_\mathrm{k}$ are based on Arnett's rule, which does not hold for SE~SNe \citep{2019ApJ...878...56K}. Therefore, the derived $M_\mathrm{ej}$ and $E_\mathrm{k}$ values should be treated with caution.}. For 68~SNe in our dataset we adopt the $v_\peak$ values reported by \citet{2019MNRAS.485.1559P}, who use the \ion{He}{1}\,$\lambda5876$ and \ion{Si}{2}\,$\lambda6355$ absorption lines to estimate the expansion velocities of SNe~IIb/Ib and Ic, respectively. We use the same lines to measure the expansion velocities of 87~SNe, which we adopt as $v_\peak$ if the epoch of the spectrum ($t_\mathrm{spec}$) is within $\pm3$\,d since $t_L^\peak$. If $t_\mathrm{spec}$ is more than three days lower (greater) than $t_L^\peak$, then we adopt the measured velocity as an upper (lower) limit for $v_\peak$. For seven SNe, where the presence of \ion{He}{1}\,$\lambda5876$ or \ion{Si}{2}\,$\lambda6355$ absorption lines in the spectra is unclear, we adopt the expansion velocities inferred from the \ion{Fe}{2}\,$\lambda5169$ absorption line. For 14~SNe without public spectra, we adopt the expansion velocities reported in the literature. We assume an error of 15\% for all expansion velocities. The $v_\peak$, $M_\mathrm{ej}$, and $E_\mathrm{k}$ estimates are collected in Columns~5, 6, and 7 of Table~\ref{table:MNi}, respectively. The reported errors on $M_\mathrm{ej}$ and $E_\mathrm{k}$ are random and do not account for systematic errors due to the use of the highly approximate equations~(\ref{eq:Mej}) and (\ref{eq:Ek}).

\section{Results}\label{sec:results}
\subsection{Distributions and correlations}\label{sec:correlations}
\begin{figure}
\includegraphics[width=1.0\columnwidth]{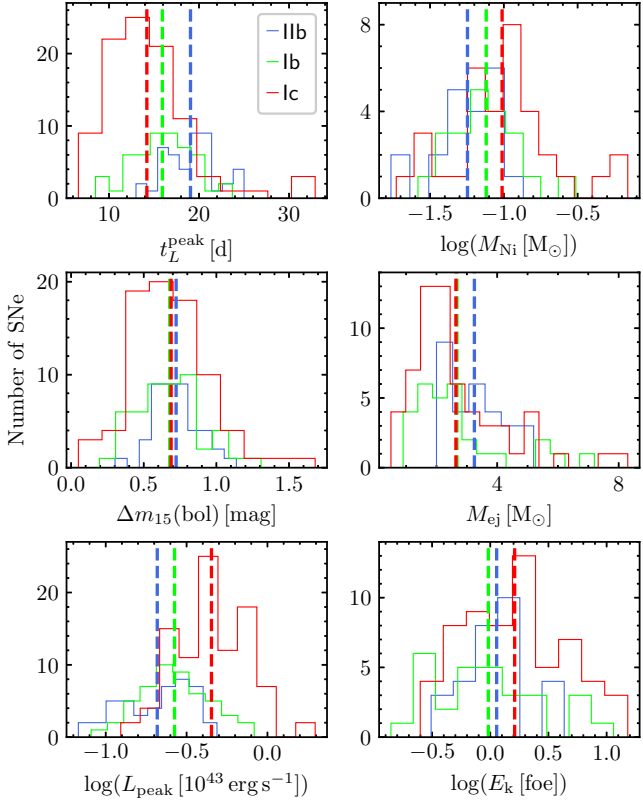}
\caption{Histograms of the bolometric light-curve properties (left-hand panels) and the explosion parameters (right-hand panels). Dashed lines indicate mean values.}
\label{fig:Lpars}
\end{figure}

Figure~\ref{fig:Lpars} shows the histograms of the bolometric light-curve properties and the explosion parameters, and Table~\ref{table:pars_stats} summarises the statistics of those distributions. The $^{56}$Ni mass distribution is discussed further in Section~\ref{sec:MNi_distribution}. The mean values of the $t_L^\peak$ and $\Delta m_{15}(\mathrm{bol})$ distributions are similar to, and $\sim$0.15\,mag lower than, respectively, the estimates inferred for the $V$-band (see Section~\ref{sec:texp}). The mean peak luminosity, mean $^{56}$Ni mass, and mean kinetic energy increase in the sequence of IIb--Ib--Ic subtypes. The median peak luminosities and mean kinetic energies are consistent with those reported by \citet{2016MNRAS.458.2973P} and \citet{2016MNRAS.457..328L}, respectively. All SE~SN subtypes have similar mean $M_\mathrm{ej}$ values around 2.6--3.2\,M$_{\sun}$, which are consistent with the mean $M_\mathrm{ej}$ of 2.8\,M$_\sun$ for all SE~SNe reported by \citet{2019MNRAS.485.1559P}. Based on the $\ssd$ values and ranges of explosion parameters and bolometric light-curve properties, we find that SE~SNe become more heterogeneous in the sequence of IIb--Ib--Ic subtypes. The relative homogeneity of SNe~IIb has been previously noted by \citet{2016MNRAS.457..328L}.

\begin{deluxetable}{lccccccc}
\tablecaption{Statistics of the bolometric light-curve properties and explosion parameters\label{table:pars_stats}}
\tablehead{
 \colhead{Type}    & \colhead{$N$} & \colhead{Min} & \colhead{Max} & \colhead{Mean} & \colhead{Median}  & \colhead{$\ssd$} & \colhead{Range}
}
\startdata 
     & \multicolumn{7}{c}{$t_L^\peak$\,[d]} \\
 IIb &  $45$ & $13.0$ & $25.0$ & $19.1$ & $19.3$  & $ 2.8$ & $12.0$ \\
 Ib  &  $50$ & $ 8.5$ & $23.7$ & $15.9$ & $16.0$  & $ 3.3$ & $15.2$ \\
 Ic  &  $96$ & $ 6.6$ & $32.9$ & $14.2$ & $13.3$  & $ 4.9$ & $26.3$ \\
 \hline
     & \multicolumn{7}{c}{$\Delta m_{15}(\mathrm{bol})$\,[mag]} \\
 IIb &  $42$ & $0.30$ & $ 1.14$ & $0.72$ & $0.71$  & $0.16$ & $0.83$ \\
 Ib  &  $49$ & $0.20$ & $ 1.31$ & $0.68$ & $0.65$  & $0.23$ & $1.11$ \\
 Ic  &  $81$ & $0.05$ & $ 1.68$ & $0.69$ & $0.67$  & $0.27$ & $1.63$ \\
 \hline
     & \multicolumn{7}{c}{$\log L_\peak$\,[$10^{43}$\,erg\,s$^{-1}$]} \\
 IIb & $45$ & $-1.17$ & $-0.31$ & $-0.68$ & $-0.63$  & $ 0.21$ & $ 0.86$ \\
 Ib  & $50$ & $-1.09$ & $-0.08$ & $-0.57$ & $-0.58$  & $ 0.22$ & $ 1.01$ \\
 Ic  & $96$ & $-0.91$ & $ 0.29$ & $-0.35$ & $-0.36$  & $ 0.22$ & $ 1.20$ \\
 \hline
     & \multicolumn{7}{c}{$M_\mathrm{Ni}\,[\mathrm{M}_{\sun}]$} \\
 IIb & $ 20$ & $0.018$ & $0.138$ & $0.065$ & $0.062$  & $ 0.03$ & $ 0.12$ \\
 Ib  & $ 21$ & $0.027$ & $0.322$ & $0.091$ & $0.079$  & $ 0.06$ & $ 0.29$ \\
 Ic  & $ 34$ & $0.019$ & $0.706$ & $0.145$ & $0.107$  & $ 0.15$ & $ 0.69$ \\
 \hline
     & \multicolumn{7}{c}{$\log M_\mathrm{Ni}\,[\mathrm{M}_{\sun}]$} \\
 IIb & $20$ & $-1.76$ & $-0.87$ & $-1.25$ & $-1.23$  & $ 0.22$ & $ 0.90$ \\
 Ib  & $21$ & $-1.58$ & $-0.51$ & $-1.12$ & $-1.12$  & $ 0.24$ & $ 1.07$ \\
 Ic  & $34$ & $-1.73$ & $-0.16$ & $-1.01$ & $-1.00$  & $ 0.35$ & $ 1.57$ \\
 \hline
     & \multicolumn{7}{c}{$M_\mathrm{ej}\,[\mathrm{M}_{\sun}]$} \\
 IIb &  $29$ & $ 2.0$ & $ 5.2$ & $ 3.2$ & $ 3.2$  & $ 0.9$ & $ 3.2$ \\
 Ib  &  $29$ & $ 0.9$ & $ 7.2$ & $ 2.7$ & $ 2.3$  & $ 1.6$ & $ 6.3$ \\
 Ic  &  $61$ & $ 0.5$ & $ 8.3$ & $ 2.6$ & $ 2.2$  & $ 1.6$ & $ 7.8$ \\
 \hline
     & \multicolumn{7}{c}{$E_\mathrm{k}\,[\mathrm{foe}]$} \\
 IIb &  $29$ & $0.31$ & $ 4.32$ & $1.41$ & $1.13$  & $1.05$ & $4.01$ \\
 Ib  &  $29$ & $0.14$ & $11.48$ & $1.87$ & $0.93$  & $2.63$ & $11.34$ \\
 Ic  &  $61$ & $0.25$ & $15.20$ & $2.80$ & $1.56$  & $3.30$ & $14.95$ \\
 \hline
     & \multicolumn{7}{c}{$\log E_\mathrm{k}\,[\mathrm{foe}]$} \\
 IIb & $29$ & $-0.51$ & $ 0.64$ & $0.05$ & $0.05$  & $0.29$ & $1.14$ \\
 Ib  & $29$ & $-0.85$ & $ 1.06$ & $-0.02$ & $-0.03$  & $0.49$ & $1.91$ \\
 Ic  & $61$ & $-0.60$ & $ 1.18$ & $0.21$ & $0.19$  & $0.46$ & $1.78$ \\
\enddata
\end{deluxetable}

For comparison with models, in Table~\ref{table:model_characteristics} we list the average explosion parameters and bolometric light-curve properties of the \textsc{CMFGEN} models for SNe~IIb, Ib, and Ic presented in \citet{2016MNRAS.458.1618D} along with the initial mass function (IMF) averaged characteristics of the \textsc{SEDONA} models for SNe~Ib/Ic reported by \citet{2021ApJ...913..145W}. The progenitors of the \citet{2016MNRAS.458.1618D} (\citealt{2021ApJ...913..145W}) models are mass donors (stripped helium stars) in close-binary systems. Based on the H/He composition of the ejecta models, \citet{2016MNRAS.458.1618D} classified 17, 6, and 4 of their 27 models as SNe~IIb, Ib, and Ic, respectively. Among the \citet{2021ApJ...913..145W} models, we select 41 with standard mass-loss rate and 
initial helium star masses of 3.3 to 8.0\,M$_\sun$, which were found to be consistent with normal SNe~Ib and Ic (i.e., SNe~Ic-BL and GRB-SNe are not included, see \citealt{2021ApJ...913..145W}). In the table we also list the average characteristics we obtain for SNe~IIb, Ib, and Ic (excluding SNe~Ic-BL) in our sample, and the corresponding mean values for SNe~Ib/Ic adopting that 35.6\% and 21.5\% of all SE~SNe are of Type~Ib and Ic (excluding Ic-BL and peculiar events), respectively \citep{2017PASP..129e4201S}.

\begin{deluxetable*}{lccccc}
\tablecaption{Average bolometric light-curve properties and explosion parameters\label{table:model_characteristics}}
\tablehead{
\nocolhead{} & \colhead{IIb} & \colhead{Ib} & \colhead{Ic} & & \colhead{Ib/Ic} 
}
\startdata 
\nocolhead{} & \multicolumn{5}{c}{This work} \\
\cline{2-6}
 $\log L_\peak$\,[$10^{43}$\,erg\,s$^{-1}$]                & $-0.68\pm0.03$ ($45$)  & $-0.57\pm0.03$ ($50$)  & $-0.40\pm0.03$ ($70$)  & & $-0.51\pm0.02$ \\
 $t_L^\peak$\,[d]              & $19.1\pm0.4$ ($45$)    & $15.9\pm0.5$ ($50$)    & $15.0\pm0.6$ ($70$)    & & $15.6\pm0.4$   \\
 $\Delta m_{15}$(bol)          & $0.72\pm0.02$ ($42$)   & $0.68\pm0.03$ ($49$)   & $0.66\pm0.04$ ($57$)   & & $0.67\pm0.03$  \\
 $M_\mathrm{ej}$\,[M$_{\sun}$] & $3.2\pm0.2$ ($29$)     & $2.7\pm0.3$ ($29$)     & $2.6\pm0.3$ ($42$)     & & $2.7\pm0.2$    \\
 $E_\mathrm{k}$\,[foe]         & $1.41\pm0.19$ ($29$)   & $1.87\pm0.49$ ($29$)   & $1.55\pm0.25$ ($42$)   & & $1.75\pm0.32$  \\
 $M_\mathrm{Ni}$\,[M$_{\sun}$] & $0.065\pm0.006$ ($20$) & $0.091\pm0.014$ ($21$) & $0.157\pm0.034$ ($25$) & & $0.116\pm0.015$\\
 $\log(L_\peak/M_\mathrm{Ni})$ & $0.52\pm0.02$ ($20$)   & $0.55\pm0.03$ ($21$)   & $0.60\pm0.06$ ($25$)   & & $0.57\pm0.03$  \\
\cline{1-6}
\nocolhead{} & \multicolumn{3}{c}{\citet{2016MNRAS.458.1618D}} & & \colhead{\citet{2021ApJ...913..145W}} \\
\cline{2-4}  \cline{6-6}
 $\log L_\peak$\,[$10^{43}$\,erg\,s$^{-1}$]                & $-0.83\pm0.04$ ($17$)  & $-0.50\pm0.10$ ($ 6$)  & $-0.62\pm0.10$ ($ 4$)  & & $-0.73$\\
 $t_L^\peak$\,[d]              & $24.1\pm0.9$ ($17$)    & $35.4\pm2.1$ ($ 6$)    & $31.8\pm1.4$ ($ 4$)    & & $21.4$\\
 $\Delta m_{15}$(bol)          & $0.66\pm0.03$ ($17$)   & $0.33\pm0.03$ ($ 6$)   & $0.41\pm0.02$ ($ 4$)   & &  --    \\
 $M_\mathrm{ej}$\,[M$_{\sun}$] & $2.4\pm0.1$ ($17$)     & $5.0\pm0.0$ ($ 6$)     & $3.6\pm0.0$ ($ 4$)     & & $2.19$ \\
 $E_\mathrm{k}$\,[foe]         & $1.80\pm0.35$ ($17$)   & $2.99\pm0.76$ ($ 6$)   & $1.88\pm0.35$ ($ 4$)   & & $1.06$\\
 $M_\mathrm{Ni}$\,[M$_{\sun}$] & $0.069\pm0.006$ ($17$) & $0.211\pm0.036$ ($ 6$) & $0.140\pm0.028$ ($ 4$) & & $0.09$\\
  $\log(L_\peak/M_\mathrm{Ni})$& $0.36\pm0.01$ ($17$)   & $0.22\pm0.01$ ($ 6$)   & $0.26\pm0.01$ ($ 4$)  & & $0.32$\\
\enddata
\tablecomments{Uncertainties are the standard error of the mean ($\ssd/\sqrt{N}$). Numbers in parentheses are the sample sizes. SNe~Ic and Ib/Ic do not include SNe~Ic-BL. Values of \citet{2021ApJ...913..145W} are IMF-averaged quantities.}
\end{deluxetable*}

The models of \citet{2016MNRAS.458.1618D} for SNe~IIb have mean $\Delta m_{15}$(bol), $E_\mathrm{k}$, and $\mni$ consistent with our estimates to $1.7\,\sigma$, while the average $t_L^\peak$ and $\log L_\peak$ are 5\,d longer and 0.15\,dex lower than our estimates, respectively. The models for SNe~Ib and Ic have average $\log L_\peak$ and $E_\mathrm{k}$ consistent with our estimates to $2.1\,\sigma$, while the mean $\Delta m_{15}$(bol) and $t_L^\peak$ are at least 0.25\,mag lower and 16.8\,d longer than our estimates, respectively. The models of \citet{2021ApJ...913..145W} for SNe~Ib/Ic have IMF-averaged $M_\mathrm{ej}$, $E_\mathrm{k}$, and $\mni$ values lower than our mean estimates but consistent with them to $2.6\,\sigma$, while the IMF-averaged $\log L_\peak$ and $t_L^\peak$ are 0.22\,dex lower and 5.8\,d longer than our mean values, respectively.

Models of \citet{2016MNRAS.458.1618D} and \citet{2021ApJ...913..145W} have mean $\log(L_\peak/\mni)$ and $t_L^\peak$ values at least $5\,\sigma$ lower and longer than observations, respectively. If the systematic differences between the average $t_L^\peak$ of models and observations are because we overestimate $t_\mathrm{expl}$, then the $\log\mni$ values computed from observations are underestimated (see Section~\ref{sec:explosion_parameters}) and therefore the observed average $\log(L_\peak/\mni)$ values are overestimated. Reducing the explosion epochs of the observed SNe~IIb, Ib, Ic, and Ib/Ic by 5.0, 19.5, 16.8, and 5.8\,d (the differences of the mean $t_L^\peak$ values for models and observations), respectively, the observed average $\log(L_\peak/\mni)$ estimates reduce to $0.49\pm0.02$, $0.44\pm0.03$, $0.51\pm0.06$, and $0.54\pm0.03$\,dex, respectively. These values are at least $4\,\sigma$ greater than those computed from models. This means that, even if our explosion epochs are overestimated, the mean ratio of peak luminosity to $^{56}$Ni mass for models is 0.13--0.25\,dex lower than observations. The latter suggests that models of \citet{2016MNRAS.458.1618D} and \citet{2021ApJ...913..145W} underestimate the peak luminosity of SE~SNe, which is consistent with findings in previous studies (e.g. \citealt{2020ApJ...890...51E}; \citealt{2021ApJ...918...89A}; \citealt{2021ApJ...913..145W}; \citealt{2022AA...657A..64S}; see Section~\ref{sec:introduction}). This is further discussed in Section~\ref{sec:decay_powered_models}.

\begin{figure}
\includegraphics[width=1.0\columnwidth]{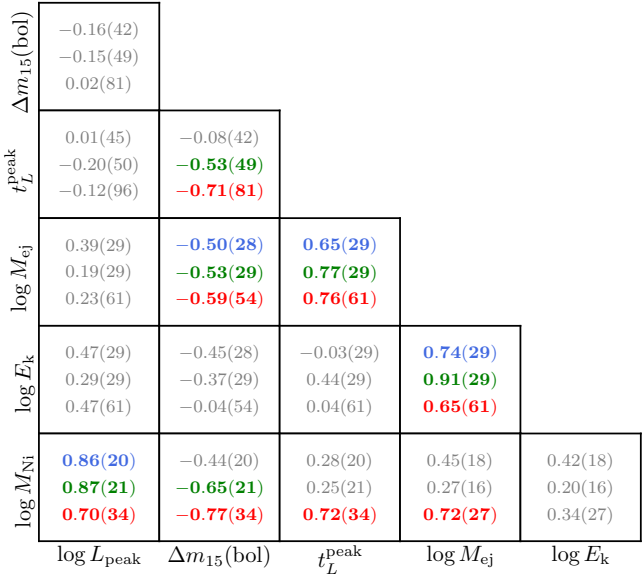}
\caption{Correlation matrix of the bolometric light-curve properties and explosion parameters. The Pearson correlation coefficient and the number of SNe (in parenthesis) are reported for SNe~IIb (top values), Ib (middle values), and Ic (bottom values). Correlations with $|r_\mathrm{P}|\geq0.5$ are indicated with bold colored numbers.} 
\label{fig:correlations}
\end{figure}

\begin{figure}
\includegraphics[width=1.0\columnwidth]{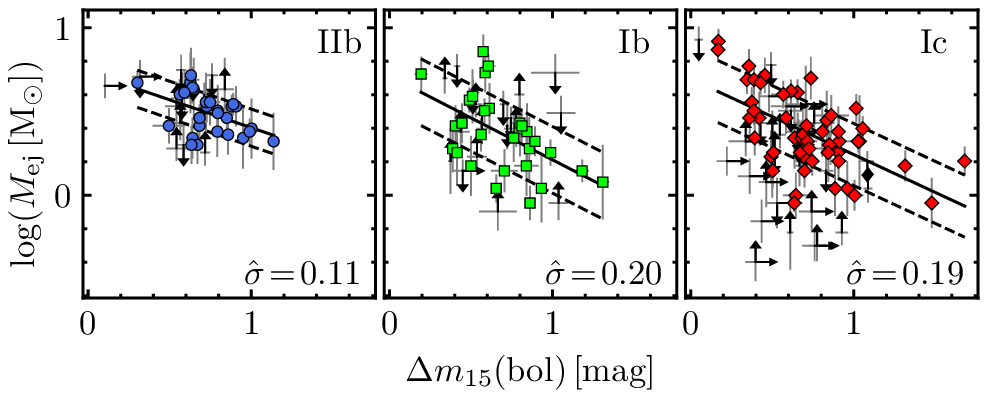}
\includegraphics[width=1.0\columnwidth]{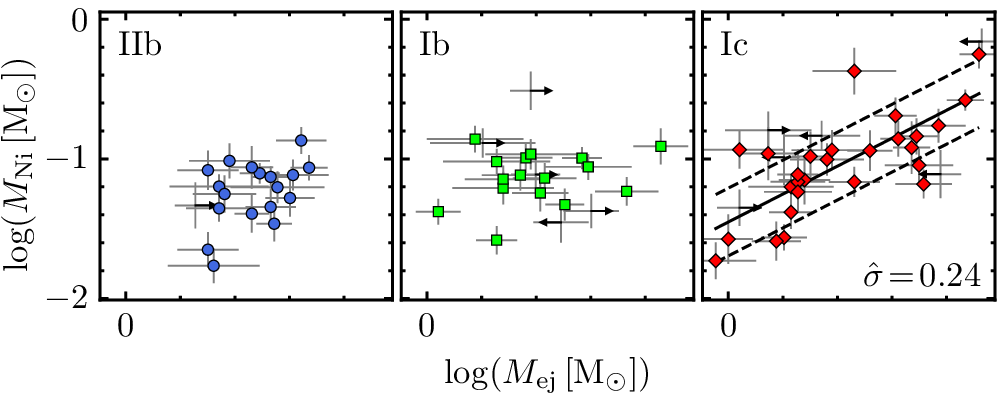}
\caption{Ejecta mass against decline rate (top panels) and $^{56}$Ni mass measured from the radioactive tail luminosity against ejecta mass (bottom panels). Solid lines correspond to linear fits, while dashed lines are the $\pm1\,\ssd$ limits. Arrows indicate lower/upper limits and error bars are $1\,\sigma$ errors.} 
\label{fig:corr_details}
\end{figure}

Figure~\ref{fig:correlations} summarizes the $r_\mathrm{p}$ values of the correlations between bolometric light-curve properties and explosion parameters for each SN subtype. We focus on correlations with $|r_\mathrm{P}|\geq0.5$ (i.e. moderate/strong correlations). For those correlations we measure $p_\mathrm{P}\leq0.003$, i.e. they are significant at the $>99.7$\% level. We recover the correlation between $\log\mni$ and $\log L_\peak$, and between $\log E_\mathrm{k}$ and $\log M_\mathrm{ej}$ reported in previous works \citep[e.g.][]{2016MNRAS.457..328L,2018AA...609A.136T,2019AA...621A..71T,2021AA...651A..81B}. We also recover the correlations between $\log\mni$ and $\log M_\mathrm{ej}$ for SNe~Ic \citep[e.g.][]{2019AA...621A..71T,2021AA...651A..81B}. The observed correlation between $\log M_\mathrm{ej}$ and $\Delta m_{15}(\mathrm{bol})$ (top panels of Figure~\ref{fig:corr_details}) is expected because the diffusion time increases with $M_\mathrm{ej}$ \citep[e.g.][]{1982ApJ...253..785A}. For SNe~Ib and Ic we detect a correlation between $t_L^\peak$ and $\Delta m_{15}(\mathrm{bol})$, and between $\log M_\mathrm{ej}$ and $t_L^\peak$. The first correlation is consistent with our findings for the $V$-band in Section~\ref{sec:texp}, while the second correlation is equivalent to the first one since $\log M_\mathrm{ej}$ and $\Delta m_{15}(\mathrm{bol})$ are correlated. For SNe~Ic we find correlations between $\log E_\mathrm{k}$ and $\log L_\peak$, $\log\mni$ and $\Delta m_{15}(\mathrm{bol})$ (also for SNe~Ib), and between $\log\mni$ and $t_L^\peak$. The latter two correlations are consistent with the picture that $\log\mni$ and $\log M_\mathrm{ej}$ correlate for SNe~Ic (bottom panels of Figure~\ref{fig:corr_details}). In our case, the latter correlation is given by
\begin{equation}
\log(\mni\,[\mathrm{M}_{\sun}]) = -1.452+\log(M_\mathrm{ej}\,[\mathrm{M}_{\sun}])\pm0.24
\end{equation}
($1\,\ssd$ error), or $\mni/M_\mathrm{ej}=4.1_{-1.8}^{+3.1}$\%.

\subsection{Peak time-luminosity relation}\label{sec:LtDM}
\begin{figure*}
\includegraphics[width=1.0\textwidth]{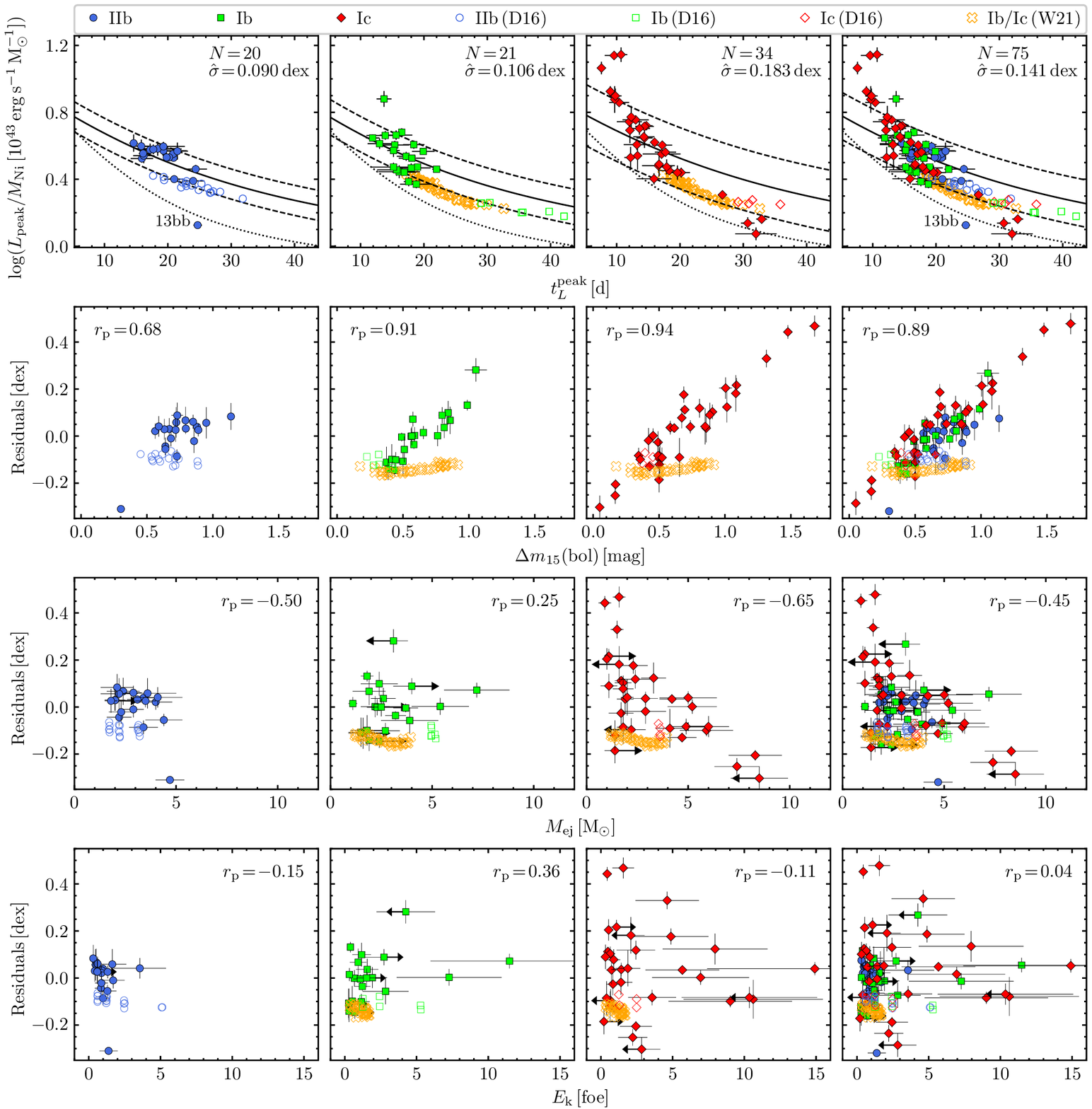}
\caption{Top row: peak time-luminosity relations for SE~SNe. Solid lines correspond to the \citet{2019ApJ...878...56K} relation (equation~\ref{eq:KK19_MNi}) fitted to observations (filled symbols), and dashed lines are $\pm1\,\ssd$ limits. Dotted lines correspond to Arnett's rule. Second, third, and fourth rows: residuals against decline rate, ejecta mass, and kinetic energy, respectively, where the Pearson correlation coefficient is indicated in each panel. Error bars are $1\,\sigma$ errors, and arrows indicate lower/upper limits. We also show the models of \citet{2016MNRAS.458.1618D} and \citet{2021ApJ...913..145W} with empty symbols.}
\label{fig:KK19}
\end{figure*}

The top panels of Figure~\ref{fig:KK19} show $\log (L_\peak/\mni)$ as a function of $t_L^\peak$ (the peak time-luminosity relation) for each SN subtype and for SE~SNe as a whole. We see that the distributions for different SN subtypes are similar. In particular, SN~IIb~2013bb (the most massive H/He-rich SNe yet found, \citealt{2019MNRAS.485.1559P}) is located close to the slow-rising ($t_L^\peak>30$\,d) SNe~Ic. This is consistent with the finding of \citet{2021ApJ...913..145W}, who suggest that SN~2013bb belongs to the SNe~Ib/Ic group but with a more massive progenitor and with a small residual H envelope. In the figure we also show the models of \citet{2016MNRAS.458.1618D} and \citet{2021ApJ...913..145W}.

Table~\ref{table:beta} lists the $\beta$ values for each SN subtype and for SE~SNe as a whole obtained by fitting the \citet{2019ApJ...878...56K} relation (equation~\ref{eq:KK19_MNi}) to our data (Column~2) and to the models of \citet{2016MNRAS.458.1618D} ($\beta_\mathrm{D}$, Column~4), along with the empirical mean $\beta$ values reported by \citet{2021ApJ...918...89A} and their $\ssd/\sqrt{N}$ errors ($\beta_\mathrm{A}$, Column~3). For completeness, we also compute $\beta$ for SNe~Ic classified as Ic-BL and for SNe~Ic, Ib/Ic, and SE~SNe without including SNe~Ic-BL. As expected, the $\beta_\mathrm{D}$ value for SE~SNe is consistent with $\beta=9/8$ reported by \citet{2019ApJ...878...56K}, which is based on the SN models of \citet{2016MNRAS.458.1618D}. Our $\beta$ estimates are, in general, greater than the $\beta_\mathrm{A}$ values but consistent with them to within 0.1--2.3\,$\sigma$. In addition, our $\beta$ values reduce but do not resolve the tension detected by \citet{2021ApJ...918...89A} between empirical $\beta$ values and those based on the models of \citet{2016MNRAS.458.1618D}. We also find a discrepancy of $6\,\sigma$ between our $\beta$ value for SNe~Ib/Ic and $\beta=1.36$ reported by \citet{2021ApJ...913..145W} for their SN~Ib/Ic models.

\begin{deluxetable}{lcccc}
\tablecaption{$\beta$ parameters for the peak time-luminosity relation\label{table:beta}}
\tablehead{  
 \colhead{Type} &  \colhead{$\beta$} &  \colhead{$\beta_\mathrm{A}$} & \colhead{$\beta_\mathrm{D}$} 
}
\startdata 
IIb                    & $0.88\pm0.07$(20) & $0.78\pm0.07$(8)  & $1.15\pm0.02$(17) \\
Ib                     & $0.90\pm0.08$(21) & $0.66\pm0.07$(8)  & $1.19\pm0.05$(6)  \\
Ic                     & $0.81\pm0.10$(34) & --                & --                \\
Ic\tablenotemark{a}    & $0.86\pm0.13$(25) & $0.88\pm0.31$(4)  & $1.15\pm0.11$(4)  \\
Ic-BL                  & $0.64\pm0.19$(9)  & $0.56\pm0.12$(7)  & --                \\
Ib/Ic\tablenotemark{a} & $0.88\pm0.08$(46) & --                & $1.18\pm0.03$(10) \\
SE\tablenotemark{a}    & $0.88\pm0.06$(66) & --                & $1.16\pm0.02$(27) \\
SE                     & $0.85\pm0.05$(75) & $0.70\pm0.07$(27) & --                \\
\enddata
\tablecomments{Uncertainties are $1\,\sigma$ errors. Numbers in parentheses are the sample sizes.}
\tablenotetext{a}{Without including SNe~Ic-BL.}
\end{deluxetable}

\begin{figure}
\includegraphics[width=\columnwidth]{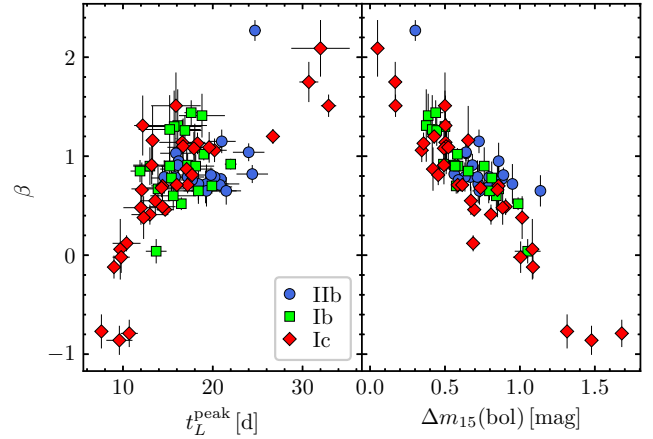}
\caption{Empirical $\beta$ values for SE~SNe against peak time (left-hand panel) and decline rate (right-hand panel).}
\label{fig:beta}
\end{figure}

As demonstrated by \citet{2019ApJ...878...56K}, their relation (shown as solid lines in Figure~\ref{fig:KK19}) provides a more accurate representation of the peak time-luminosity relation than Arnett's rule (dotted lines). However, we notice that SNe with $t_L^\peak$ lower and greater than 15\,d are located systematically above and below the \citet{2019ApJ...878...56K} relation, respectively, indicating that $\beta$ varies significantly from one SN to another. This can be seen in the left-hand panel of Figure~\ref{fig:beta}, which shows $\beta$ for each SN computed with equation~(\ref{eq:KK19_MNi}) against peak time. The $\beta$ values range from $-0.86$ to 2.27, and tend to decrease as $t_L^\peak$ decreases. In particular, eight SNe have $\beta\lesssim0$ where $\beta$ values $\leq0$ are  physically meaningless. This, along with the fact that the distribution of our SNe in the peak time-luminosity space is not well described by equation~(\ref{eq:KK19_MNi}) with a single $\beta$ value, casts doubt on the usefulness of the \citet{2019ApJ...878...56K} relation for estimating $^{56}$Ni masses of SE~SNe.

The second, third, and fourth rows of Figure~\ref{fig:KK19} show the residuals of the \citet{2019ApJ...878...56K} relation against decline rate, ejecta mass, and kinetic energy, respectively. We find strong correlations between residuals and decline rate for SNe~IIb ($r_\mathrm{p}=0.68$), SNe~Ib ($r_\mathrm{p}=0.91$), SNe~Ic ($r_\mathrm{p}=0.94$), and all SE~SNe ($r_\mathrm{p}=0.89$), and between residuals and ejecta mass for SNe~Ic ($r_\mathrm{p}=-0.65$). On the other hand, the residuals for the models of \citet{2016MNRAS.458.1618D} and \citet{2021ApJ...913..145W}\footnote{Since \citet{2021ApJ...913..145W} do not provide $\Delta m_{15}(\mathrm{bol})$, we use their $\Delta m_{15}(V)$ values and the transformation $\Delta m_{15}(\mathrm{bol})=-0.10+0.93\Delta m_{15}(V)$ obtained with our SN sample.} show only a small variation with $\Delta m_{15}(\mathrm{bol})$ or $M_\mathrm{ej}$. The right-hand panel of Figure~\ref{fig:beta} shows $\beta$ against decline rate, where we find strong correlations between both parameters ($r_\mathrm{p}$ of $-0.67$, $-0.91$, $-0.94$, and $-0.88$ for SNe~IIb, Ib, Ic and all SE~SNe, respectively). This confirms that $\beta$ is not constant for SE~SNe, but instead decreases with decline rate. Therefore, the \citet{2019ApJ...878...56K} relation with a single $\beta$ value is not very useful to estimate individual $^{56}$Ni masses of SE~SNe. Despite this, using an empirical average $\beta$ value it would still be possible to estimate the mean $^{56}$Ni mass of a SN sample but with a large error. 
Given the strength of the correlation between the observed residuals and decline rates for all SE~SNe, we attempt to include the decline rate as an additional parameter in the calibration of the peak time-luminosity relation.

\begin{figure}
\includegraphics[width=\columnwidth]{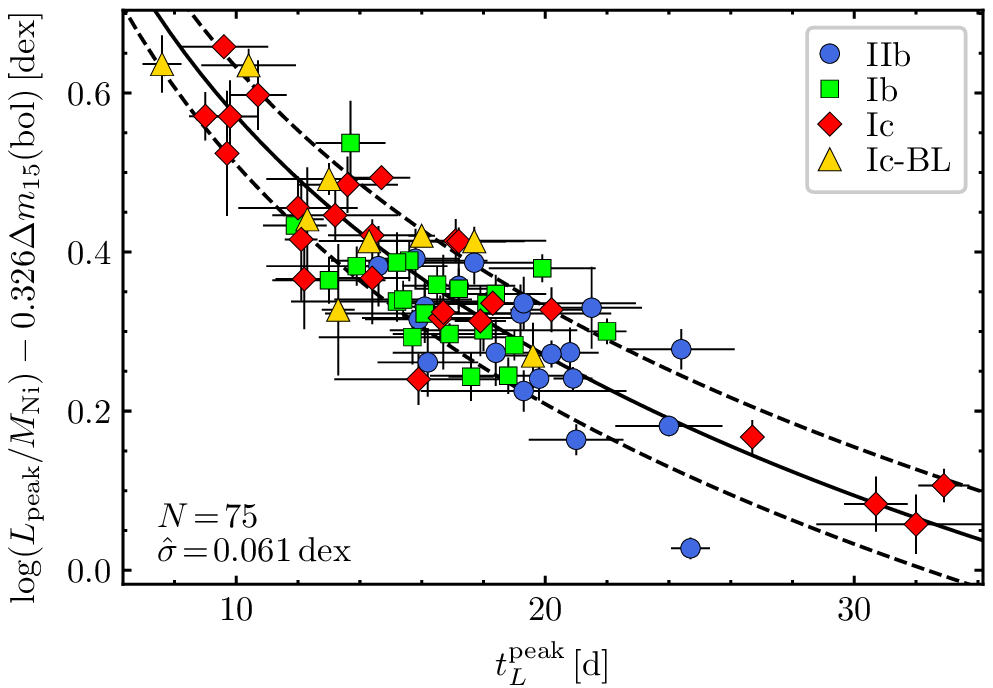}
\includegraphics[width=\columnwidth]{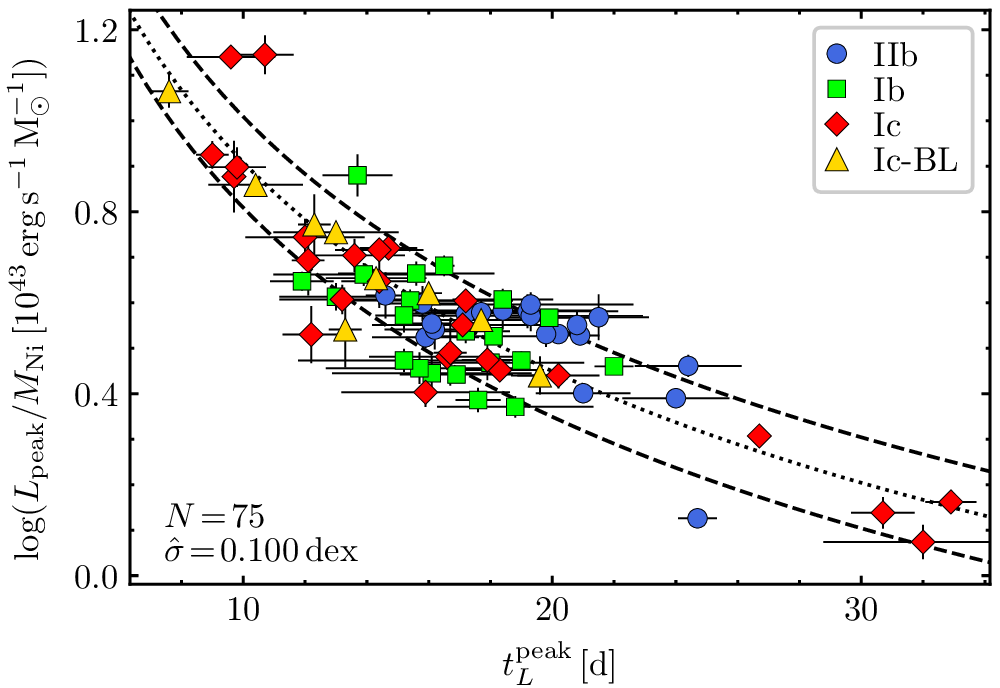}
\caption{Peak time-luminosity relation for SE~SNe, corrected (top panel) and uncorrected for decline rate (bottom panel). 
Solid and dotted lines are the best fits given by equations~(\ref{eq:NLT_Dm}) and ~(\ref{eq:NLT_tp}), respectively, while dashed lines are $\pm1\,\ssd$ limits around the fits.}
\label{fig:MNi_from_peak}
\end{figure}

The top panel of Figure~\ref{fig:MNi_from_peak} shows the peak time-luminosity relation for SE~SNe corrected for the decline rate, which we parameterize as
\begin{equation}\label{eq:NLT_Dm}
\log(L_\peak/\mni)-b\Delta m_{15}(\mathrm{bol}) =a-\log\left(\frac{t_L^\peak}{10\,\text{d}}\right),
\end{equation} 
where $a=0.571\pm0.020$, $b=0.326\pm0.028$, and $\sigma_0=0.044$\,dex. By applying the correction for decline rate, the $\ssd$ value reduces from 0.141\,dex (see top-right panel of Figure~\ref{fig:KK19}) to 0.061\,dex. We also propose an alternative calibration for the peak time-luminosity relation uncorrected for decline rate, given by
\begin{equation}\label{eq:NLT_tp}
\log(L_\peak/\mni) = \log\left[\frac{2(1+t_L^\peak/t_*)}{(t_L^\peak/t_*)^2}\right].
\end{equation}
with $t_*=15.74\pm0.28$\,d, $\sigma_0=0.081$\,dex, and $\ssd=0.100$\,dex. This calibration, shown in the bottom-panel of Figure~\ref{fig:MNi_from_peak}, is useful for those SNe without $\Delta m_{15}(\mathrm{bol})$ estimates. We note that, in the peak time-luminosity relations shown in Fig.~\ref{fig:MNi_from_peak}, the SNe classified as Ic-BL (yellow triangles) are not distinct from the rest of the SNe~Ic.

\begin{figure}
\includegraphics[width=\columnwidth]{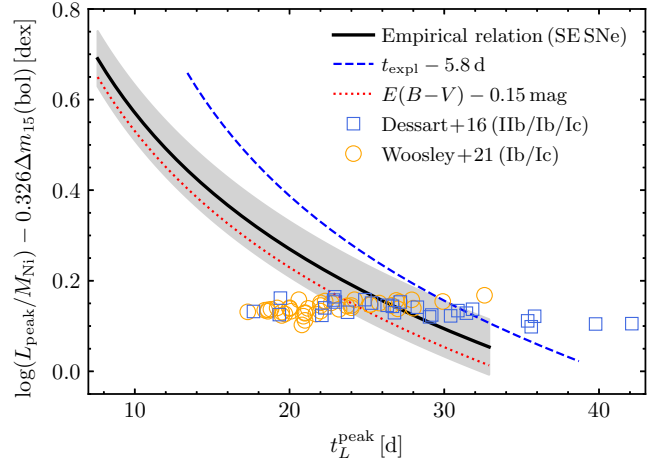}
\caption{Empirical peak time-luminosity relation corrected for decline rate (solid black line) along with its $\pm1\,\ssd$ region (gray region). The red dotted (blue dashed) line shows the effect on the empirical relation of decreasing $E(B-V)$ by 0.15\,mag ($t_\mathrm{expl}$ by 5.8\,d). Models of \citet{2016MNRAS.458.1618D} and \citet{2021ApJ...913..145W} are shown with empty symbols.}
\label{fig:PLR_obs_models}
\end{figure}

Figure~\ref{fig:PLR_obs_models} shows the empirical peak time-luminosity relation for SE~SNe corrected for decline rate (solid black line) along with the location of models of \citet{2016MNRAS.458.1618D} and \citet{2021ApJ...913..145W}. We see that unlike the empirical relation, which decreases logarithmically with peak time, the $\log(L_\peak/\mni)-0.326\Delta m_{15}(\mathrm{bol})$ estimates for models remain nearly constant with an average value of $0.138\pm0.002$\,dex ($\ssd/\sqrt{N}$ error). In the four-parameter space shown in the figure, models of \citet{2016MNRAS.458.1618D} and \citet{2021ApJ...913..145W} are consistent with observations only for SNe with $t_L^\peak\sim27$\,d. Since 98\% of the SNe in our sample have $t_L^\peak<27$\,d, we can state that models of \citealt{2016MNRAS.458.1618D} and \citealt{2021ApJ...913..145W} in general overestimate (underestimate) $\mni$ ($L_\peak$) for SE~SNe for a given peak time, decline rate, and peak luminosity ($^{56}$Ni mass). Note that if we reduce $t_\mathrm{expl}$ by 5.8\,d to have an average $t_L^\peak$ for SNe~Ib/Ic equal to that for models of \citet{2021ApJ...913..145W} (see Table~\ref{table:model_characteristics}), then the empirical relation becomes steeper (blue dashed line) and the discrepancy between observations and models increases. The figure also shows the effect on the empirical relation of decreasing $E(B-V)$ by 0.15\,mag (the median $E(B-V)$ of our SN sample), which produces a vertical displacement of only $-0.04$\,dex. Therefore, the calibration of the peak time-luminosity relation is little affected by changes in $E(B-V)$.

\subsection{$^{56}$Ni mass distribution}\label{sec:MNi_distribution}
\subsubsection{Mean $^{56}$Ni mass}
The $\log\mni$ values computed with equation~(\ref{eq:NLT_Dm}) or (\ref{eq:NLT_tp}) ($\log\mni^\peak$) are listed in Column~9 of Table~\ref{table:MNi}. We adopt the weighted average of $\log\mni^\peak$ and $\log\mni$ measured from the radioactive tail as the final $\log\mni$, which are collected in Column~10 of Table~\ref{table:MNi}. The $\mni$ values and their errors, given by
\begin{equation}
\mni = 10^{\log\mni+0.5\phi}
\end{equation}
and
\begin{equation}
\sigma_{\mni}=\mni\sqrt{10^\phi-1}
\end{equation}
respectively \citep{2021MNRAS.505.1742R} with $\phi=\ln(10)\sigma_{\log\mni}^2$, are listed in Column~11 of Table~\ref{table:MNi}.

\begin{figure}
\includegraphics[width=1.0\columnwidth]{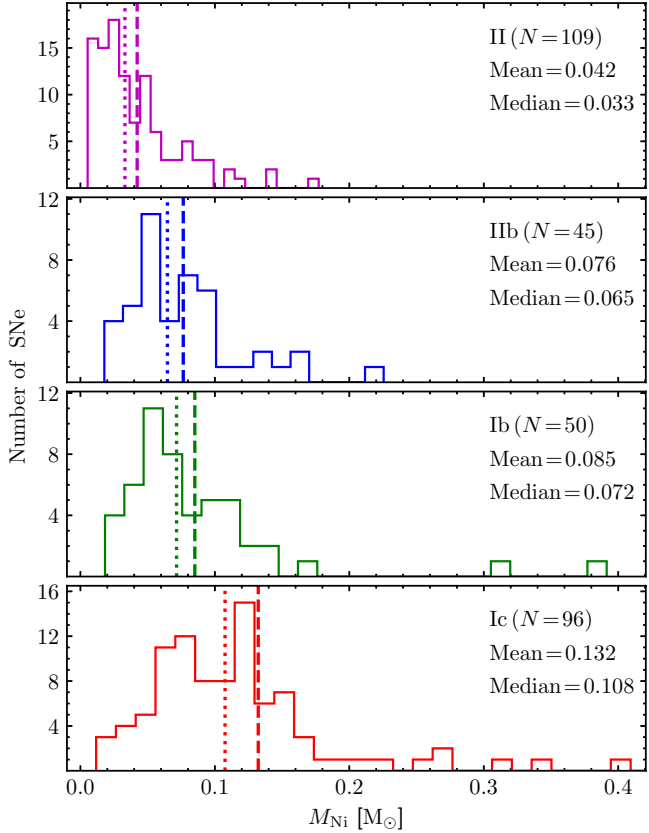}
\caption{$^{56}$Ni mass distribution for each SN subtype. Dashed and dotted lines indicate mean and median $^{56}$Ni mass values, respectively. The four SN~Ic events with $^{56}$Ni masses between 0.45 and 0.70\,M$_{\sun}$ are not shown for clarity. For comparison, we include the distribution for the SN~II sample of \citet{2021MNRAS.505.1742R}.} 
\label{fig:MNi_hist_ecdf}
\end{figure}

Figure~\ref{fig:MNi_hist_ecdf} shows the $\mni$ histograms for each SN subtype, while Table~\ref{table:mean_MNi} summarizes the statistics of those distributions. To evaluate possible systematics on the derived mean $\mni$ resulting from reddening effects, we re-compute the mean $\mni$ using only SNe with $E_{B-V}$ lower than the median $E_{B-V}$ for each SN subtype (Column~9). The resulting mean values are consistent with the previous result to within $0.7\,\sigma$. We also compare the mean $\mni$ values with those computed using $\mni$ from the peak (Column~10) and the radioactive tail (Column~11) luminosity. These estimates are again consistent at $1.3\,\sigma$.

\begin{deluxetable*}{ccccccccccc}
\tablecaption{Statistics of the $^{56}$Ni mass distributions\label{table:mean_MNi}}
\tablehead{
 \colhead{Type} & \colhead{$N$} & \colhead{Min} & \colhead{Max} & \colhead{Median} & \colhead{Mean} & \colhead{$\ssd$} & \colhead{$\sigma$} & \multicolumn{3}{c}{Mean}\\
  \cline{9-11} 
 \nocolhead{}   & \nocolhead{} & \nocolhead{} & \nocolhead{} & \nocolhead{} & \nocolhead{} & \nocolhead{} & \nocolhead{} & \colhead{$\mni$ with low ${E_{B-V}}$\tablenotemark{a}}& \colhead{Peak $\mni$} & \colhead{Tail $\mni$}
}
\startdata
IIb & $45$ & $0.018$ & $0.226$ & $0.065$ & $0.076$ & $0.043$ & $0.006$ & $0.070\pm0.007$ & $0.077\pm0.007$ & $0.065\pm0.006$ \\
Ib  & $50$ & $0.018$ & $0.391$ & $0.072$ & $0.085$ & $0.065$ & $0.009$ & $0.095\pm0.017$ & $0.085\pm0.009$ & $0.091\pm0.014$ \\
Ic  & $96$ & $0.012$ & $0.704$ & $0.108$ & $0.132$ & $0.113$ & $0.012$ & $0.130\pm0.018$ & $0.135\pm0.012$ & $0.145\pm0.026$ \\
\enddata
\tablecomments{Values are in M$_{\sun}$ units. $\sigma=\ssd/\sqrt{N}$ is the standard error of the mean.}
\tablenotetext{a}{Mean $^{56}$Ni mass and $\sigma$ error estimated with SNe having $E_{B-V}$ lower that the median $E_{B-V}$.}
\end{deluxetable*}

\subsubsection{Selection bias correction}\label{sec:selection_bias} 
Our SN sample, collected from the literature and the ZTF BTS, is potentially affected by selection bias. To correct for this bias, we use the VL samples given in Section~\ref{sec:EBV_accuracy} as approximations for complete samples.

\begin{figure}
\includegraphics[width=1.0\columnwidth]{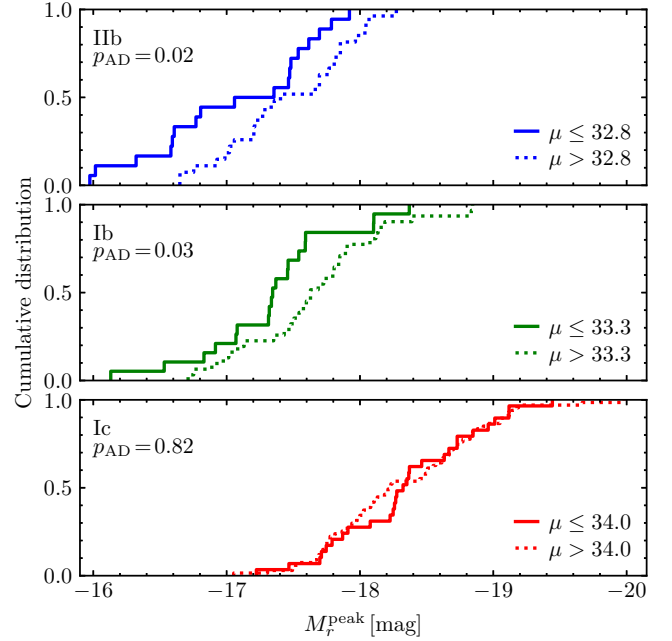}
\caption{Cumulative distributions for the $M_r^\peak$ values in the volume-limited (solid lines) and non-volume-limited (dotted lines) samples.} 
\label{fig:CDF_Mr_peak}
\end{figure}

Figure~\ref{fig:CDF_Mr_peak} shows the cumulative distributions for the $M_r^\peak$ values in the VL samples (solid lines) and those for the SNe at $\mu>\mu_\mathrm{VL}$, which we refer as the non-VL (NVL) samples. Using the $k$-sample AD test we obtain $p_\mathrm{AD}$ values of 0.02, 0.03, and 0.82 for the VL and NVL $M_r^\peak$ distributions of SNe~IIb, Ib, and Ic, respectively. For SNe~Ic the null hypothesis that the VL and NVL samples are drawn from a common $M_r^\peak$ distribution cannot be rejected at the 82\% significance level, so we expect that the mean $^{56}$Ni masses of the VL and NVL samples be statistically consistent between them. Indeed, the mean $\mni$ values of the latter samples ($0.116\pm0.018$ and $0.139\pm0.015$\,M$_{\sun}$, respectively) are consistent within $1\,\sigma$. We therefore adopt the mean $^{56}$Ni mass of the full sample ($\langle\mni\rangle=0.132\pm0.012$\,M$_{\sun}$) as representative of a VL sample. In the case of SNe~IIb and Ib, the mean $\mni$ value of the NVL sample is 2.5 and $2.0\,\sigma$ greater than for the VL sample, respectively, indicating the need for selection-bias correction. 

To correct the SNe~IIb and Ib samples for selection bias, we proceed as in \citet{2021MNRAS.505.1742R}. First, we divide both the VL and the NVL sample into absolute magnitude groups, and assume that the brightest groups are less affected by selection bias. Then, in order to have the same ratio between the number of SNe in the brightest group ($N_\mathrm{bight}$) and in a dimmer group ($N_\mathrm{dim}$) for the VL and NVL sample, we have to include $M$ dim SNe to the NVL sample, such that
\begin{equation}\label{eq:number_ratio}
\left(\frac{N_\mathrm{bight}}{N_\mathrm{dim}}\right)_\mathrm{VL} = \left(\frac{N_\mathrm{bight}}{N_\mathrm{dim}+M}\right)_\mathrm{NVL}.
\end{equation}

In our case, given the low number of SNe in the VL samples, we divide the VL and NVL samples into two groups containing SNe brighter and dimmer than the mean $M_r^\peak$ value for the SNe in the VL samples ($-17.059$ for SNe~IIb and $-17.340$ for SNe~Ib). For SNe~IIb (Ib) we obtain $N_\mathrm{bight}/N_\mathrm{dim}$ ratios of 10/8 (10/9) and 20/7 (24/7) for the VL and NVL sample, respectively. Therefore, following equation~(\ref{eq:number_ratio}), we have to include 9 (15) dim SNe from the VL sample to the NVL sample of SNe~IIb (Ib).

The mean $^{56}$Ni mass corrected for selection bias can be written as
\begin{equation}
\langle\mni\rangle=\langle\mni\rangle_\mathrm{unc}-\mathrm{sbc},
\end{equation}
where sbc is the selection bias correction, given by
\begin{equation}
\mathrm{sbc}=\frac{M}{N+M}(\langle\mni\rangle_\mathrm{unc}-\langle\mni\rangle_{M})
\end{equation}
\citep{2021MNRAS.505.1742R}, while $\langle\mni\rangle_\mathrm{unc}$ and $\langle\mni\rangle_M$ are the mean $^{56}$Ni mass computed with the full SN sample (of size $N$) and with the $M$ dim SNe, respectively. Performing $10^5$ simulations, where the $M$ SNe are randomly selected from the VL dim sample, for SNe~IIb and Ib we compute a sbc value of $0.010\pm0.001$ and $0.003\pm0.001$\,M$_{\sun}$, respectively. Therefore for these two SN subtypes we obtain $\langle\mni\rangle$ estimates of $0.066\pm0.006$ and $0.082\pm0.009$\,M$_{\sun}$, respectively.

We find that the $\langle\mni\rangle$ values for SNe~IIb and Ib are consistent to within $1.5\,\sigma$, and the individual $\mni$ values may be drawn from a common distribution ($p_\mathrm{AD}=0.62$). This has been previously suggested by \citet{2021ApJ...918...89A} based on 8~SNe~IIb and 8~SNe~Ib. For SNe~Ic we find that their $\langle\mni\rangle$ value is $4.9\,\sigma$ and $3.3\,\sigma$ greater than those of SNe~IIb and Ib, respectively. Specifically, we can reject the hypothesis that the $\mni$ values of SNe~Ib and Ic are drawn from a common distribution ($p_\mathrm{AD}=0.02$), contrary to the findings of \citet{2019AA...628A...7A} and \citet{2021ApJ...918...89A}. In the first work the author used a collection of $^{56}$Ni masses generally measured with Arnett's rule, and hence those estimates are suspect. In the second work, the results are based on small numbers (8~SNe~Ib and 4~SNe~Ic). We also compute $\langle\mni\rangle$ for SNe~Ic-BL (26~SNe) and for Ic without Ic-BL (70~SNe), obtaining values of $0.138\pm0.019$ and $0.130\pm0.014$\,M$_{\sun}$, respectively, that are consistent at $0.3\,\sigma$. Thus, in terms of the $^{56}$Ni mass, there is no systematic difference between SNe~Ic-BL and other SNe~Ic.

Our $\langle\mni\rangle$ values for SNe~IIb, Ib, and Ic are 1.8, 2.2, and 3.6 times greater, respectively, than the value for SNe~II reported by \citet{2021MNRAS.505.1742R} ($0.037\pm0.005$\,M$_{\sun}$) at significance levels of $3.7\,\sigma$, $4.4\,\sigma$, and $7.3\,\sigma$, respectively. This finding is again consistent with previous works \citep[e.g.][]{2015arXiv150602655K,2019AA...628A...7A,2020AA...641A.177M,2020MNRAS.496.4517S,2021MNRAS.505.1742R,2021ApJ...918...89A}.

\subsection{Mean iron yield}
Using the relative fractional SN rates provided in \citet{2017PASP..129e4201S}, the mean $^{56}$Ni mass of SE and CC~SNe is given by
\begin{equation}
\langle\mni\rangle_\mathrm{SE}=0.36\langle\mni\rangle_\mathrm{IIb}+0.356\langle\mni\rangle_\mathrm{Ib}+0.284\langle\mni\rangle_\mathrm{Ic}
\end{equation}
and
\begin{equation}
\langle\mni\rangle_\mathrm{CC}=0.696(\pm0.067)\langle\mni\rangle_\mathrm{II}+0.304(\pm0.05)\langle\mni\rangle_\mathrm{SE},
\end{equation}
respectively. With our $\langle\mni\rangle$ estimates for SNe~IIb, Ib, and Ic, and $\langle\mni\rangle_\mathrm{II}=0.039\pm0.005$\,M$_{\sun}$ (which includes long-rising and CSM-interacting SNe) reported in \citet{2021MNRAS.505.1742R}, we obtain $\langle\mni\rangle_\mathrm{SE}$ and $\langle\mni\rangle_\mathrm{CC}$ of $0.090\pm0.005$ and $0.055\pm0.006$\,M$_{\sun}$, respectively. Of the mean $^{56}$Ni mass produced by CC~SNe, 50\% is produced by SE~SNe and 50\% by SNe~II. Finally, assuming a $M_\mathrm{Fe}$ to $\mni$ ratio of $1.07\pm0.04$ \citep[e.g.][]{2021MNRAS.505.1742R}, we obtain a mean iron yield of $0.097\pm0.007$ and $0.058\pm0.007$\,M$_{\sun}$ for SE and CC~SNe, respectively. Our mean iron yield for CC~SNe is 22\% lower than the estimate of 0.074\,M$_{\sun}$ that \citet{2017ApJ...848...25M} derived, based on previous estimates of $\langle\mni\rangle$ for the various SN types.

The distances we have used in this work and those used in \citet{2021MNRAS.505.1742R} are based on a Cepheid-calibrated Hubble parameter $H_0=74$~km\,s$^{-1}$\,Mpc$^{-1}$. For a different choice of $H_0$, our $^{56}$Ni mass and iron yield measurements will scale as ${(74/H_0)}^{2}$.

\section{Discussion}\label{sec:discussion}
\subsection{Comparison with other works}\label{sec:comparison_with_other_works}
We now compare the $\mni$ values we have computed from the radioactive tail ($\mni^\tail$) with $^{56}$Ni masses reported in the literature. We do not include SN samples with $^{56}$Ni masses computed from pseudo-bolometric light curves \citep[e.g.][]{2013MNRAS.434.1098C,2019MNRAS.485.1559P,2020AA...641A.177M} because, given that those light curves are not corrected for the unobserved UV and IR flux, the inferred $\mni$ values are in principle lower limits. From the literature we select $^{56}$Ni masses reported by \citet{2016MNRAS.458.2973P} and \citet{2016MNRAS.457..328L}, computed with Arnett's rule and the Arnett model, respectively, the $\mni$ values computed with the \citet{2019ApJ...878...56K} relation (K\&K) by \citet{2021ApJ...918...89A}, and the $\mni^\tail$ estimates reported by \citet{2020MNRAS.496.4517S} and \citet{2021ApJ...918...89A}. In \citet{2016MNRAS.458.2973P}, peak luminosities were calculated using fluxes integrated over bands from $u/U$ to $K$ and assuming a 10\% contribution from the unobserved UV and IR flux, while $^{56}$Ni masses were computed with Arnett's rule.
In \citet{2016MNRAS.457..328L}, luminosities were estimated with the BC technique and the BC calibration of \citet{2014MNRAS.437.3848L,2016MNRAS.457..328L}, while the $^{56}$Ni masses were estimated with an updated version of the Arnett model given by \citet{2008MNRAS.383.1485V}. In \citet{2020MNRAS.496.4517S}, luminosities were calculated via flux integration over bands from $u/U$ to $H/K$, along with corrections for the missing UV and IR flux. To compute $\mni^\tail$, \citet{2020MNRAS.496.4517S} replaced the deposition function (equation~\ref{eq:fdep}) with a more versatile function (with two free parameters). In \citet{2021ApJ...918...89A}, luminosities were estimated with the BC technique and the BC calibration of \citet{2014MNRAS.437.3848L,2016MNRAS.457..328L}. To calculate $^{56}$Ni masses with the K\&K relation, \citet{2021ApJ...918...89A} used their empirically calibrated $\beta$ values for each SN subtype, while to compute $\mni^\tail$ the authors neglected the term describing the escape of positrons (equation~\ref{eq:q_pos}). In addition, 20~SNe in the present work have $^{56}$Ni masses estimated by comparing hydrodynamical models with bolometric light curves (16~SNe) or pseudo-bolometric light curves with fluxes integrated from $u/U$ to $K$ (SNe~1993J, 1994I, 2008D, and 2011fu). These SNe, along with the reported $\mu$, $E_{B-V}^\mathrm{MW}$, $E_{B-V}$, $R_V$, $t_\mathrm{expl}$, $L_\peak$, and inferred $^{56}$Ni masses are collected in Table~\ref{table:MNi_hydro}. For the comparison with the $\mni$ values from the literature, we rescale our $\mni^\tail$ values using the distances, reddenings, and explosion epochs adopted in the respective works. For the comparison we also have to compensate for differences in luminosity due to differences in the techniques used to calculate luminosity. For this, we divide our $\mni^\tail$ values by the ratio of $L_\peak$ from the literature to our $L_\peak$ values (which we rescale using the distances and reddenings from the literature).

\begin{figure}
\includegraphics[width=1.0\columnwidth]{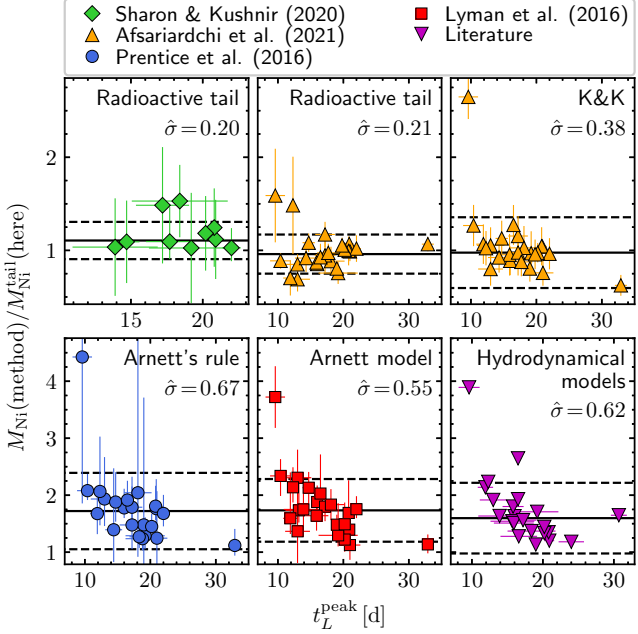}
\caption{Ratios of the $^{56}$Ni masses estimated with different methods in the literature to our recalibrated $\mni^\tail$ values against peak time. Solid lines are median values, dashed lines are $\pm1\,\ssd$ limits, and error bars are $1\,\sigma$ errors.} 
\label{fig:MNi_comparison}
\end{figure}

Figure~\ref{fig:MNi_comparison} shows the ratios of the $^{56}$Ni masses computed in the literature with the methods mentioned above to the $\mni^\tail$ values estimated here, against peak times. The median ratios of the $\mni^\tail$ values ($\ssd$ values) reported by \citet{2020MNRAS.496.4517S} ($N=10$) and \citet{2021ApJ...918...89A} ($N=24$) to those estimated here are of 1.11 (0.20) and 0.96 (0.21), respectively. Our $\mni^\tail$ estimates are, on average, consistent to within $1\,\ssd/\sqrt{N}$ with those computed with the methodology of \citet{2021ApJ...918...89A}. The $\mni^\tail$ values of \citet{2020MNRAS.496.4517S} are, on average, $1.7\,\ssd/\sqrt{N}$ greater than our estimates. This difference, which is due to the deposition function adopted by \citet{2020MNRAS.496.4517S}, will be discussed further in Section~\ref{sec:fdep}.
The median ratio of the $^{56}$Ni masses computed with the K\&K relation by \citet{2021ApJ...918...89A} ($N=24$) to our $\mni^\tail$ estimates is of 0.98, which is consistent with unity. This is expected because the $\beta$ values used by \citet{2021ApJ...918...89A} to compute $\mni$ were empirically calibrated against $\mni^\tail$. We see, however, that the K\&K relation tends to overestimate (underestimate) the $^{56}$Ni masses of SNe with peak times lower (greater) than 18\,d, which is expected given that the peak time-luminosity relation of SE~SNe is not well described by the \citet{2019ApJ...878...56K} relation (see Section~\ref{sec:LtDM}).

In the bottom panels of Figure~\ref{fig:MNi_comparison} we see that the $^{56}$Ni mass ratios for Arnett's rule ($N=22$), the Arnett model ($N=23$), and hydrodynamical models ($N=22$) are greater than unity, and tend to increase as peak times decrease. This behavior in the case of Arnett's rule can also be seen in top panels of Figure~\ref{fig:KK19}, where the difference between the observed $\log(L_\peak/\mni)$ values and Arnett's rule is greater for SNe with short peak times. The median $^{56}$Ni mass ratios ($\ssd$ values) for the three methods mentioned above are of 1.72 (0.67), 1.73 (0.55) and 1.60 (0.62), respectively, which are at least $4.5\,\ssd/\sqrt{N}$ greater than unity. This indicates that the Arnett model and Arnett's rule systematically overestimate the $^{56}$Ni mass of SE~SNe typically by 70\%, while hydrodynamical models typically overestimate it by 60\%. To quantify more accurately the typical overestimation of $\mni$ by Arnett's rule, we estimate $^{56}$Ni masses using $\mni=L_\peak/(q_\gamma(t_L^\peak)+q_\text{pos}(t_L^\peak))$ and our $t_L^\peak$ and $\log L_\peak$ values listed in Table~\ref{table:MNi}, and compare them with our $\mni^\tail$ estimates. We find a median $^{56}$Ni mass ratio of $1.72\pm0.06$ ($\ssd/\sqrt{N}$ error), which is equal to our previous estimate and confirms that Arnett's rule overestimate $\mni$ of SE~SNe typically by 70\%. We also find a mean $^{56}$Ni mass ratio of $1.77\pm0.06$, which is lower than the average overestimation factor of $\sim2$ found by \citet{2021ApJ...918...89A} from observations, and greater than the average overestimation factor of 1.30--1.41 suggested by \citet{2016MNRAS.458.1618D} and \citet{2021ApJ...913..145W} based on their models.

Table~\ref{table:mean_MNi_lit} summarizes the mean $^{56}$Ni masses (uncorrected for selection bias) we obtain for SNe~IIb, Ib, Ic without Ic-BL, and Ic-BL, along with the values reported by \citet{2019MNRAS.485.1559P}, \citet{2019AA...628A...7A}, and \citet{2021ApJ...918...89A}. In these three works, respectively, the authors computed mean $^{56}$Ni masses using the largest samples of SE~SNe to date with $^{56}$Ni masses computed from SNe analyzed homogeneously (80~SNe), with $\mni$ values collected from the literature (143~SNe), and with $\mni^\tail$ estimates (27~SNe). Our mean $^{56}$Ni masses are statistically consistent to those of \citet{2021ApJ...918...89A}, while the mean $^{56}$Ni masses of \citet{2019AA...628A...7A} are 1.5--3.7 times larger than our estimates. This is in part because the $^{56}$Ni masses collected and used by \citet{2019AA...628A...7A} were generally estimated with Arnett's rule. We find a good agreement between our mean values and those of \citet{2019MNRAS.485.1559P}, even though their $\mni$ estimates were computed with Arnett's rule and the Arnett model, which overestimate $\mni$ typically by 70\%. This is because the $L_\peak$ estimates of \citet{2019MNRAS.485.1559P}, which only include fluxes in the wavelength range 400--1000\,nm, are on average 1.71 times lower than our $L_\peak$ values. Therefore, the underestimated $L_\peak$ values and the overestimated $^{56}$Ni masses in \citet{2019MNRAS.485.1559P} offset each other, resulting in mean $^{56}$Ni masses which are, on average, similar to our results.

\begin{deluxetable}{lccccc}
\tablecaption{Mean $^{56}$Ni masses reported in the literature.\label{table:mean_MNi_lit}}
\tablehead{
 \colhead{Type} & \colhead{$N$} & \colhead{Mean $\mni$ (M$_\sun$)} & & \colhead{$N$} & \colhead{Mean $\mni$ (M$_\sun$)} 
}
\startdata
 \nocolhead{} & \multicolumn{2}{c}{This work} & & \multicolumn{2}{c}{\citet{2021ApJ...918...89A}} \\ 
 \cline{2-3} \cline{5-6}
 IIb   & 45 & $0.076\pm0.006$ & &  8 & $0.060\pm0.007$ \\
 Ib    & 50 & $0.085\pm0.009$ & &  8 & $0.110\pm0.039$ \\
 Ic    & 70 & $0.130\pm0.014$ & &  4 & $0.200\pm0.110$ \\
 Ic-BL & 26 & $0.138\pm0.019$ & &  7 & $0.150\pm0.026$ \\
 \nocolhead{} & \multicolumn{2}{c}{\citet{2019AA...628A...7A}} & &\multicolumn{2}{c}{\citet{2019MNRAS.485.1559P}} \\
 \cline{2-3} \cline{5-6}
 IIb   & 27 & $0.124\pm0.012$ & & 21 & $0.070\pm0.030$ \\
 Ib    & 33 & $0.199\pm0.025$ & & 25 & $0.090\pm0.060$ \\
 Ic    & 48 & $0.198\pm0.020$ & & 19 & $0.110\pm0.090$ \\
 Ic-BL & 32 & $0.507\pm0.072$ & & 11 & $0.150\pm0.070$ \\
\enddata
\tablecomments{Uncertainties are the standard error of the mean. SNe~Ic do not include SNe~Ic-BL.}
\end{deluxetable}

\subsection{Comparison with neutrino-driven explosion models}
Various neutrino-driven explosion models have predicted maximum possible values for the synthesized $^{56}$Ni mass in CC~SN explosions, of 0.15 to 0.226\,M$_\sun$ \citep[e.g.][]{2012ApJ...757...69U,2015ApJ...801...90P,2016ApJ...821...38S,2019MNRAS.483.3607S,2020ApJ...890...51E,2021ApJ...913..145W}. Although in our sample there are some SNe with $^{56}$Ni masses $1\,\sigma$ greater than these theoretical limits, we include them when computing mean $^{56}$Ni masses, as there is no observational evidence supporting the existence of such upper limits (e.g. differences in any other characteristics of such high-yield SNe). Based on spherically symmetric neutrino-driven explosion models, \citet{2021ApJ...913..145W} reported an IMF-averaged $^{56}$Ni yield of $0.09$\,M$_\sun$ for models consistent with SNe~Ib/Ic (excluding SNe~Ic-BL), whose $\mni$ do not exceed 0.15\,M$_\sun$. In our sample, there are two SNe~Ib and five SNe~Ic (excluding SNe~Ic-BL) with $^{56}$Ni masses exceeding 0.15\,M$_\sun$ by $>1\,\sigma$. If we nonetheless exclude those events, we obtain mean $^{56}$Ni masses of $0.074\pm0.005$ and $0.105\pm0.008$\,M$_\sun$ for SNe~Ib and Ic, respectively. Accounting for their relative rates (see Section~\ref{sec:correlations}), for SNe Ib/Ic we obtain a mean $\mni$ of $0.086\pm0.004$\,M$_\sun$, which is consistent with the value reported by \citet{2021ApJ...913..145W}.

\subsection{Systematics}
\subsubsection{Sample completeness}
In Section~\ref{sec:selection_bias} we used VL samples as a reference to evaluate the selection bias correction. However, a volume-limited sample is merely an approximation for a complete sample. In particular, our VL samples could underestimate the fraction of faint ($M_r^\peak>-16.5$) SNe~IIb/Ib due to Malmquist bias. In addition, \citet{2021ApJ...922..141O} show that if SE~SNe with $\mni<0.01$ (corresponding to $M_r^\peak>-15.5$) exist, then they will be missing from samples collected from the literature. If the fraction of faint SE~SNe is intrinsically low, then their omission from the sample should not have a severe effect on the mean $^{56}$Ni and iron yields. However, if faint SE~SNe are actually common, then the mean $^{56}$Ni and iron yields we report should be considered upper limits.

Recently, \citet{2022MNRAS.515..897R} computed progenitor luminosities ($L_\mathrm{prog}$) for a sample of 112 SNe~II and compared them to the luminosities of red supergiants (RSGs, identified as the progenitors of SNe~II, e.g. \citealt{2009MNRAS.395.1409S}) in LMC, SMC, M31, and M33. In particular, after correcting for selection bias, \citet{2022MNRAS.515..897R} found a smaller fraction of SN~II with $L_\mathrm{prog}<10^{4.6}$\,L$_\sun$ compared to RSG samples. This is most likely because the VL sample used to evaluate the selection bias correction is not complete for low-luminosity SNe~II. Given that these SNe produce a low amount of $^{56}$Ni and given that the SN sample of \citet{2022MNRAS.515..897R} is virtually identical to that of \citet{2021MNRAS.505.1742R}, the mean $^{56}$Ni mass we adopted for SNe~II and the $\langle\mni\rangle_\mathrm{CC}$ value we compute are potentially overestimated. To roughly quantify this overestimation, we calculate the fraction of missing SNe~II with $L_\mathrm{prog}<10^{4.6}$\,L$_\sun$ in the selection-bias corrected sample of \citet{2022MNRAS.515..897R} by assuming that all RSGs with luminosity between $10^{4.361}$ and $10^{5.091}$\,L$_\sun$ (the minimum and maximum $L_\mathrm{prog}$ value in the SN sample of \citealt{2022MNRAS.515..897R}, respectively) explode as SNe~II. Using the the RSG samples used in \citet{2022MNRAS.515..897R}, we find that the fraction of missing low-luminosity SNe~II is 0.29, while the mean $^{56}$Ni mass of the SNe~II with $L_\mathrm{prog}<10^{4.6}$\,L$_\sun$ is $0.011$\,M$_\sun$. Therefore, if all RSGs with luminosity between $10^{4.361}$ and $10^{5.091}$\,L$_\sun$ indeed explode as SNe~II,
then the mean $^{56}$Ni mass for SNe~II is reduced by 0.008\,M$_\sun$, while our $\langle\mni\rangle_\mathrm{CC}$ value reduces by 0.0057\,M$_\sun$. Since the possible overestimation of $\langle\mni\rangle_\mathrm{CC}$ is similar to the $\langle\mni\rangle_\mathrm{CC}$ error of $0.006$\,M$_\sun$, this uncertainty do not have a severe effect on the mean $^{56}$Ni and iron yields of CC~SNe.

\subsubsection{Deposition function}\label{sec:fdep}
An important assumption we make in this work is that the deposition function of SE~SNe is well represented by the formula of \citet{1997ApJ...491..375C} (equation~\ref{eq:fdep}, hereafter $f_\mathrm{dep}^\mathrm{C\&W}$), which was derived analytically under simplistic but reasonable assumptions (at least to first order). Recently, \citet{2020MNRAS.496.4517S} proposed
\begin{equation}\label{eq:fdep_SK20}
 f_\mathrm{dep}^\mathrm{S\&K}(t,t_\mathrm{esc},n)=(1+(t/t_\mathrm{esc})^n)^{-2/n}
\end{equation}
as a more appropriate deposition function for SNe. This function is more versatile than $f_\mathrm{dep}^\mathrm{C\&W}$ because it includes a second free parameter, $n$, which controls the sharpness of the transition between regions optically thick ($f_\mathrm{dep}=1$) and thin ($f_\mathrm{dep}=(t_\mathrm{esc}/t)^2$ for $t\gg t_\mathrm{esc}$) to $\gamma$-rays. To evaluate whether $f_\mathrm{dep}^\mathrm{S\&K}$ provides a better fit to observations in the radioactive tail than $f_\mathrm{dep}^\mathrm{C\&W}$, we compute $\log\mni^\tail$ using the methodology described in Section~\ref{sec:MNi_tail} with $f_\mathrm{dep}^\mathrm{S\&K}$ as deposition function, and compare the $\ssd$ estimates around the $\log\mni^\tail$ values that maximize the posterior probability ($\ssd_{\log\mni^\tail}$, e.g. the $\ssd$ value in the bottom panel of Figure~\ref{fig:2009jf_MNi}) with the $\ssd$ estimates computed with $f_\mathrm{dep}^\mathrm{C\&W}$. Among the 67~SNe that require the inclusion of a deposition function, 28 of them have $n\leq1$, which is not a reasonable value because the derivative of equation~(\ref{eq:fdep_SK20}) diverges at $t\rightarrow 0$ \citep[see][]{2020MNRAS.496.4517S}.

\begin{figure}
\includegraphics[width=1.0\columnwidth]{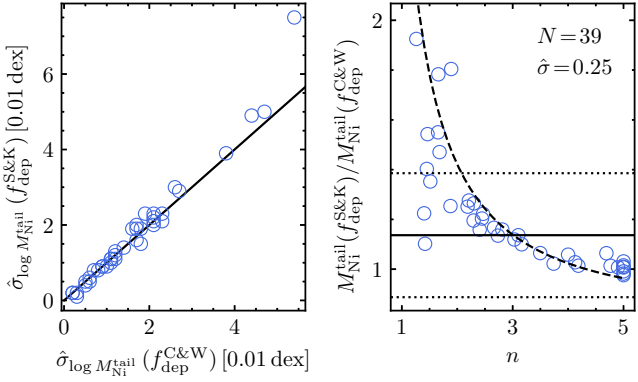}
\caption{Left: $\ssd$ values of the $\log\mni^\tail$ estimates calculated with $f_\mathrm{dep}^\mathrm{S\&K}$ against those computed with $f_\mathrm{dep}^\mathrm{C\&W}$ for 39~SNe with $n>1$. The solid line is a one-to-one correspondence. Right: ratios of the $^{56}$Ni masses estimated with $f_\mathrm{dep}^\mathrm{S\&K}$ to those computed with $f_\mathrm{dep}^\mathrm{C\&W}$ against $n$. The solid line is the median value and dotted lines are $\pm1\ssd$ limits. The dashed line is $f_\mathrm{dep}^\mathrm{C\&W}/f_\mathrm{dep}^\mathrm{S\&K}$ with $t_\mathrm{esc}/t=1.37$ for $f_\mathrm{dep}^\mathrm{C\&W}$ and $t_\mathrm{esc}/t=1.22$ for $f_\mathrm{dep}^\mathrm{S\&K}$.}
\label{fig:ssd_MNi_ratio}
\end{figure}

The left-hand panel of Figure~\ref{fig:ssd_MNi_ratio} shows the $\ssd_{\log\mni^\tail}$ values for the $^{56}$Ni masses computed using $f_\mathrm{dep}^\mathrm{S\&K}$ against those calculated using $f_\mathrm{dep}^\mathrm{C\&W}$ for the 39~SNe with $n>1$. We find that the distribution can be expressed as a straight line with a slope of unity and a $y$-intercept of 0.001\,dex, meaning that the $\ssd_{\log\mni^\tail}$ values for the $^{56}$Ni masses computed using $f_\mathrm{dep}^\mathrm{S\&K}$ are slightly greater than those calculated with $f_\mathrm{dep}^\mathrm{C\&W}$. Therefore, for SE~SNe with $n>1$, $f_\mathrm{dep}^\mathrm{S\&K}$ fits observations in the radioactive tail as well as, but not better than, $f_\mathrm{dep}^\mathrm{C\&W}$.

The right-hand panel of Figure~\ref{fig:ssd_MNi_ratio} shows the ratios of the $\mni^\tail$ values calculated with $f_\mathrm{dep}^\mathrm{S\&K}$ to those computed with $f_\mathrm{dep}^\mathrm{C\&W}$ against $n$. The observed dependence of these ratios on $n$, especially for $n>2$, arises because in the radioactive tail $L\approx\mni\,q_\gamma\, f_\mathrm{dep}$ (equations~\ref{eq:tail}--\ref{eq:qt}), so the $^{56}$Ni mass ratio is $\approx f_\mathrm{dep}^\mathrm{C\&W}/f_\mathrm{dep}^\mathrm{S\&K}$. Indeed, the distribution is close to $f_\mathrm{dep}^\mathrm{C\&W}/f_\mathrm{dep}^\mathrm{S\&K}$ (dashed line) with $t_\mathrm{esc}/t=1.37$ for $f_\mathrm{dep}^\mathrm{C\&W}$ and $t_\mathrm{esc}/t=1.22$ for $f_\mathrm{dep}^\mathrm{S\&K}$, corresponding to the median $t_\mathrm{esc}/t$ values for the 39~SNe with $n>1$. The median $^{56}$Ni mass ratio is $1.14\pm0.04$ ($\ssd/\sqrt{N}$ error), which is consistent with the overestimation factor that we obtain in Section~\ref{sec:comparison_with_other_works} and indicates that $\mni^\tail$ values computed with $f_\mathrm{dep}^\mathrm{S\&K}$ are typically 14\% greater than those computed with $f_\mathrm{dep}^\mathrm{C\&W}$. In the case that $f_\mathrm{dep}^\mathrm{S\&K}$ is the true deposition function of SE~SNe, then our $\langle\mni\rangle_\mathrm{CC}$ estimate increases by 0.004\,M$_\sun$, which is lower than the $\langle\mni\rangle_\mathrm{CC}$ error of 0.006\,M$_\sun$. Thus, a systematic uncertainty of 14\% on $^{56}$Ni mass due to the adopted deposition function does not have a severe effect on the the mean $^{56}$Ni and iron yields of CC~SNe.

\subsection{Implications for radioactive $^{56}$Ni-powered models\label{sec:decay_powered_models}}
As we found in Section~\ref{sec:comparison_with_other_works}, the Arnett model, Arnett's rule, and hydrodynamical models typically overestimate the $^{56}$Ni masses of SE~SNe by a factor of 1.60--1.73, so these models underestimate $L_\peak$ for a given $^{56}$Ni mass typically by  60--70\%. In addition, as mentioned in Section~\ref{sec:LtDM}, models of \citet{2016MNRAS.458.1618D} and \citet{2021ApJ...913..145W} underestimate $L_\peak$ for a given $^{56}$Ni mass, decline rate, and peak time. In particular, for a median $t_L^\peak$ of 15.9\,d (the median for our SN sample), these models underestimate $L_\peak$ typically by 70\%. Given that the Arnett model, Arnett's rule, hydrodynamical models used in the references listed in Table~\ref{table:MNi_hydro}, and models of \citet{2016MNRAS.458.1618D} and \citet{2021ApJ...913..145W} assume that the SN light curves at peak are powered only by the radioactive $^{56}$Ni decay chain, we can state that radioactive $^{56}$Ni-powered models typically underestimate peak luminosities for SE~SNe by 60--70\%.

A more direct way to evaluate whether a given set of models underestimates the light at peak is by comparing the absolute magnitude of the brightest model with those of observed SE~SNe. In the case of the SN~Ib/Ic models of \citet{2021ApJ...913..145W}, the brightest model have $M_r^\peak=-17.8$\,mag (e.g. \citealt{2022AA...657A..64S}). Among the 48~SNe~Ib and 65~SNe~Ic (without Ic-BL) in our sample with $\mni$ that not exceed 0.15\,M$_\sun$ by $>1\,\sigma$, 27\% and 71\% have $M_r^\peak<-17.8$, respectively, while if we use only SNe with $E_{B-V}$ lower than the median $E_{B-V}$ for each SN subtype, the percentages reduces to 17\% and 56\%, respectively. This indicates that an important fraction of SNe~Ib and Ic are brighter than predicted by the models of \citet{2021ApJ...913..145W}. To estimate lower limits for these fractions, we use $M_r^\peak$ values uncorrected for host-galaxy reddening ($M_{r,\mathrm{unc}}^\peak$), finding 6~SNe~Ib and 14~SNe~Ic with $M_{r,\mathrm{unc}}^\peak<-17.8$. Therefore, at least 12\% of SNe~Ib and 22\% of SNe~Ic are brighter than predicted by the models of \citet{2021ApJ...913..145W}. These percentages are lower than the 36\% reported by \citet{2022AA...657A..64S}, which is most likely due to their sample is more affected by selection bias than our sample.

It is important to note that the \citet{2019ApJ...878...56K} relation does not depend only on $\beta$ but also, and most importantly, on the heating source. For the analysis of SE~SNe, in this work and in the literature \citep[e.g.][]{2019ApJ...878...56K,2021ApJ...918...89A,2021ApJ...913..145W} the radioactive $^{56}$Ni decay chain is adopted as heating source. Therefore, the accurate fit of the \citet{2019ApJ...878...56K} relation to the peak time-luminosity relation for the numerical models of \citet{2016MNRAS.458.1618D} and \citet{2021ApJ...913..145W} with a single $\beta$ value (e.g. Figure~15 of \citealt{2019ApJ...878...56K} and Figure~A1 of \citealt{2021ApJ...913..145W}) is expected because those models are powered by the radioactive $^{56}$Ni decay chain. The fact that the \citet{2019ApJ...878...56K} relation for radioactive $^{56}$Ni-powered transients with a single $\beta$ value does not fit observations suggests that SE~SNe at peak may not be powered only by the radioactive $^{56}$Ni decay chain. This evidence, along with the fact that radioactive $^{56}$Ni-powered models typically underestimate peak luminosities suggests the possible presence of an additional power source contributing to the luminosity of SE~SNe at peak. In \citet{Rodriguez_nature} we use the bolometric light curves and $^{56}$Ni masses computed in this work to show that, in all or most SE~SNe, a central engine in addition to radioactive $^{56}$Ni decay operates on timescales of hours to days after the explosion. The central engine contributes significantly to the luminosity at those times, and its properties appear consistent with those of magnetars.

\subsection{Reddening versus EW$_\mathrm{NaID}$ relation}
\begin{figure}
\includegraphics[width=1.0\columnwidth]{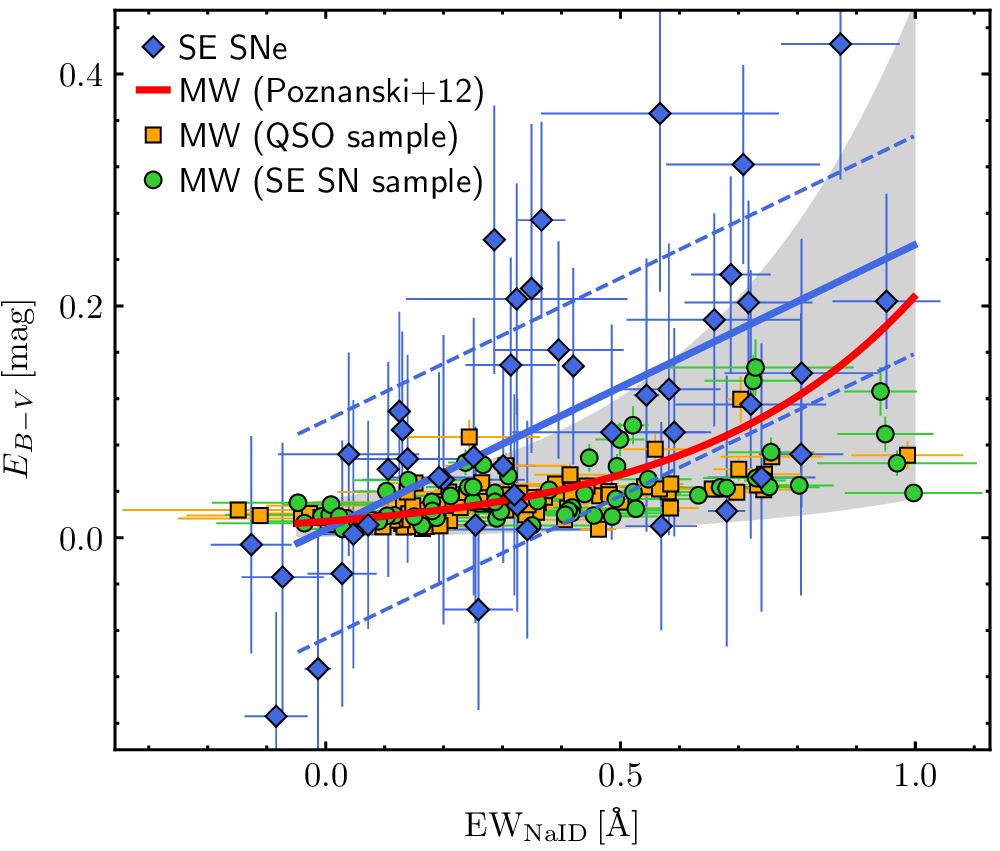}
\includegraphics[width=1.0\columnwidth]{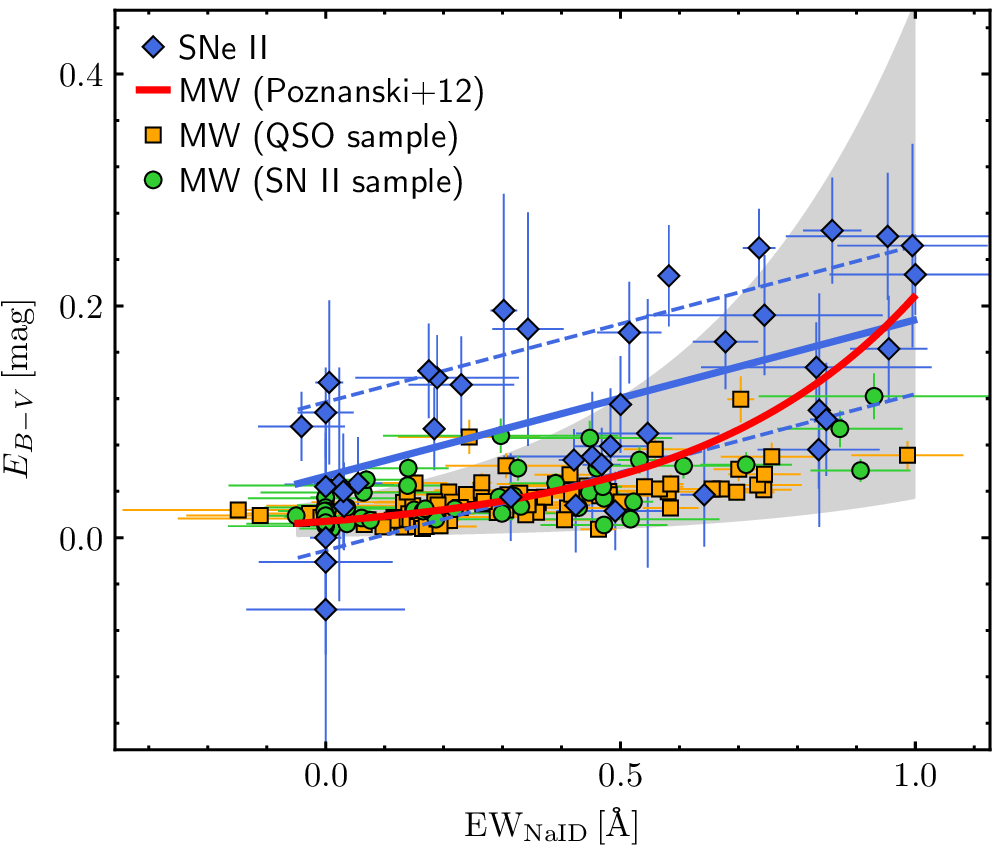}
\caption{Top panel: MW (green circles) and SE~SN (blue diamonds) reddenings against their respective \ion{Na}{1}\,D EWs measured from SE~SN spectra. The blue solid line is the $E_{B-V}$ versus $\mathrm{EW_{NaID}}$ relation for SE~SNe, while dashed lines are the $\pm1\,\ssd$ limits. Bottom panel: similar to the top panel but for SNe~II instead of SE~SNe. The red line and the gray region are the relation between MW reddening and $\mathrm{EW_{NaID}}$ reported by \citet{2012MNRAS.426.1465P} and its $\pm1\,\ssd$ region, respectively. Orange squares are the MW reddenings and $\mathrm{EW_{NaID}}$ values for the QSO sample of \citet{2012MNRAS.426.1465P}. Error bars are $1\,\sigma$ errors.}
\label{fig:EGBV_vs_EW}
\end{figure}

The top panel of Figure~\ref{fig:EGBV_vs_EW} shows the $E_{B-V}$ versus $\mathrm{EW_{NaID}}$ relation we obtained in Section~\ref{sec:EhBV} for SE~SNe (blue solid line) together with the relation of \citet{2012MNRAS.426.1465P} for the MW (red line) for $\mathrm{EW_{NaID}}<1$\,\AA. The panel also shows the MW \ion{Na}{1}\,D EWs reported by \citet{2012MNRAS.426.1465P} for a sample of high-resolution spectra of quasars (QSOs) and their MW reddenings from \citet{2011ApJ...737..103S} (orange squares). Our relation for SE~SNe provides $E_{B-V}$ values on average 0.05\,mag higher than those calculated with the \citet{2012MNRAS.426.1465P} relation. We also see that almost all MW values for the QSO sample are below our relation for SE~SNe, indicating a systematic difference between the latter relation and that of the MW. To evaluate a possible underestimation of our $\mathrm{EW_{NaID}}$ values computed from low-resolution spectra (e.g. due to contamination by host galaxy light, \citealt{2011MNRAS.415L..81P}), we also measure MW \ion{Na}{1}\,D EWs from the same spectra (Column~2 of Table~\ref{table:EhBV}). Those values along with the respective MW reddenings (Column~4 of Table~\ref{table:SN_sample}) are plotted in the panel as green circles. To represent the dependence of MW reddening on $\mathrm{EW_{NaID}}$ we use the linear relation $E_{B-V}=a+b\,\mathrm{EW_{NaID}}[\mathring{\mathrm{A}}]$, whose parameters for the QSO and SE~SN samples are listed in Table~\ref{table:EBV_NaID_pars}. For both samples we find strong correlations ($r_\mathrm{P}\approx0.6$ and $p_\mathrm{P}<0.001$) with parameters consistent within their errors. This means that our $\mathrm{EW_{NaID}}$ measurements from low-resolution spectra are consistent with those computed by \citet{2012MNRAS.426.1465P} from high-resolution spectra.

\begin{deluxetable}{lccccc}
\tablecaption{$E_{B-V}$ versus $\mathrm{EW_{NaID}}$ relation parameters\label{table:EBV_NaID_pars}}
\tablehead{
 \colhead{Dust source\tablenotemark{a}} & \colhead{$N$}   & \colhead{$a$}        & \colhead{$b$}         & \colhead{$\ssd$}  & \colhead{$r_\mathrm{P}$}
}  
\startdata
MW (QSOs)   &  $86$ & $0.014(2)$  & $0.048(6)$  & $0.015$ & $0.64$ \\
MW (SE~SNe) &  $62$ & $0.018(3)$  & $0.050(10)$ & $0.024$ & $0.61$ \\ 
MW (SNe~II) &  $54$ & $0.019(2)$  & $0.055(9)$  & $0.018$ & $0.68$ \\
MW (All)    & $202$ & $0.017(1)$  & $0.048(5)$  & $0.019$ & $0.63$ \\
SN II hosts &  $41$ & $0.053(14)$ & $0.135(28)$ & $0.065$ & $0.62$ \\
SE SN hosts &  $45$ & $0.007(24)$ & $0.246(54)$ & $0.094$ & $0.60$ \\
\enddata
\tablecomments{Numbers in parentheses are $1\,\sigma$ errors in units of 0.001.} 
\tablenotetext{a}{For the MW, in parentheses are indicated the spectra used to measure $\mathrm{EW_{NaID}}$.}
\end{deluxetable}

We have repeated the above analysis, but now using the SN~II sample of \citet{2021MNRAS.505.1742R}. The bottom panel of Figure~\ref{fig:EGBV_vs_EW} shows the MW (green circles) and SN~II (blue diamonds) reddenings against their respective \ion{Na}{1}\,D EWs, measured here from SN~II spectra (values reported in Table~\ref{table:EhBV_SNII}). We find a strong correlation between $E_{B-V}$ and $\mathrm{EW_{NaID}}$ for the MW and SNe~II, whose parameters are listed in Table~\ref{table:EBV_NaID_pars}. The parameters of the MW relation computed with the SN~II sample are consistent within their errors with those calculated with the QSO and SE~SNe samples. Combining those three samples we obtain a slope of $0.048\pm0.005$ for the MW $E_{B-V}$ versus $\mathrm{EW_{NaID}}$ relation, which is $3.0\,\sigma$ and $3.6\,\sigma$ lower than the slope for SNe~II and SE~SNe, respectively. If we use $R_V=3.1$ to compute $E_{B-V}$ in Section~\ref{sec:EhBV}, the slope of the relation for SE~SNe reduces to $0.234\pm0.055$, which is $3.4\,\sigma$ greater than the slope for the MW. Since the methods to compute $E_{B-V}$ for SE~SNe and SNe~II are systematically different, we can reject the possibility that the steeper relation between $E_{B-V}$ and $\mathrm{EW_{NaID}}$ for SNe~II and SE~SN compared to that of the MW is due to an overestimation of the SN reddenings. The evidence therefore suggests higher dust-to-gas ratios and/or lower neutral sodium to total gas ratios for regions where SE~SNe and SNe~II explode, compared to the diffuse interstellar medium of our Galaxy. For this reason, the relation of \citet{2012MNRAS.426.1465P} should not be used to measure $E_{B-V}$ from $\mathrm{EW_{NaID}}$ for SE~SNe and SNe~II. Instead, we recommend using equation~(\ref{eq:Eh_NaID}) and $E_{B-V}=0.053+0.135\,\mathrm{EW_{NaID}}[\mathring{\mathrm{A}}]$, respectively.

\subsection{LSST}
Within a few years, the Rubin Observatory Legacy Survey of Space and Time (LSST) will become the main source of photometric transient and variable data. Given the large number of CC~SNe that are expected to be discovered per year (around $10^5$, \citealt{2009JCAP...01..047L}), light-curve classifiers will be required to determine the class of those transients. In the case of SE~SNe, current light-curve classifiers achieve an accuracy of 50\% for the ZTF \citep[light curves in two bands sampled with a cadence of 3\,d,][]{2021AJ....161..141S} and of 74\% for light curves in four bands sampled with a cadence of 7--10\,d \citep{2020PASJ...72...89T}. A contamination of about 25\% is clearly not negligible. Therefore, if the classification accuracy does not improve much for the LSST (which will observe with six non-coeval photometric bands, with a cadence that is  yet to be decided), then the number of SNe from LSST useful to improve the mean iron yield of SE~SNe will be limited by the capability of carrying out spectroscopic classifications.

\section{Conclusions}\label{sec:conclusions}
We have carried out a systematic analysis of 191~SE~SN events, with the aim to estimate the mean $^{56}$Ni and iron yields of SE~SNe, and of CC~SNe in general. We have used color-curves to infer host galaxy reddenings and the representative $R_V$ value for each SN subtype. We have derived BCs for SNe~IIb, Ib, and Ic, with suitable data for this, which we have then used to construct bolometric light curves. Using those light curves, we have calculated $^{56}$Ni masses from the radioactive tail, and calibrated the relations between $^{56}$Ni mass and peak time/luminosity. Finally, we have used those relations to compute $^{56}$Ni masses for the entire sample.

Our main conclusions are the following:
\begin{enumerate}
\item For SNe~IIb, Ib, and Ic we find representative $R_V$ values of $2.6\pm0.4$, $2.7\pm0.5$, and $3.8\pm0.4$, respectively. Although SNe Ic may display a larger $R_V$ compared to SNe~IIb and Ib, we cannot rule out that the $R_V$ values of all SE~SNe are drawn from a common distribution. We further find that the host-galaxy reddening distributions for SNe~II and IIb are statistically similar, the said distribution for SNe~Ib is similar to those of SNe~II, IIb, and Ic, but the distribution for SNe~Ic is statistically different from those for SNe~II and SNe~IIb.

\item We find that the equation of \citet{2019ApJ...878...56K} for radioactive $^{56}$Ni-powered transients with a single $\beta$ value, which allows to relate peak time and luminosity to the $^{56}$Ni mass, is not very useful to estimate individual $^{56}$Ni masses for SE~SNe. The reason is that $\beta$ is not constant but decreases with decline rate, varying significantly between SNe. Instead, we have derived an empirically calibrated relation between peak time, peak luminosity, and $^{56}$Ni mass. This relation allows estimating $^{56}$Ni masses with a precision of 24\%. Furthermore, we have shown that the $^{56}$Ni mass correlates not only with the peak time and luminosity, but also with the decline rate. The correlation that we infer allows estimating $^{56}$Ni masses with an improved precision of 14\%.

\item We derive mean $^{56}$Ni masses of $0.066\pm0.006$, ${0.082\pm0.009}$, and $0.132\pm0.012$\,M$_{\sun}$ for SNe~IIb, Ib, and Ic, respectively. The $^{56}$Ni mass distributions of SNe~IIb and Ib are statistically similar, while SNe~Ic synthesize systematically more $^{56}$Ni than SNe~IIb and Ib. The mean $^{56}$Ni mass of each of these SE~SN subtypes is significantly greater than that of SNe~II.

\item For SE~SNe as a whole, we obtain mean $^{56}$Ni and iron yields of $0.090\pm0.005$ and $0.097\pm0.007$\,M$_{\sun}$, respectively. Combined with the recent mean $^{56}$Ni mass for SNe~II found by \citet{2021MNRAS.505.1742R}, we derive mean $^{56}$Ni and iron yields for CC~SNe as a whole, of $0.055\pm0.006$ and $0.058\pm0.007$\,M$_{\sun}$, respectively. Iron production in CC~SNe is split about 50-50 among SE~SNe and SNe~II.

\item Radioactive $^{56}$Ni-powered models, like those of \citet{2016MNRAS.458.1618D}, \citet{2021ApJ...913..145W}, and different hydrodynamical models in the literature, typically underestimate the peak luminosity for SE~SNe by 60--70\%. A possible explanation is the presence of a power source other than radioactive 56Ni decay that contributes to the luminosity at peak.

\item From analysing the correlation between reddening and $\mathrm{EW_{NaID}}$, we have found that the regions in which SE~SNe and SNe II explode have higher dust-to-gas ratios and/or lower neutral sodium to total gas ratios, compared to the diffuse interstellar medium of our Galaxy.
\end{enumerate}

Even though for some SE~SNe the radioactive $^{56}$Ni-powered models can provide $^{56}$Ni masses consistent with $\mni^\tail$, these models, in general, overestimate $\mni$. Therefore, they only provide an upper limit for $\mni$. For accurate estimations of $^{56}$Ni masses of SE~SNe, we recommend using the luminosity in the radioactive tail and equation~(\ref{eq:MNi_eq}) and/or the peak time, peak luminosity, and decline rate along with equation~(\ref{eq:NLT_Dm}) (or equation~\ref{eq:NLT_tp} if decline rate is not available).

\begin{acknowledgments}
We thank A. Aryan, R. Chornock, A. Gangopadhyay, D. K. Sahu, and  M. Stritzinger for sharing spectra with us.
This paper is part of a project that has received funding from the European Research Council (ERC) under the European Union's Seventh Framework Programme, Grant agreement No. 833031 (PI: Dan Maoz) and grant agreement No. 818899 (PI: Ehud Nakar).
Based on observations obtained with the Samuel Oschin 48-inch Telescope at the Palomar Observatory as part of the Zwicky Transient Facility project. ZTF is supported by the National Science Foundation under Grant No. AST-1440341 and a collaboration including Caltech, IPAC, the Weizmann Institute for Science, the Oskar Klein Center at Stockholm University, the University of Maryland, the University of Washington, Deutsches Elektronen-Synchrotron and Humboldt University, Los Alamos National Laboratories, the TANGO Consortium of Taiwan, the University of Wisconsin at Milwaukee, and Lawrence Berkeley National Laboratories. Operations are conducted by COO, IPAC, and UW.
The work made use of Swift/UVOT data reduced by P. J. Brown and released in the Swift Optical/Ultraviolet Supernova Archive (SOUSA). SOUSA is supported by NASA's Astrophysics Data Analysis Program through grant NNX13AF35G.
This research has made use of the NASA/IPAC Extragalactic Database (NED) which is operated by the Jet Propulsion Laboratory, California Institute of Technology, under contract with the National Aeronautics and Space Administration. 
This work has made use of the Weizmann Interactive Supernova Data Repository (\url{https://www.wiserep.org}).
This research has made use of the Spanish Virtual Observatory (\url{https://svo.cab.inta-csic.es}) project funded by MCIN/AEI/10.13039/501100011033/ through grant PID2020-112949GB-I00.
\end{acknowledgments}

\software{\textsc{emcee} \citep{2013PASP..125..306F},  
          \textsc{alr} \citep{2019MNRAS.483.5459R}
          \textsc{loess} \citep{Cleveland_etal1992}
          }



\appendix

\section{Synthetic magnitudes and effective wavelengths}\label{sec:syn_mag}
Given a SED $f_{\lambda}$ (in erg\,s$^{-1}$\,cm$^{-2}\,\mbox{\normalfont\AA}^{-1}$), we can compute the synthetic magnitude in the $x$-band using
\begin{equation}\label{eq:syn_mag}
m_x=-2.5\log{\int d\lambda S_{x,\lambda} \frac{\lambda f_{\lambda}}{hc}} + \mathrm{ZP}_{\mathrm{mag},x}
\end{equation}
\citep[e.g.][]{2001PhDT.......173H}. Here, $\lambda$ is the observed wavelength (in \AA), $S_{x,\lambda}$ is the peak-normalized $x$-band transmission function (considering a photon-counting detector), ${hc=1.986\times10^{-8}}$\,erg\,\AA, and $\mathrm{ZP}_{\mathrm{mag},x}$ is the zero-point for the magnitude system.

To convert magnitudes to monochromatic fluxes $\bar{f}_x$ (in erg\,s$^{-1}$\,cm$^{-2}$\,\AA$^{-1}$), we use ${f_{\lambda}=\bar{f}_x}$ in equation~(\ref{eq:syn_mag}), thus obtaining equation~(\ref{eq:f_eff}), where
\begin{equation}
\mathrm{ZP}_{\mathrm{flux},x}=2.5\log{\int d\lambda S_{x,\lambda} \frac{\lambda}{hc} } +\mathrm{ZP}_{\mathrm{mag},x}.
\end{equation}

The $\mathrm{ZP}_{\mathrm{mag},x}$ and $\mathrm{ZP}_{\mathrm{flux},x}$ values for the Johnson-Kron-Cousins $BV\!RI$, Sloan $gri$, ZTF $r$ ($r_\ztf$), and 2MASS $J\!H\!K$ bands were reported in \citet{2021MNRAS.505.1742R}. Here we compute the corresponding values for the Johnson $U$, ZTF $g$ ($g_\ztf$), CSP $Y$, and Swift UVOT $w2$, $w1$, and $u$ bands.

To compute $\mathrm{ZP}_{\mathrm{mag},x}$ for $w2$, $w1$, $u$, $U$, and $Y$ bands in the Vega system, we use equation~(\ref{eq:syn_mag}) along with the Vega SED published by \citet{2004AJ....127.3508B}\footnote{\url{https://ssb.stsci.edu/cdbs/current_calspec/alpha_lyr_stis_010.fits}} as $f_\lambda$. We adopt the transmission functions given in \citet{2005PASP..117..810S} for the $U$-band, in the CSP webpage\footnote{\url{https://csp.obs.carnegiescience.edu/data/filters}} for the $Y$-band, and in the SVO Filter Profile Service\footnote{\url{http://svo2.cab.inta-csic.es/theory/fps/}} \citep{2012ivoa.rept.1015R,2020sea..confE.182R} for the UVOT filters. We also adopt ${U=0.02}$ \citep{1996AJ....111.1748F}, ${Y=0.0}$ \citep{2006PASP..118....2H}, and $m_x=0.0$ for the UVOT filters \citep{2008MNRAS.383..627P}. To calculate $\mathrm{ZP}_{\mathrm{mag},g_\ztf}$ in the AB system, we use equation~(\ref{eq:syn_mag}), ${f_{\lambda}=1/\lambda^2}$ and ${m_x=-2.408}$. To construct the $g_\ztf$ transmission function, we use equation~(1) of \citet{2007MNRAS.376.1301P}, adopting the corresponding ZTF filter transmission\footnote{Since they consider an energy-counting detector, we have to divide the transmissions by $\lambda$ before to use them in equation~(\ref{eq:syn_mag}).\label{ec}} and CCD quantum efficiency,\footnote{\url{https://github.com/ZwickyTransientFacility/ztf_information}} the atmospheric extinction at Palomar Observatory of \citet{1975ApJ...197..593H} (assuming an airmass of 1.2) combined with atmospheric telluric lines, a standard aluminium reflectivity curve, and a constant lens throughput.

Table~\ref{table:zp_and_leff} collects the $\mathrm{ZP}_{\mathrm{mag},x}$ and $\mathrm{ZP}_{\mathrm{flux},x}$ values for the $w2$, $w1$, $u$, $U$, $g_\ztf$, and $Y$ bands, along with the estimates reported in \citet{2021MNRAS.505.1742R}.

\begin{deluxetable}{c c c c c}
\tablecaption{Properties of the filters used in this work\label{table:zp_and_leff}}
\tablehead{
 \colhead{$x$} &\colhead{System} & \colhead{$\mathrm{ZP}_{\mathrm{mag},x}$} & \colhead{$\mathrm{ZP}_{\mathrm{flux},x}$} & \colhead{$\bar{\lambda}_x$} \\
 \nocolhead{}  & \nocolhead{}    & \colhead{(mag)}                          & \colhead{(mag)}                           & \colhead{(\AA)}       
}
\startdata
 $w2       $ & Vega   & $13.837$ & $-20.677$ & \nodata      \\
 $w1       $ & Vega   & $14.092$ & $-20.956$ & \nodata      \\
 $u\tablenotemark{a}      $ & Vega   & $14.126$ & $-21.099$ & $ 3530\pm50$ \\
 $U        $ & Vega   & $14.229$ & $-20.953$ & $ 3630\pm40$ \\
 $B        $ & Vega   & $15.300$ & $-20.462$ & $ 4430\pm50$ \\
 $g        $ & AB     & $15.329$ & $-20.770$ & $ 4760\pm50$ \\
 $g_\ztf   $ & AB     & $15.382$ & $-20.802$ & $ 4850\pm50$ \\
 $V        $ & Vega   & $14.865$ & $-21.074$ & $ 5500\pm30$ \\
 $r        $ & AB     & $14.986$ & $-21.361$ & $ 6190\pm30$ \\
 $r_\ztf   $ & AB     & $15.212$ & $-21.436$ & $ 6400\pm30$ \\
 $R        $ & Vega   & $15.053$ & $-21.629$ & $ 6480\pm50$ \\
 $i        $ & AB     & $14.710$ & $-21.780$ & $ 7450\pm30$ \\
 $I        $ & Vega   & $14.538$ & $-22.354$ & $ 8030\pm70$ \\
 $Y        $ & Vega   & $13.933$ & $-23.133$ & $10440\pm60$ \\
 $J        $ & Vega   & $13.729$ & $-23.787$ & $12260\pm60$ \\
 $H        $ & Vega   & $13.412$ & $-24.886$ & $16470\pm20$ \\
 $K        $ & Vega   & $12.691$ & $-25.948$ & $21550\pm80$ \\
\enddata
\tablenotetext{a}{Swift $u$-band.}
\end{deluxetable}

The monochromatic fluxes (equation~\ref{eq:f_eff}) are associated to effective wavelengths $\bar{\lambda}_x$, given by
\begin{equation}\label{eq:leff}
\bar{\lambda}_x=\frac{\int d\lambda S_{x,\lambda} \lambda^2 f_{\lambda}}{\int d\lambda S_{x,\lambda} \lambda f_{\lambda}}
\end{equation}
\citep[e.g.][]{2012PASP..124..140B}, which depend not only on the transmission function but also on the SED of the observed object. To estimate $\bar{\lambda}_x$ for SE~SNe, we employ the spectra of the SNe in our sample as a proxy for $f_\lambda$. Since the wavelength range of our spectra do not cover the UV filters, for those bands we adopt a Planck function as $f_\lambda$ with temperatures between 3400 and 19000\,K, which produces intrinsic $\bv$ colors similar to those observed for SE~SNe (between 0.0 and 1.5\,mag). In addition, we include the effect of reddening using $E_{B-V}^\mathrm{MW}$ and $E_{B-V}$ between $0.0$ and $0.4$\,mag, and host galaxy $R_V$ ($R_V^\mathrm{host}$) between $2.1$ and $4.1$. Column~5 of Table~\ref{table:zp_and_leff} lists the $\bar{\lambda}_x$ values and their $1\,\ssd$ errors for $u/U$, optical, and IR bands.

In the case of the $w1$ and $w2$ bands, their transmission functions have a red tail (e.g. Figure~3 in \citealt{2014ApSS.354...89B}), and so their $\bar{\lambda}_x$ values have a strong dependence on the SED shape. Indeed, we find that the dependence of $\bar{\lambda}_x$ on the intrinsic $\bv$ color and reddening can be expressed as
\begin{equation}\label{eq:leff_w2w1}
\bar{\lambda}_x=l_0+l_1(B-V)_0+l_2(B-V)_0^2+\delta\lambda_\mathrm{red},
\end{equation}
where
\begin{equation}
\delta\lambda_\mathrm{red}=l_3E_{B-V}^\mathrm{MW} +l_4(1-l_5\ln{R_V^\mathrm{host}})E_{B-V}
\end{equation}
is the shift on $\bar{\lambda}_x$ induced by reddening.

\begin{figure}
\includegraphics[width=1.0\columnwidth]{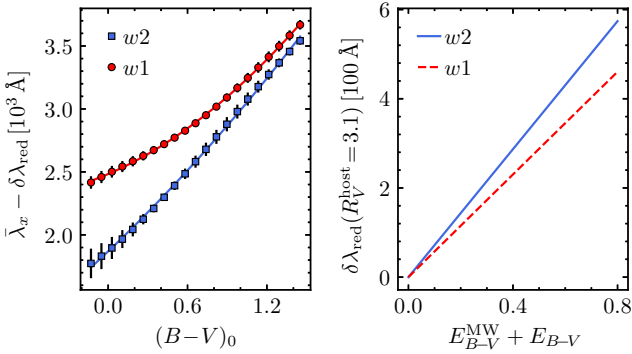}
\caption{Left panel: ${\bar{\lambda}_x-\delta\lambda_\mathrm{red}}$ for the $w2$ and $w1$ filters as a function of $(\bv)_0$ for a Planck function. Left panel: shift on $\bar{\lambda}_x$ induced by reddening with $R_V^\mathrm{host}=3.1$ as a function of the total reddening.} 
\label{fig:l_eff_swift}
\end{figure}

Table~\ref{table:UVOT} lists the $l_i$ parameters for the $w2$ and $w1$ filters, while Figure~\ref{fig:l_eff_swift} shows the correlation between ${\bar{\lambda}_x-\delta\lambda_\mathrm{red}}$ and $(\bv)_0$ (left-hand panel), and between $\delta\lambda_\mathrm{red}$ for $R_V^\mathrm{host}=3.1$ and the total reddening (right-hand panel). As we can see, $\bar{\lambda}_x$ increases as the SED becomes redder or more reddened by dust.

\begin{deluxetable}{cccccccc}
\tablecaption{$\bar{\lambda}_x$ parameters for the $w2$ and $w1$ filters\label{table:UVOT}}
\tablehead{
 \colhead{$x$} & \colhead{$l_0$} & \colhead{$l_1$} & \colhead{$l_2$} & \colhead{$l_3$} & \colhead{$l_4$} & \colhead{$l_5$} & \colhead{$\ssd$} \\
 \nocolhead{}  & \colhead{(\AA)} & \colhead{(\AA)} & \colhead{(\AA)} & \colhead{(\AA)} & \colhead{(\AA)} &  \nocolhead{}     & \colhead{(\AA)}
}
\startdata
$w2$ & $1869$ & $ 999$ & $ 122$ & $ 718$ & $1337$ & $0.40$ & $50$ \\
$w1$ & $2483$ & $ 452$ & $ 255$ & $ 576$ & $ 913$ & $0.32$ & $30$ \\
\enddata
\end{deluxetable}

\section{Extinction and $K$-corrections}\label{sec:AK_correction}
The $x$-band magnitude corrected for effects of MW reddening $E_{B-V}^\mathrm{MW}$, heliocentric redshift $z$, and host galaxy reddening $E_{B-V}$, is given by
\begin{equation}
m_{x,0} = m_x-A_x^\mathrm{h}-K_x-A_x^\mathrm{MW}.
\end{equation}
Here, $m_{x}$ is the apparent magnitude,
\begin{equation}\label{eq:Ah}
A_x^\mathrm{h}=2.5\log\left[\frac{\int{d\lambda S_{x,\lambda}\lambda f_{\lambda}}}{\int{d\lambda S_{x,\lambda}\lambda f_{\lambda} 10^{-0.4R_{\lambda} E_{B-V}}}}\right]
\end{equation}
is the host galaxy broad-band extinction,
\begin{equation}\label{eq:Ki}
K_x=-2.5\log(1+z)+K^s_{x}
\end{equation}
is the $K$-correction, with
\begin{equation}\label{eq:Ks}
 K^s_x=2.5\log\left[\frac{\int{d\lambda S_{x,\lambda}\lambda f_{\lambda}10^{-0.4R_{\lambda} E_{B-V}}}}{\int{d\lambda' S_{x,\lambda}\lambda' f_{\lambda'}10^{-0.4R_{\lambda'} E_{B-V}}}}\right],
\end{equation}
and
\begin{equation}\label{eq:AG}
A_x^\mathrm{MW}=2.5\log\left[\frac{\int{d\lambda S_{x,\lambda}\lambda' f_{\lambda'} 10^{-0.4R_{\lambda'} E_{B-V}}}}{\int{d\lambda S_{x,\lambda }\lambda' f_{\lambda'} 10^{-0.4(R_{\lambda'} E_{B-V}+R_{\lambda}^\mathrm{MW} E_{B-V}^\mathrm{MW})}}}\right]
\end{equation}
is the MW broad-band extinction. In the previous expressions, $\lambda'\!=\!\lambda/(1+z)$ is the wavelength in the SN rest frame, while $R_{\lambda}$ and $R_{\lambda}^\mathrm{MW}$ are the extinction curve for the host galaxy and the MW, respectively. We adopt the \citet{1999PASP..111...63F} extinction curve with $R_V=3.1$ for the MW, and with $R_V$ values of $2.6$, $2.7$, and $3.7$ for the host galaxies of SNe~IIb, Ib, and Ic, respectively (see Section~\ref{sec:EhBV}).

To compute $A_x^\mathrm{h}$, $K^s_x$, and $A_x^\mathrm{MW}$ we use the spectra of the SNe in our sample as $f_\lambda$. From those spectra we select those having differences between synthetic and observed colors lower than 0.1\,mag, and then we correct them for redshift and reddening. Since  $A_x^\mathrm{h}$, $K^s_x$, and $A_x^\mathrm{MW}$ have dependencies on $E_{B-V}$, $z$, and/or $E_{B-V}^\mathrm{MW}$, we compute those corrections using values from uniform distributions. For SNe~IIb, Ib, and Ic we use ($E_{B-V}^\mathrm{MW},z,E_{B-V}$) between zero and ($0.4,0.06,0.4$), ($0.3,0.08,0.6$), and ($0.3,0.13,0.7$), respectively.

\begin{figure*}
\includegraphics[width=0.32\textwidth]{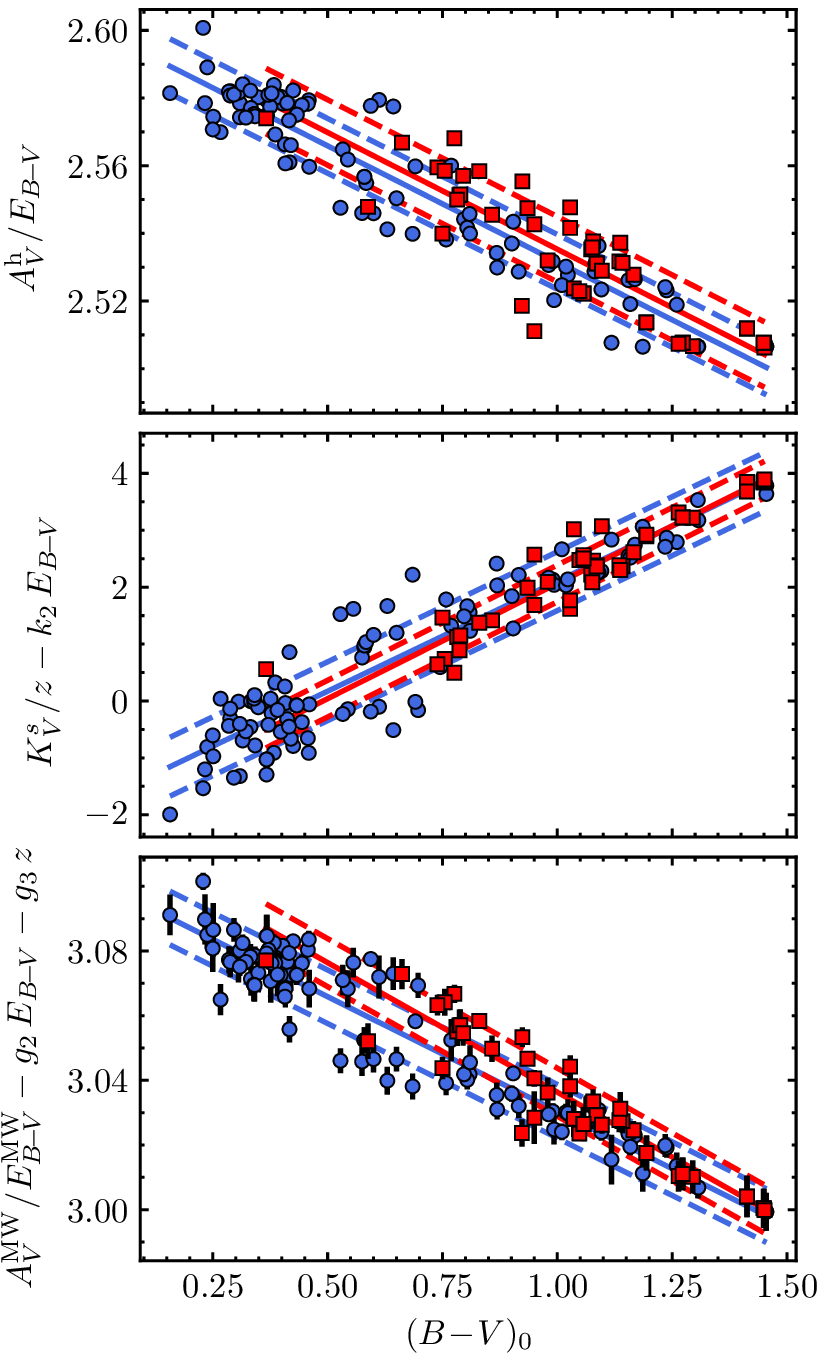}
\includegraphics[width=0.32\textwidth]{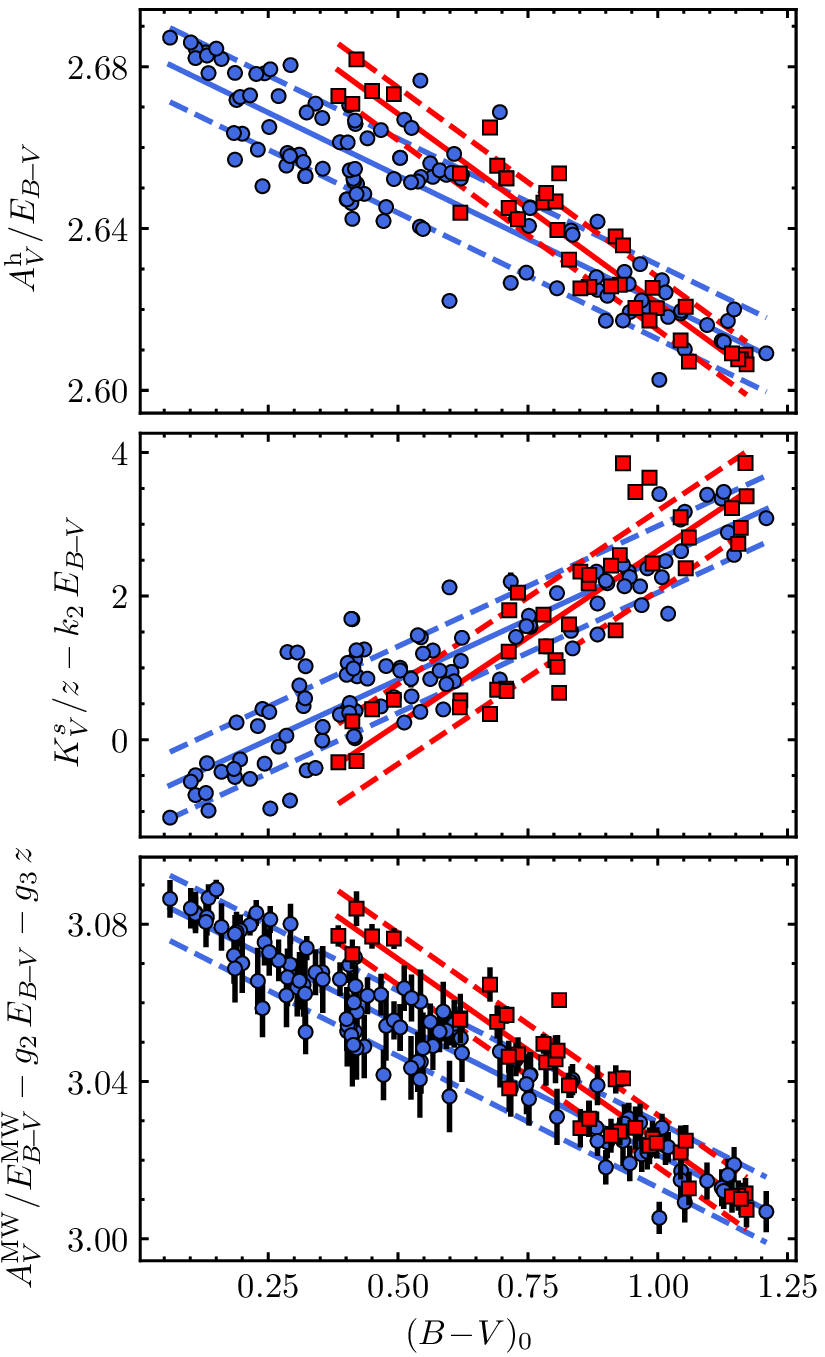}
\includegraphics[width=0.32\textwidth]{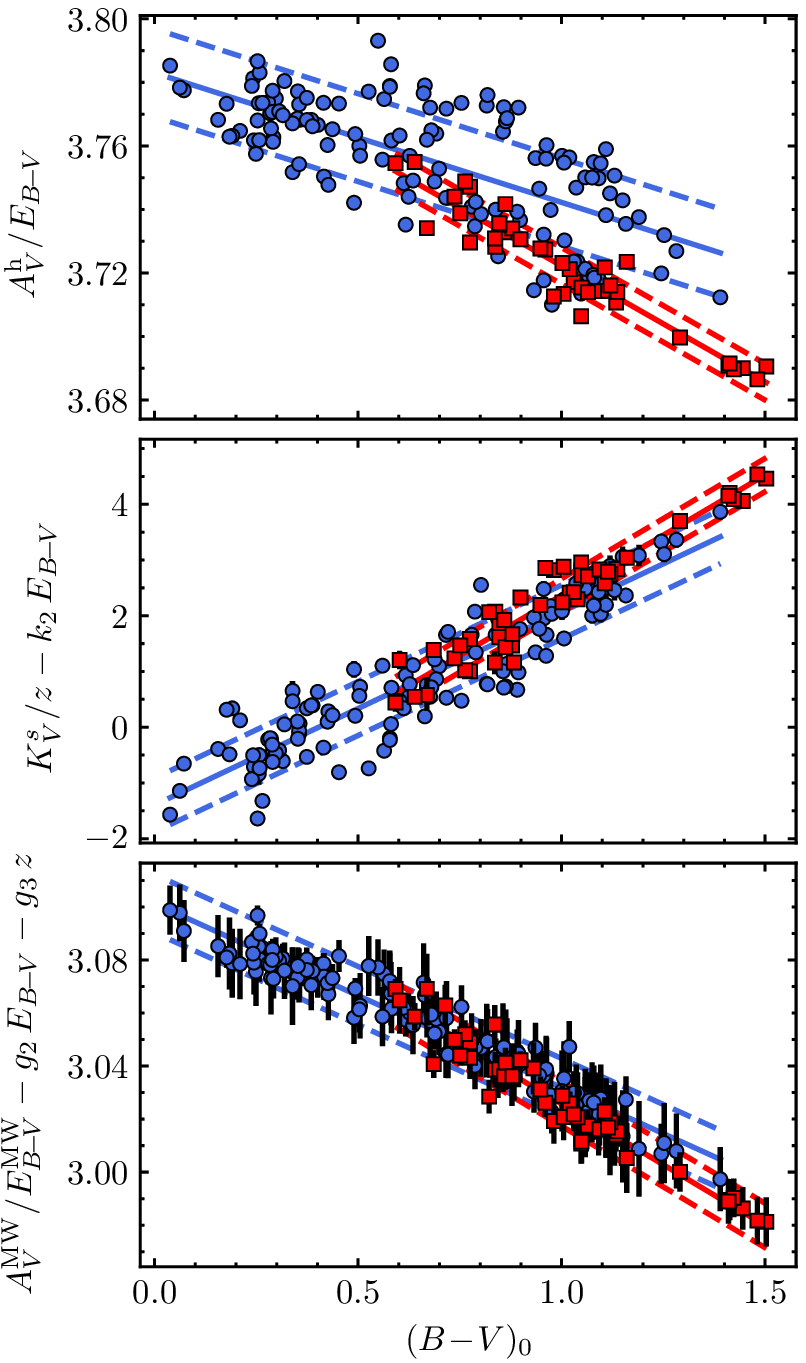}
\caption{$A_V^\mathrm{h}/E_{B-V}$ (top panels), $K_V^s/z$ (middle panels), and $A_V^\mathrm{MW}/E_{B-V}^\mathrm{MW}$ (bottom panels) as a function of $(B-V)_0$ for SNe~IIb (left-hand panels), Ib (middle-hand panels), and Ic (right-hand panels) at times before (blue circles) and after (red squares) the maximum of the $\bv$ color. Solid lines are straight line fits, while dashed lines are the $1\,\ssd$ limits around the fits.} 
\label{fig:AKA_V}
\end{figure*}

The top panels of Figure~\ref{fig:AKA_V} shows the ratios between the $A_V^\mathrm{h}$ values and the input $E_{B-V}$ against $(\bv)_0$, where blue (red) symbols correspond to $t-t_V^\peak$ epochs before (after) the time of maximum $\bv$ color ($t_{B-V}^\mathrm{max}$). We see that $A_V^\mathrm{h}/E_{B-V}$ decreases as $(B-V)_0$ increases, which is consistent with the fact that $\bar{\lambda}_V$ increases as the SN becomes redder. In general, we find that the quantity $A_x^\mathrm{h}/E_{B-V}$ is linearly correlated with the intrinsic color $c_0$, i.e.,
\begin{equation}\label{eq:Ah_E}
A_x^\mathrm{h}/E_{B-V}=h_{x,c,0}+h_{x,c,1}c_0.
\end{equation}
In the case of $K_x^s$, we find that $K_x^s/z$ correlates with $c_0$ and $E_{B-V}$ following
\begin{equation}\label{eq:Ks_z}
K_x^s/z = k_{x,c,0}+k_{x,c,1}c_0+k_{x,c,2} E_{B-V},
\end{equation}
while for $A_x^\mathrm{MW}/E_{B-V}^\mathrm{MW}$ we find a correlation with $c_0$, $E_{B-V}$, and $z$, given by
\begin{equation}\label{eq:AG_E}
A_x^\mathrm{MW}/E_{B-V}^\mathrm{MW}=g_{x,c,0}+g_{x,c,1}c_0+g_{x,c,2} E_{B-V}+g_{x,c,3} z.
\end{equation}

Tables~\ref{table:Ah_pars}, \ref{table:Ks_pars}, and \ref{table:AG_pars} collect the parameters of the $A_x^\mathrm{h}/E_{B-V}$, $K_x^s/z$, and $A_x^\mathrm{MW}/E_{B-V}^\mathrm{MW}$ terms, respectively, for different optical bands and colors. Solid lines in Figure~\ref{fig:AKA_V} correspond to the calibrations for $x=V$ and $c=\bv$ for epochs before (blue) and after (red) $t_{B-V}^\mathrm{max}$. Since IR spectra are scarce, for each IR band we compute only one set of $A_x^\mathrm{h}/E_{B-V}$, $K_x^s/z$, and $A_x^\mathrm{MW}/E_{B-V}^\mathrm{MW}$ calibrations for all phases, whose parameters are listed in Table~\ref{table:IR_cor}.

Using equation~(\ref{eq:Ah_E}), (\ref{eq:Ks_z}), and (\ref{eq:AG_E}), we can compute $A_x^\mathrm{h}$, $K^s_x$, and $A_x^\mathrm{MW}$ through
\begin{equation}
A_x^\mathrm{h}=(A_x^\mathrm{h}/E_{B-V})E_{B-V},
\end{equation}
\begin{equation}
K_x=-2.5\log(1+z)+(K_x^s/z)z,
\end{equation}
and
\begin{equation}
A_x^\mathrm{MW}=(A_x^\mathrm{MW}/E_{B-V}^\mathrm{MW})E_{B-V}^\mathrm{MW},
\end{equation}
respectively. The $c_0$ value, necessary to evaluate $A_x^\mathrm{h}/E_{B-V}$, $K_x^s/z$, and $A_x^\mathrm{MW}/E_{B-V}^\mathrm{MW}$, is given by
\begin{eqnarray}
&c_0=\left[c  - (g_{c,c,0}+g_{c,c,2} E_{B-V}+g_{c,c,3} z)E_{B-V}^\mathrm{MW}\right. \nonumber \\
&\left.-(k_{c,c,0}+k_{c,c,2} E_{B-V})z -h_{c,c,0} E_{B-V}\right]/(1 \nonumber \\
&+h_{c,c,1} E_{B-V}+k_{c,c,1} z+g_{c,c,1}E_{B-V}^\mathrm{MW}),
\end{eqnarray}
where $c=x_1-x_2$ is the observed color, while ${q_{c,c,i}=q_{x_1,c,i}-q_{x_2,c,i}}$ for $q=h,k,g$ and $i=1,2,3$.

\section{Magnitude transformations}\label{sec:mag_conv}
To compute conversions between Sloan and natural ZTF $gr$ photometry, we use synthetic magnitudes (equation~\ref{eq:syn_mag}) calculated from the same spectra used in Appendix~\ref{sec:AK_correction}. We find that $x-x_\ztf$ (hereafter $\Delta x$) for $x=g,r$ can be parametrized as a function of the time since explosion ($t$) as
\begin{eqnarray}
\Delta x = \left\{
\begin{array}{lr}
 \delta_{x,0}+\Psi_{x,a}(t), & \,\,t\leq t_x\\
 \delta_{x,0}+\Psi_{x,a}(t_x)+\Psi_{x,b}(t-t_x),   & t > t_x
\end{array}\right. .
\end{eqnarray}
Here $\delta_{x,0}$ is the zero-point for the calibration, $t_x$ is a reference time to be determined, while
\begin{equation}
\Psi_{x,a}(t)=a_{x,1}\frac{t}{100\,\mathrm{d}}+a_{x,2}\left(\frac{t}{100\,\mathrm{d}}\right)^2
\end{equation}
and
\begin{equation}
\Psi_{x,b}(t-t_x)=b_{x,1}\frac{t-t_x}{100\,\mathrm{d}}+b_{x,2}\left(\frac{t-t_x}{100\,\mathrm{d}}\right)^2
\end{equation}
describe the dependence of $\Delta x$ on $t$ before and after $t_x$, respectively.

From a sample of $N$ SNe, we estimate $t_x$ and the parameters of $\Psi_{x,a}$ and $\Psi_{x,b}$ minimizing
\begin{eqnarray}\label{eq:Sloan_ZTF}
s^2&=&\sum_{j=1}^N\left[\sum_k^{t_{j,k} \leq t_x}\left[\Delta x_{j,k}-\delta_{x,j}-\Psi_{x,a}(t_{j,k})\right]^2\right.\nonumber\\ 
    &+&\left.\sum_k^{t_{j,k} > t_x}\left[\Delta x_{j,k}-\delta_{x,j}-\Psi_{x,a}(t_x)-\Psi_{x,b}(t_{j,k}-t_x)\right]^2\right],
\end{eqnarray}
where $\delta_{x,j}$ is the zero-point for the $j$-th SN. We then adopt the mean and $\ssd$ of the $\delta_{x,j}$ estimates as $\delta_{x,0}$ and its error, respectively.

\begin{figure}
\includegraphics[width=\columnwidth]{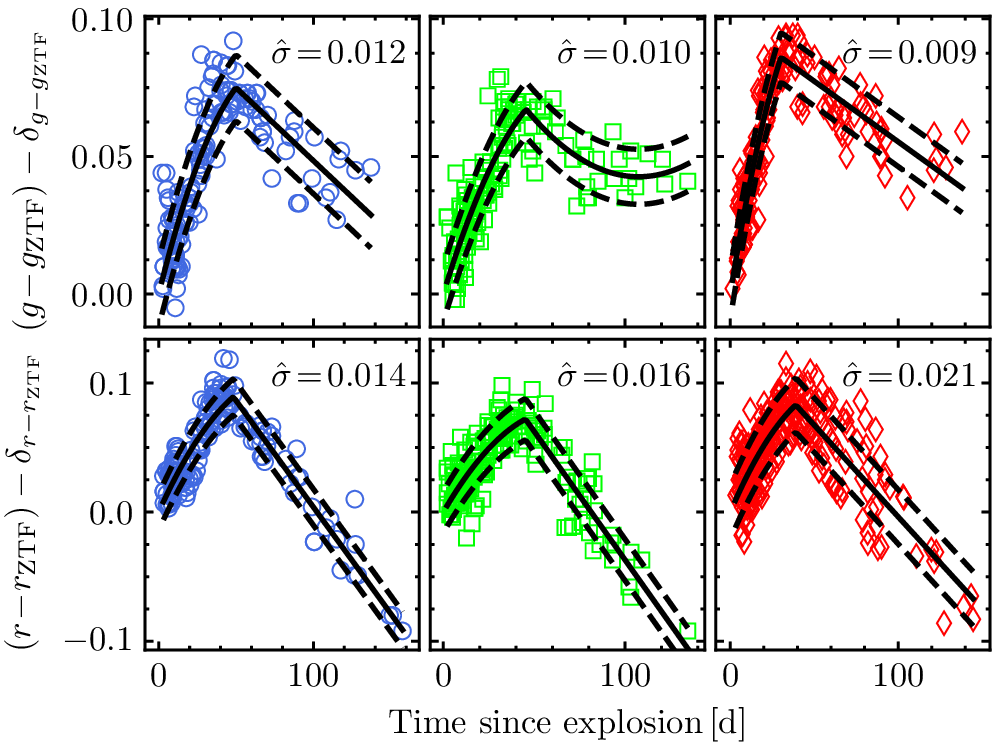}
\includegraphics[width=\columnwidth]{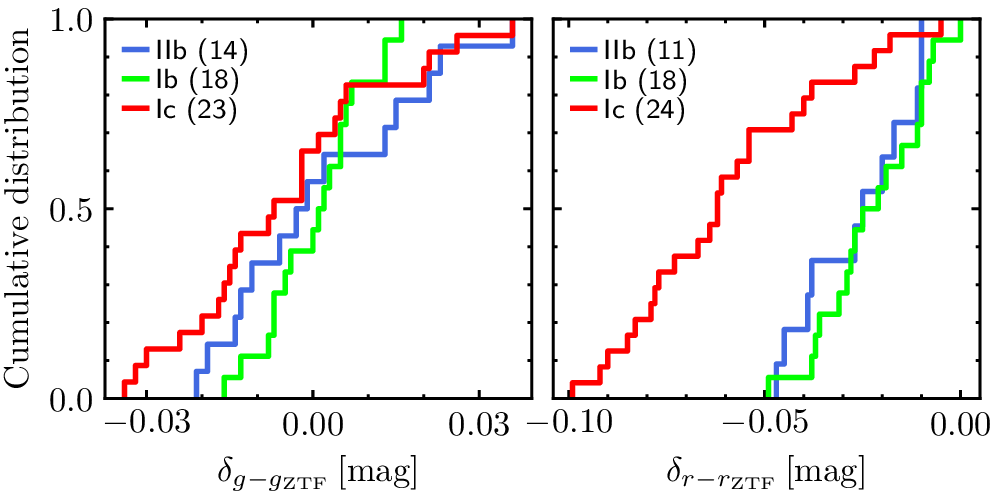}
\caption{Top and middle rows: shapes of the magnitude transformation between Sloan and ZTF systems for SNe~IIb (blue circles), Ib (green squares), and Ic (red diamonds). Solid lines are the best fits to the data, while dashed lines are the $1\,\ssd$ limits around the fits. Bottom panels: Cumulative distributions for the zero-points of the $g-g_\ztf$ (left panel) and $r-r_\ztf$ (right panel) calibrations.} 
\label{fig:Psi_Sloan_ZTF}
\end{figure}

Figure~\ref{fig:Psi_Sloan_ZTF} shows the results of the minimization of equation~(\ref{eq:Sloan_ZTF}) for $g-g_\ztf$ (top panels) and $r-r_\ztf$ (middle panels), where the $\delta_{x,j}$ values are shown in the bottom panels. Table~\ref{table:Sloan_to_ZTF} lists the parameters for the Sloan/ZTF magnitude transformations, along with the time range where they are valid (Column~11). The errors are $\leq0.025$\,mag.

\begin{deluxetable*}{ccccccccccc}
\tablecaption{Parameters for the transformation between Sloan and ZTF $gr$ magnitudes\label{table:Sloan_to_ZTF}}
\tablehead{
 \colhead{$\Delta x$} & \colhead{Type} & \colhead{$a_{x,1}$} & \colhead{$a_{x,2}$}& \colhead{$t_x$} & \colhead{$b_{x,1}$} & \colhead{$b_{x,2}$}& \colhead{$\ssd$}  & \colhead{$\delta_{x,0}$}   & \colhead{$N$} & \colhead{$t$ range}\\
 \nocolhead{}          & \nocolhead{} & \colhead{(mag)} & \colhead{(mag)} & \colhead{(d)} & \colhead{(mag)} & \colhead{(mag)} & \colhead{(mag)} & \colhead{(mag)} & \nocolhead{} & \colhead{(d)}
}
\startdata 
$g-g_\ztf$ & IIb  & $0.226$ & $-0.152$ & $49.9$ & $-0.053$ & $0.0$   & $0.012$ & $ 0.002\pm0.017$ & $14$  & $2.0$--$137.2$ \\
           & Ib   & $0.217$ & $-0.153$ & $45.7$ & $-0.079$ & $0.064$ & $0.010$ & $ 0.001\pm0.009$ & $18$  & $2.1$--$134.5$ \\
           & Ic   & $0.384$ & $-0.329$ & $30.3$ & $-0.044$ & $0.0$   & $0.009$ & $-0.005\pm0.019$ & $23$  & $1.3$--$138.1$ \\
$r-r_\ztf$ & IIb  & $0.266$ & $-0.168$ & $48.4$ & $-0.165$ & $0.0$   & $0.014$ & $-0.026\pm0.014$ & $11$  & $3.1$--$157.6$ \\
           & Ib   & $0.243$ & $-0.185$ & $45.2$ & $-0.199$ & $0.0$   & $0.016$ & $-0.022\pm0.013$ & $18$  & $2.1$--$134.5$ \\
           & Ic   & $0.304$ & $-0.234$ & $39.0$ & $-0.142$ & $0.0$   & $0.021$ & $-0.060\pm0.025$ & $24$  & $3.0$--$144.7$ \\
\enddata
\end{deluxetable*}

\section{Equivalent Width}\label{sec:pEW}
By definition, the equivalent width (EW) of a spectral line with flux $f_\lambda$ at wavelength $\lambda$ is
\begin{equation}\label{eq:pEW}
\mathrm{EW} = \int_{\lambda_b}^{\lambda_r}\left(1-\frac{f_\lambda}{f_{c,\lambda}}\right) d\lambda,
\end{equation}
where $f_{c,\lambda}$ is the flux of the continuum at $\lambda$, while $\lambda_b$ and $\lambda_r$ are the blue and red endpoints of the line profile, respectively. For a spectrum with fluxes $f_i$ at wavelengths $\lambda_i$ and a spectral dispersion $h$, we evaluate equation~(\ref{eq:pEW}) using the trapezoidal rule, obtaining
\begin{equation}\label{eq:EW}
\mathrm{EW}=h\left[N-1-\frac{n_b+n_r}{2}-\sum_{i=b+1}^{r-1}n_i\right].
\end{equation}
Here
\begin{equation}\label{eq:ni}
n_i=f_i/f_{c,\lambda_i}
\end{equation}
is the continuum normalized flux, $b$ and $r$ are the indexes of the $\lambda_i$ elements corresponding to $\lambda_b$ and $\lambda_r$, respectively, and $N$ is the number of pixels in the line region ($\lambda_b\leq\lambda_i\leq\lambda_r$).

To estimate $f_{c,\lambda}$, we first define the continuum region as the $N_r$ pixels at the left of $\lambda_r$ and the $N_b$ pixels at the right of $\lambda_b$. Typically we adopt $N_r=N_b=N$, but we also choose $N_r$ and $N_b$ in order not to include nearby absorption lines and/or bad pixels. Then we fit a low-order polynomial (typically of order two) to the points in the continuum region, which we use as $f_{c,\lambda}$. We also measure the $\ssd$ dispersion around the continuum fit ($\ssd_{f_c}$), and adopt it as the error in the continuum flux.

The error on EW, computed by error propagation, is
\begin{equation}\label{eq:eEW}
\sigma_{\mathrm{EW}} = h\left[\frac{\sigma_{n_b}^2+\sigma_{n_r}^2}{4}+\sum_{i=b+1}^{r-1} \sigma_{n_i}^2\right]^{1/2},
\end{equation}
where
\begin{equation}\label{eq:error_ni}
\sigma_{n_i}^2 = n_i^2\left[\left(\sigma_{f_i}/f_i\right)^2+(\ssd_{f_c}^2+\delta_{f_{c}}^2)/f_{c,i}^2\right].
\end{equation}
Here $\sigma_{f_i}$ is the error in the line flux, which we assume equal to $\ssd_{f_c}$ for weak lines, and $\delta_{f_c}$ is the error induced by the uncertainty on the determination of the continuum. To estimate $\delta_{f_c}$ we vary the extremes of the continuum region by $\pm1$\,pixel, calculate the EW for each configuration, and compute the $\ssd$ value of the EW estimates ($\ssd_\mathrm{EW}$). Then, $\delta_{f_c}$ is such that equation~(\ref{eq:eEW}) is equal to $\ssd_\mathrm{EW}$ for $\ssd_{f_c}=\sigma_{f_i}=0$, so
\begin{equation}
\delta_{f_c}=\frac{\ssd_\mathrm{EW}}{h}\left[\frac{1}{4}\left(\frac{n_b^2}{f_{c,b}^2}+\frac{n_r^2}{f_{c,r}^2}\right)+\sum_{i=b+1}^{r-1} \frac{n_i^2}{f_{c,i}^2}\right]^{-1/2}.
\end{equation}
In particular, for the case of weak lines, equation~(\ref{eq:eEW}) can be written as
\begin{equation}\label{eq:eEW_simple}
\sigma_{\mathrm{EW}} = \sqrt{2h(\lambda_r-\lambda_b-\mathrm{EW})/\mathrm{SNR}^2+\ssd_\mathrm{EW}^2},
\end{equation}
where we adopt the ratio between the mean of the continuum flux and $\ssd_{f_c}$ as the signal to noise ratio (SNR).

\begin{figure}
\includegraphics[width=1.0\columnwidth]{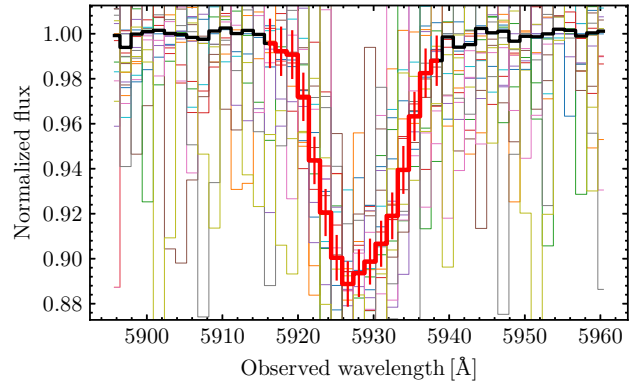}
\caption{Continuum normalized spectra of SN~2004gq around the host galaxy \ion{Na}{1}\,D line (thin colored lines) and the corresponding weighted-average spectrum (thick line). The thick red line indicates the line region, and vertical error bars are $1\,\sigma$ errors.}
\label{fig:normalized_spectra}
\end{figure}

To measure EW from a sample of spectra, we combine those obtained with the same instrument configuration (i.e. with the same spectral resolution and dispersion). This improves the SNR, lessens $\ssd_\mathrm{EW}$, and therefore reduces $\sigma_{\mathrm{EW}}$. First, for the $j$-th instrument configuration, we linearly interpolate the spectra in a range between $\lambda_b-hN_b$ and $\lambda_r+hN_r$ with a spectral dispersion $h$. Next, for each spectrum we determine the continuum and compute $n_i$ and $\sigma_{n_i}$ using equation~(\ref{eq:ni}) and (\ref{eq:error_ni}), respectively. Then, we combine the $n_i$ values for each $\lambda_i$ using the weighted average along with the \citet{1977eda..book.....T} rule to identify and discard outliers. Figure~\ref{fig:normalized_spectra} shows as an example the continuum normalized spectra of SN~2004gq observed with the FAST spectrograph around the host galaxy \ion{Na}{1}\,D line (thin colored lines), along with the combined spectrum (thick line). After that, we compute $\mathrm{EW}_j$ and $\sigma_{\mathrm{EW}_j}$ in the combined spectrum using equation~(\ref{eq:EW}) and (\ref{eq:eEW}), respectively, and repeat the whole process for all the available instrument configurations. Finally, using the set of $\mathrm{EW}_j\pm\sigma_{\mathrm{EW}_j}$ values, we compute EW maximizing the log-likelihood of a constant model ($y_j=\mathrm{EW}_j$ and $\bar{y}=\mathrm{EW}$ in equation~\ref{eq:lnL}). The error on EW is given by
\begin{equation}
\sigma_{\mathrm{EW}}=\left[\sum_j 1/(\sigma_{\mathrm{EW}_j}^2+\sigma_0^2)\right]^{-1/2}.
\end{equation}

It is worth mentioning that the expression for the EW error given in equation~(6) of \citet{2006AN....327..862V} is incorrect. The problem originates from their equation~(4), where they use $\ssd_{f_c}$ instead of $\ssd_{f_c}/\sqrt{N}$ as the error of the mean continuum flux. The same occurs for the error of the mean line flux (they use $\ssd_f$ instead of $\ssd_f/\sqrt{N}$). Including the missing $1/\sqrt{N}$ factor and assuming $\ssd_f=\ssd_{f_c}$, equation~(6) of \citet{2006AN....327..862V} becomes
\begin{equation}
\sigma_\mathrm{EW}=\sqrt{2h(\lambda_r-\lambda_b-\mathrm{EW})}/\mathrm{SNR},
\end{equation}
which corresponds to equation~(\ref{eq:eEW_simple}) for $\ssd_\mathrm{EW}=0$.

\section{Reference list for the data}\label{sec:all_refs}
Here the references for Tables~\ref{table:SN_sample}, \ref{table:mu_values}, \ref{table:t_values}, and \ref{table:EhBV}.

(1) \citet{1993IAUC.5773....1P}; (2) \citet{1993IAUC.5780....2T}; (3) \citet{1994AJ....107.1022R}; (4) \citet{1994AJ....107.1453B}; (5) \citet{1996AJ....112..732R}; (6) \citet{1997AJ....113..231W}; (7) \citet{2002AJ....123..753M}; (8) \citet{1996AJ....111..327R}; (9) \citet{1999AJ....117..736Q}; (10) \citet{1998Natur.395..670G}; (11) \citet{1999PASP..111..964M}; (12) \citet{2001ApJ...555..900P}; (13) \citet{2002AA...386..944S}; (14) \citet{2011AJ....141..163C}; (15) \citet{2011MNRAS.411.2726B}; (16) \citet{2002AJ....124.2100S}; (17) \citet{2003ApJ...592..467Y}; (18) \citet{2003MNRAS.340..375P}; (19) \citet{2003PASP..115.1220F}; (20) \citet{2006ApJ...644..400T}; (21) \citet{2008MNRAS.383.1485V}; (22) \citet{2014ApJS..213...19B}; (23) \citet{2006MNRAS.371.1459T}; (24) \citet{2011ApJ...741...97D}; (25) \citet{2018AA...609A.134S}; (26) \citet{2018PASP..130f4002S}; (27) \citet{2014ApJ...787..157P}; (28) \citet{2009ApJ...696..713S}; (29) \citet{2009AA...508..371H}; (30) \citet{2014ApJ...790..120C}; (31) \citet{2009ApJ...697..676S}; (32) \citet{2013MNRAS.434.2032R}; (33) \citet{2009ApJ...692L..84M}; (34) \citet{2009ApJ...702..226M}; (35) \citet{2008MNRAS.389..955P}; (36) \citet{2009ApJ...704L.118R}; (37) \citet{2009PZ.....29....2T}; (38) \citet{2011MNRAS.413.2140T}; (39) \citet{2011ApJ...728...14P}; (40) \citet{2011MNRAS.416.3138V}; (41) \citet{2014ApJ...792....7F}; (42) \citet{2012ApJ...749L..28V}; (43) \citet{2014AA...562A..17E}; (44) \citet{2014ApSS.354...89B}; (45) \citet{2015AA...580A.142E}; (46) \citet{2013ApJ...767...71M}; (47) \citet{2013MNRAS.431..308K}; (48) \citet{2015MNRAS.454...95M}; (49) \citet{2014MNRAS.439.1807B}; (50) \citet{2021AA...651A..81B}; (51) \citet{2020MNRAS.497.3542P}; (52) \citet{2016AA...593A..68F}; (53) \citet{2015RAA....15..225L}; (54) \citet{2013ApJ...770L..38M}; (55) \citet{2021MNRAS.507.1229P}; (56) \citet{2012ApJ...760L..33B}; (57) \citet{2019MNRAS.485.1559P}; (58) \citet{2014AJ....147...37V}; (59) \citet{2014MNRAS.445.1647M}; (60) \citet{2016MNRAS.460.1500S}; (61) \citet{2016ApJ...821...57D}; (62) \citet{2020AA...634A..21S}; (63) \citet{2014MNRAS.445.1932S}; (64) \citet{2016ApJ...825L..22F}; (65) \citet{2018ApJ...863..109Z}; (66) \citet{2018MNRAS.475.2591S}; (67) \citet{2018Sci...362..201D}; (68) \citet{2017MNRAS.471.2463B}; (69) \citet{2020MNRAS.497.3770G}; (70) \citet{2018MNRAS.476.3611G}; (71) \citet{2021ApJ...909..100S}; (72) \citet{2016AA...592A..89T}; (73) \citet{2021MNRAS.505.2530A}; (74) \citet{2018MNRAS.473.3776K}; (75) \citet{2018MNRAS.478.4162P}; (76) \citet{2019ApJ...883..147T}; (77) \citet{2020PZ.....40....1T}; (78) \citet{2017ApJ...836L..12T}; (79) \citet{2018Natur.554..497B}; (80) \citet{2018ApJ...860...90V}; (81) \citet{2019ApJ...871..176X}; (82) \citet{2021MNRAS.502.3829T}; (83) \citet{2021MNRAS.501.5797B}; (84) \citet{2019PASP..131g8001G}; (85) \citet{2019PASP..131a8002B}; (86) \citet{2019PASP..131a8003M}; (87) \citet{2021MNRAS.504.2073K}; (88) \citet{2021ApJ...908..232R}; (89) \citet{2020ApJ...902...86H}; (90) \citet{2021MNRAS.506.1832M}; (91) \citet{2011ApJ...743..176G}; (92) \citet{2010ApJ...718.1118D}; (93) \citet{2017AJ....154...51M}; (94) \citet{2016ApJ...826...56R}; (95) \citet{2017ApJ...836...74J}; (96) \citet{2006ApJS..165..108S}; (97) \citet{2015AA...574A..60T}; (98) \citet{2011ApJ...742L..18A}; (99) \citet{2015ApJ...811..117S}; (100) \citet{1993AA...280L..11V}; (101) \citet{1995ApJ...444..165H}; (102) \citet{2002PASJ...54..899T}; (103) \citet{2008ApJ...673L.155V}; (104) \citet{2012ApJ...748L..11R}; (105) \citet{2013ApJ...775L...7C}; (106) \citet{2018MNRAS.480.2072K}.

\section{Tables}

\startlongtable
\begin{deluxetable*}{l l l c C c l}
\tablenum{F1}
\tablecaption{SN sample\label{table:SN_sample}}
\tablehead{
\colhead{SN} & \colhead{Type} & \colhead{Host galaxy} & \colhead{$E_{B-V}^\mathrm{MW}$} & \colhead{$cz$} & \colhead{Filters} & \colhead{References} \\    
\nocolhead{} & \nocolhead{} & \nocolhead{} & \colhead{(mag)} & \colhead{(km\,s$^{-1}$)} & \nocolhead{} & \nocolhead{} 
}
\startdata
1993J\tablenotemark{a}    & IIb      & M81                       & 0.069 &   -34 &  $U\!BV\!RI$ $J\!H\!K\!$                & 1, 2, 3, 4, 5, 6, 7 \\
1994I                     &  Ic      & M51a                      & 0.031 &   463 &  $U\!BV\!RI$                            & 8                   \\
1996cb                    & IIb      & NGC 3510                  & 0.026 &   713 &  $BV\!R$                                & 9                   \\
1998bw\tablenotemark{a}   & Ic (BL)  & ESO 184-G82               & 0.049 &  2599 &  $U\!BV\!RI$ $J\!H\!K\!$                & 10, 11, 12, 13, 14  \\
1999dn\tablenotemark{a}   &  Ib      & NGC 7714                  & 0.045 &  2798 &  $U\!BV\!RI$ $J\!H\!K\!$                & 15                  \\
1999ex                    &  Ib      & IC 5179                   & 0.017 &  3422 &  $U\!BV\!RI$                            & 16                  \\
2002ap\tablenotemark{a}   & Ic (BL)  & M74                       & 0.062 &   657 &  $U\!BV\!RI$ $J\!H\!K\!$                & 17, 18, 19, 20      \\
2003jd                    & Ic (BL)  & MCG -01-59-21             & 0.038 &  5654 &  $BV\!RI$                               & 21, 22              \\
\enddata
\tablecomments{References for the data are in Appendix~\ref{sec:all_refs}.}
\tablenotetext{a}{SNe used to compute BCs.}
\end{deluxetable*}

\begin{deluxetable*}{l l c c c c c c}
\tablenum{F2}
\tablecaption{Host galaxy distance moduli\tablenotemark{*}\label{table:mu_values}}
\tablehead{
\colhead{Host galaxy} & \colhead{SN} & \colhead{$\mu_{\mathrm{CPL}}$} & \colhead{$\mu_{\mathrm{TRGB}}$} & \colhead{$\mu_{\mathrm{TF}}$} & \colhead{$\mu_z$} & \colhead{$\mu_{\mathrm{SVF}}$} & \colhead{$\mu$}
}
\startdata
M81      & 1993J  & $27.87\pm0.14^{91}$ & $27.86\pm0.06^{92}$ & $27.61\pm0.35$ & \nodata & \nodata & $27.86\pm0.05$ \\
M51a     & 1994I  & \nodata             & $29.67\pm0.07^{93}$ & \nodata        & \nodata & \nodata & $29.67\pm0.07$ \\
         & 2011dh & \nodata             & \nodata             & \nodata        & \nodata & \nodata & $29.67\pm0.07$ \\
M74      & 2002ap & \nodata             & $29.95\pm0.08^{93}$ & \nodata        & \nodata & \nodata & $29.95\pm0.08$ \\
NGC 4038 & 2004gt & $31.29\pm0.11^{94}$ & $31.68\pm0.04^{95}$ & \nodata        & \nodata & \nodata & $31.63\pm0.04$ \\
         & 2013dk & \nodata             & \nodata             & \nodata        & \nodata & \nodata & $31.63\pm0.04$ \\
NGC 4414 & 2013df & $31.65\pm0.17^{96}$ & \nodata             & $30.95\pm0.35$ & \nodata & \nodata & $31.52\pm0.15$ \\
M100     & 2020oi & $31.18\pm0.05^{96}$ & \nodata             & \nodata        & \nodata & \nodata & $31.18\pm0.05$ \\
\enddata
\tablecomments{Column~1: SN host galaxy names. Column~2: SN names, Columns~3 and 4: Cepheids and TRGB distances from the literature, respectively. Column~5: Tully-Fisher distances from EDD. Column~6: distances compute from recessional redshifts. Column~6: distances inferred from smoothed velocity fields \citep{2020AJ....159...67K}. Column~8: adopted distance moduli. References for the data (superscript numbers) are in Appendix~\ref{sec:all_refs}.}
\tablenotetext{*}{This table is available in its entirety in machine-readable form.}
\end{deluxetable*}

\begin{deluxetable*}{lcccccccccc}
\tablenum{F3}
\tablecaption{Maximum-light epochs\tablenotemark{*}\label{table:tmax}}
\tablehead{
\colhead{SN} & \colhead{$t_B^\peak$} & \colhead{$t_g^\peak$} & \colhead{$t_{g_\ztf}^\peak$} & \colhead{$t_V^\peak$} & \colhead{$t_r^\peak$} & \colhead{$t_{r_\ztf}^\peak$} & \colhead{$t_R^\peak$} & \colhead{$t_i^\peak$} & \colhead{$t_I^\peak$} & \colhead{$\langle t_V^\peak\rangle$}
}
\startdata
1993J  & $49093.5(3)$ & \nodata & \nodata & $49095.1(2)$ & \nodata & \nodata & $49095.4(2)$ & \nodata & $49096.1(3)$ & $49095.0(2)$ \\
1994I  & $49450.0(2)$ & \nodata & \nodata & $49450.8(1)$ & \nodata & \nodata & $49451.1(1)$ & \nodata & $49451.7(1)$ & $49450.7(1)$ \\
1996cb & $50449.9(1)$ & \nodata & \nodata & $50451.7(1)$ & \nodata & \nodata & $50453.4(2)$ & \nodata & \nodata      & $50451.7(1)$ \\
\enddata
\tablecomments{Epochs are in MJD. Numbers in parentheses are $1\,\sigma$ errors in units of the last significant digits.}
\tablenotetext{*}{This table is available in its entirety in machine-readable form.}
\end{deluxetable*}

\begin{deluxetable*}{lcccccccccc}
\tablenum{F4}
\tablecaption{Decline rate parameters\tablenotemark{*}\label{table:Dm15}}
\tablehead{
\colhead{SN} & \colhead{$\Delta m_{15}(B)$} & \colhead{$\Delta m_{15}(g)$} & \colhead{$\Delta m_{15}(g_\ztf)$} & \colhead{$\Delta m_{15}(V)$} & \colhead{$\Delta m_{15}(r)$} & \colhead{$\Delta m_{15}(r_\ztf)$} & \colhead{$\Delta m_{15}(R)$} & \colhead{$\Delta m_{15}(i)$} & \colhead{$\Delta m_{15}(I)$} & \colhead{$\langle\Delta m_{15}(V)\rangle$}
}
\startdata
1993J  & 1.631(19) & \nodata & \nodata & 1.015(17) & \nodata & \nodata & 0.725(19) & \nodata & 0.571(17) & 1.017(14) \\
1994I  & 2.033(52) & \nodata & \nodata & 1.710(15) & \nodata & \nodata & 1.417(17) & \nodata & 1.095(18) & 1.703(14) \\
1996cb & 1.365(13) & \nodata & \nodata & 0.825(12) & \nodata & \nodata & 0.555(13) & \nodata & \nodata   & 0.829(12) \\
\enddata
\tablecomments{Values are in magnitude units. Numbers in parentheses are $1\,\sigma$ errors in units of the last significant digits.}
\tablenotetext{*}{This table is available in its entirety in machine-readable form.}
\end{deluxetable*}

\begin{deluxetable*}{lcccCC}
\tablenum{F5}
\tablecaption{SN explosion epochs\tablenotemark{*}\label{table:t_values}}
\tablehead{
\colhead{SN} & \colhead{$t_\mathrm{non-det}$ (MJD)} & \colhead{$t_\mathrm{detect}$ (MJD)} & \colhead{Phase source} & \colhead{$t_\mathrm{expl}^\peak$ (MJD)} & \colhead{$t_\mathrm{expl}$ (MJD)}
}
\startdata
2006fo & 53975.410 & 53994.430 & 97  & 53988.0\pm3.2 & 53987.8\pm3.0 \\
2011dh & 55712.275 & 55712.893 & 98  & 55713.6\pm3.5 & 55712.6\pm0.2 \\
2013bb & 56368.373 & 56369.320 & 99 & 56375.0\pm3.6 & 56368.9\pm0.3 \\
\enddata
\tablecomments{Column~1: SN names. Column~2, 3, and 4: SN last non-detection epochs, first detection epochs, and their references (codes are in Appendix~\ref{sec:all_refs}.), respectively. Column~5: explosion epochs estimated with maximum-light epochs. Column~6: explosions epochs.}
\tablenotetext{*}{This table is available in its entirety in machine-readable form.}
\end{deluxetable*}

\begin{deluxetable*}{lCccCcCc}
\tablenum{F6}
\tablecaption{Vertical intercepts of the color curves\tablenotemark{*}\label{table:delta}}
\tablehead{
\colhead{SN} & \colhead{$\delta_{B-V}$} & \colhead{$\delta_{g-i}$} & \colhead{$\delta_{V-r}$} & \colhead{$\delta_{V-R}$} & \colhead{$\delta_{V-i}$} & \colhead{$\delta_{V-I}$} & \colhead{$\delta_{(g-r)_\ztf}$}
}
\startdata
1993J  & 0.500(44) & \nodata & \nodata & 0.240(45) & \nodata & 0.338(28) & \nodata \\
1994I  & 0.857(53) & \nodata & \nodata & 0.230(36) & \nodata & 0.446(72) & \nodata \\
1996cb & 0.295(30) & \nodata & \nodata & 0.130(33) & \nodata & \nodata   & \nodata \\
\enddata
\tablecomments{Values are in magnitude units. Numbers in parentheses are $1\,\sigma$ errors in units of the last significant digits.}
\tablenotetext{*}{This table is available in its entirety in machine-readable form.}
\end{deluxetable*}

\begin{deluxetable*}{lcccccc}
\tablenum{F7}
\tablecaption{\ion{Na}{1}\,D equivalent widths and host galaxy reddenings\tablenotemark{*}\label{table:EhBV}}
\tablehead{
\colhead{SN} & \colhead{$\mathrm{EW_{NaID}^{MW}}$ (\AA)} & \colhead{$\mathrm{EW_{NaID}}$ (\AA)} & \colhead{$\tilde{E}_{B-V}$ (mag)} & \colhead{$E_{B-V}^\mathrm{CC}$ (mag)} & \colhead{$E_{B-V}^\mathrm{NaID}$ (mag)} & \colhead{$E_{B-V}$ (mag)}
}
\startdata
1993J     & $ 0.447\pm0.007^{100}$ & $ 0.251\pm0.010^{100}$ & $ 0.487\pm0.034$ & $ 0.070\pm0.119$ & $0.062\pm0.094$ & $ 0.065\pm0.074$ \\
1994I     & $ 0.180\pm0.005^{101}$ & $ 4.100\pm0.010^{101}$ & $ 0.788\pm0.039$ & $ 0.233\pm0.092$ & \nodata         & $ 0.233\pm0.092$ \\
2002ap    & $ 0.494\pm0.004^{102}$ & $ 0.106\pm0.002^{102}$ & $ 0.614\pm0.040$ & $ 0.059\pm0.092$ & $0.026\pm0.094$ & $ 0.043\pm0.066$ \\
2007gr    & $ 0.310\pm0.000^{103}$ & $ 0.130\pm0.000^{103}$ & $ 0.648\pm0.014$ & $ 0.093\pm0.084$ & $0.032\pm0.094$ & $ 0.066\pm0.063$ \\
2011dh    & $ 0.250\pm0.003^{104}$ & $ 0.286\pm0.007^{104}$ & $ 0.674\pm0.018$ & $ 0.257\pm0.115$ & $0.070\pm0.094$ & $ 0.145\pm0.073$ \\
iPTF13bvn & $ 0.250\pm0.000^{105}$ & $ 0.420\pm0.000^{105}$ & $ 0.512\pm0.011$ & $ 0.148\pm0.084$ & $0.103\pm0.094$ & $ 0.128\pm0.063$ \\
2016gkg   & $ 0.290\pm0.000^{79}$  & $ 0.680\pm0.030^{78}$  & $ 0.440\pm0.021$ & $ 0.023\pm0.116$ & $0.167\pm0.095$ & $ 0.109\pm0.073$ \\
2017ein   & $ 0.150\pm0.040^{106}$ & $ 1.441\pm0.067$       & $ 0.942\pm0.010$ & $ 0.387\pm0.083$ & \nodata         & $ 0.387\pm0.083$ \\
\enddata
\tablecomments{References for the $\mathrm{EW_{NaID}}$ values from the literature (superscript numbers) are in Appendix~\ref{sec:all_refs}.}
\tablenotetext{*}{This table is available in its entirety in machine-readable form.}
\end{deluxetable*}

\begin{deluxetable*}{cccccccccc}
\tablenum{F8}
\tablecaption{Absolute $r$-band magnitudes at peak\label{table:Mrpeak}}
\tablehead{
\colhead{SN} & \colhead{$M_r^\peak$} & \colhead{SN} & \colhead{$M_r^\peak$} & \colhead{SN} & \colhead{$M_r^\peak$} & \colhead{SN} & \colhead{$M_r^\peak$} & \colhead{SN} & \colhead{$M_r^\peak$} 
}
\startdata
1993J     & $-17.482$ & 2007C     & $-17.592$ & iPTF12hvv & $-18.729$ & 2019bjv   & $-17.744$ & 2020mob   & $-17.021$ \\
1994I     & $-17.711$ & 2007Y     & $-16.131$ & 2013F     & $-18.701$ & 2019bwi   & $-18.370$ & 2020nke   & $-17.083$ \\
1996cb    & $-16.593$ & 2007ag    & $-18.085$ & 2013ak    & $-17.537$ & 2019cda   & $-17.775$ & 2020qrx   & $-18.925$ \\
1998bw    & $-18.569$ & 2007ce    & $-19.011$ & 2013bb    & $-16.650$ & 2019dgz   & $-17.849$ & 2020rlg   & $-18.449$ \\
1999dn    & $-16.916$ & 2007cl    & $-17.892$ & 2013df    & $-17.464$ & 2019gqd   & $-19.056$ & 2020rwz   & $-17.307$ \\
1999ex    & $-17.345$ & 2007gr    & $-17.534$ & 2013dk    & $-19.189$ & 2019gwc   & $-18.225$ & 2020sai   & $-17.629$ \\
2002ap    & $-17.870$ & 2007hn    & $-18.226$ & 2013ek    & $-18.119$ & 2019hjg   & $-17.494$ & 2020sbw   & $-17.665$ \\
2003jd    & $-18.614$ & 2007kj    & $-16.889$ & 2013ge    & $-17.721$ & 2019hsx   & $-18.826$ & 2020scb   & $-18.371$ \\
\enddata
\end{deluxetable*}

\begin{deluxetable*}{ccccccccccc}
\tablenum{F9}
\tablecaption{Parameters of the BC calibrations\label{table:BC_cal}}
\tablehead{
\colhead{$x$} & \colhead{$x-y$}  & \colhead{$a_1$}    & \colhead{$a_2$}    & \colhead{$a_3$}    & \colhead{$a_4$}    & \colhead{$\ssd$}  & \colhead{$\mathrm{ZP}$}       & \colhead{$N$}  & \colhead{$t_{x-y}^\mathrm{max}$ (d)} & \colhead{$t-t_\mathrm{expl}$ range (d)}
}
\startdata
\multicolumn{11}{c}{SNe~IIb} \\
 $B     $ & $B-V          $ & $ 0.206$ & $-0.878$ & $ 0.546$ & $-1.348$ & $0.047$ & $11.274\pm0.053$ & $15$ & $18.8$ & $3.4,119.6$ \\
 $B     $ & $B-r          $ & $-0.182$ & $-0.362$ & $ 0.278$ & $-0.961$ & $0.042$ & $11.391\pm0.052$ & $13$ & $18.2$ & $3.4,105.3$ \\
 $B     $ & $B-R          $ & $-0.203$ & $-0.255$ & $ 0.433$ & $-0.939$ & $0.051$ & $11.410\pm0.036$ & $ 8$ & $16.6$ & $3.4,119.6$ \\
 $B     $ & $B-i          $ & $-0.338$ & $-0.177$ & $ 0.396$ & $-0.872$ & $0.029$ & $11.322\pm0.027$ & $13$ & $18.0$ & $3.4,105.3$ \\
 $B     $ & $B-I          $ & $-0.328$ & $-0.131$ & $ 1.087$ & $-1.089$ & $0.056$ & $11.442\pm0.056$ & $ 8$ & $17.2$ & $3.4,119.6$ \\
 $g     $ & $g-r          $ & $-0.093$ & $-0.681$ & $ 0.128$ & $-0.934$ & $0.053$ & $11.531\pm0.061$ & $13$ & $21.7$ & $3.4,105.3$ \\
 $g     $ & $g-R          $ & $ 0.159$ & $-0.604$ & $ 0.234$ & $-0.791$ & $0.057$ & $11.467\pm0.059$ & $ 6$ & $21.8$ & $3.4,105.1$ \\
\enddata
\tablecomments{$\mathrm{BC}=\mathrm{ZP}+a_1(x-y)+a_2(x-y)^2$ if $t-t_y^\peak\leq t_{x-y}^\mathrm{max}$, otherwise $\mathrm{BC}=\mathrm{ZP}+a_3+a_4(x-y)$.}
\end{deluxetable*}

\begin{deluxetable*}{lcccccc}
\tablenum{F10}
\tablecaption{Bolometric light-curve properties, ejecta velocities, and explosion parameters\tablenotemark{*}\label{table:MNi}}
\tablehead{
\colhead{SN} & \colhead{$t_L^\peak$} & \colhead{$\log L_\peak$} & \colhead{$\Delta m_\mathrm{15}(\mathrm{bol})$} & \colhead{$v_\peak$} &\colhead{$M_\mathrm{ej}$} & \colhead{$E_\mathrm{k}$} \\
    & \colhead{(d)} & \colhead{(dex)} & \colhead{(mag)} & \colhead{(km\,s$^{-1}$)} & \colhead{(M$_\sun$)} & \colhead{(foe)} \\
 \colhead{(1)} & \colhead{(2)} & \colhead{(3)} & \colhead{(4)} & \colhead{(5)} & \colhead{(6)} & \colhead{(7)}
}
\startdata
1993J     & $20.9\pm0.1$ & $-0.601\pm0.002$ & $   0.880\pm 0.010$ & $    7000$ & $   3.4\pm0.5$ & $    0.99\pm0.45$ \\
1994I     & $ 9.6\pm0.1$ & $-0.589\pm0.007$ & $   1.478\pm 0.020$ & $    9000$ & $   0.9\pm0.3$ & $    0.44\pm0.24$ \\
1996cb    & $21.0\pm0.2$ & $-0.943\pm0.002$ & $   0.727\pm 0.007$ & $    7000$ & $   3.4\pm0.7$ & $    0.99\pm0.47$ \\
1998bw    & $16.0\pm0.4$ & $-0.235\pm0.005$ & $   0.615\pm 0.018$ & $   15000$ & $   4.2\pm0.6$ & $    5.68\pm2.56$ \\
1999dn    & $16.3\pm0.8$ & $-0.831\pm0.007$ & $   0.405\pm 0.022$ & $    9000$ & $   2.6\pm0.7$ & $    1.27\pm0.64$ \\
\hline
SN & $\log\mni^\tail$ & $t_\mathrm{esc}$ & $\log( L_\peak/\mni^\tail)$ & $\log\mni^\peak$ & $\log\mni$ & $\mni$ \\
   & (dex) & (d) & (dex) & (dex) & (dex) & (M$_\sun$) \\
 (1) & (8) & (9) & (10) & (11) & (12) & (13) \\
\enddata
\tablecomments{Column~1: SN name. Columns~2, 3, and 4: peak time, peak luminosity, and decline rate, respectively. Columns~5, 6, and 7: ejecta velocity at $t_L^\peak$, ejecta mass, and kinetic energy, respectively. Columns~8 and 9: $^{56}$Ni mass and $\gamma$-ray escape time, respectively, measured from the luminosity in the radioactive tail. Column~10: $\log L_\peak$ minus $\log\mni^\tail$. Column~11: $^{56}$Ni mass computed from peak time, peak luminosity (and decline rate if it is available), and the empirical correlation we present in Section~4.2. Column~12 and 13: Weighted average of $\log\mni^\tail$ and $\log\mni^\peak$, and the corresponding $^{56}$Ni mass, respectively. $L_\peak$ and $\mni$ are in units of $10^{43}$\,erg\,s$^{-1}$ and M$_\sun$, respectively. Uncertainties are $1\,\sigma$ errors, while lower and upper $t_\mathrm{esc}$ errors are 16th and 84th percentiles.}
\tablenotetext{*}{This table is available in its entirety in machine-readable form.}
\end{deluxetable*}

\begin{deluxetable*}{lcccccccc}
\tablenum{F11}
\tablecaption{SNe with $^{56}$Ni masses estimated with hydrodynamical models.\label{table:MNi_hydro}}
\tablehead{
 \colhead{SN} & \colhead{$\mu$} & \colhead{$E_{B-V}^\mathrm{MW}$} & \colhead{$E_{B-V}$} & \colhead{$R_V$} & \colhead{$t_\mathrm{expl}$} & \colhead{$L_\peak$} & \colhead{$\mni$}     & \colhead{Reference} \\
 \nocolhead{} & \colhead{(mag)} & \colhead{(mag)}   & \colhead{(mag)} & \nocolhead{} & \colhead{(MJD)} & \colhead{($10^{43}$\,erg\,s$^{-1}$)}  & \colhead{(M$_\sun$)} & \nocolhead{} 
}
\startdata
 1993J     & 27.80 & 0.030 & 0.0   & \nodata & 49074.1 &  0.21 & 0.078 & \citet{1996AA...306..219U} \\ 
 1994I     & 29.60 & 0.030 & 0.270 & 3.1     & 49438.5 &  0.23 & 0.070 & \citet{2006MNRAS.369.1939S} \\ 
 1998bw    & 32.89 & 0.016 & 0.0   & \nodata & 50928.9 &  1.00 & 0.400 & \citet{2001ApJ...550..991N} \\ 
 2004gq    & 32.00 & 0.065 & 0.108 & 2.4     & 53346.9 &  0.20 & 0.110 & \citet{2018AA...609A.136T} \\ 
 2004gt    & 31.83 & 0.041 & 0.428 & 2.5     & 53342.8 &  0.31 & 0.160 & \citet{2018AA...609A.136T} \\ 
 2006T     & 32.50 & 0.066 & 0.323 & 1.3     & 53757.6 &  0.23 & 0.120 & \citet{2018AA...609A.136T} \\ 
 2006ep    & 33.96 & 0.032 & 0.118 & 5.1     & 53975.5 &  0.24 & 0.120 & \citet{2018AA...609A.136T} \\ 
 2006fo    & 34.59 & 0.026 & 0.156 & 4.1     & 53983.4 &  0.41 & 0.250 & \citet{2018AA...609A.136T} \\
\enddata
\end{deluxetable*}

\begin{deluxetable*}{lcccc}
\tablenum{F12}
\tablecaption{\ion{Na}{1}\,D equivalent widths and reddenings for SNe~II\tablenotemark{*}\label{table:EhBV_SNII}}
\tablehead{
\colhead{SN} & \colhead{$\mathrm{EW_{NaID}^{MW}}$ (\AA)} & \colhead{${E_{B-V}^\mathrm{MW}}$ (mag)\tablenotemark{a}} & \colhead{$\mathrm{EW_{NaID}}$}  & \colhead{${E_{B-V}}$ (mag)\tablenotemark{b}}
}
\startdata
1990K       & \nodata          & \nodata         & $ 1.000\pm0.144$ & $ 0.227\pm0.034$ \\
1991al      & $ 0.470\pm0.031$ & $0.044\pm0.007$ & $ 0.421\pm0.024$ & $ 0.067\pm0.026$ \\
1992ba      & $ 0.069\pm0.075$ & $0.050\pm0.008$ & $-0.041\pm0.072$ & $ 0.096\pm0.029$ \\
\enddata
\tablenotetext{a}{Values from \citet{2011ApJ...737..103S} with a random error of 16\%.}
\tablenotetext{b}{Values from \citet{2021MNRAS.505.1742R}.}
\tablenotetext{*}{This table is available in its entirety in machine-readable form.}
\end{deluxetable*}

\begin{deluxetable*}{llccccccccccc}
\tablenum{F13}
\caption{Host galaxy broad-band extinction parameters for optical bands\tablenotemark{*}\label{table:Ah_pars}}
\tablehead{
        &  &       & \multicolumn{4}{c}{$t-t_{x_2}^\peak\leq t_c^\mathrm{max}$} & & \multicolumn{4}{c}{$t-t_{x_2}^\peak>t_c^\mathrm{max}$} & \\
         \cline{4-7}  \cline{9-12} 
\colhead{Type} &\colhead{$x$} & \colhead{$c$}     & \colhead{$h_{x,c,0}$} & \colhead{$h_{x,c,1}$} & \colhead{$\ssd$}& \colhead{$N$} & & \colhead{$h_{x,c,0}$} & \colhead{$h_{x,c,1}$} &  \colhead{$\ssd$} & \colhead{$N$} & \colhead{$t-t_{x_2}^\peak$} \\
      &    & \colhead{($x_1-x_2$)} &                   &                   &             &     & &                     &                     &                 &     & \colhead{range (d)}
}
\startdata
IIb & $B     $ & $B\!-\!V       $ & $ 3.497$ & $-0.123$ & $0.010$ & $ 99$ & & $ 3.466$ & $-0.104$ & $0.009$ & $ 46$ & $-17,117$ \\
    & $B     $ & $B\!-\!r       $ & $ 3.496$ & $-0.099$ & $0.008$ & $ 78$ & & $ 3.458$ & $-0.078$ & $0.004$ & $ 30$ & $-18,106$ \\
    & $B     $ & $B\!-\!R       $ & $ 3.498$ & $-0.084$ & $0.009$ & $ 48$ & & $ 3.461$ & $-0.068$ & $0.007$ & $ 27$ & $-18,106$ \\
\enddata
\tablecomments{The $t_c^\mathrm{max}$ values are listed in Column~11 of Table~\ref{table:BC_cal}.}
\tablenotetext{*}{This table is available in its entirety in machine-readable form.}
\end{deluxetable*}

\begin{deluxetable*}{llccccccccccccc}
\tablenum{F14}
\tablecaption{$K$-correction parameters for optical bands\tablenotemark{*}\label{table:Ks_pars}}
\tablehead{
        &  &       & \multicolumn{5}{c}{$t-t_{x_2}^\peak\leq t_c^\mathrm{max}$} & & \multicolumn{5}{c}{$t-t_{x_2}^\peak>t_c^\mathrm{max}$} & \\
                 \cline{4-8}  \cline{10-14} 
 \colhead{Type} &\colhead{$x$} & \colhead{$c$} & \colhead{$k_{x,c,0}$} & \colhead{$k_{x,c,1}$} & \colhead{$k_{x,c,2}$} & \colhead{$\ssd$}& \colhead{$N$} & & \colhead{$k_{x,c,0}$} & \colhead{$k_{x,c,1}$} & \colhead{$k_{x,c,2}$} & \colhead{$\ssd$} & \colhead{$N$} & \colhead{$t-t_{x_2}^\peak$} \\
 &    &  \colhead{($x_1-x_2$)} &                   &                   &                   &             &     &     &                &                     &                     &                &     & \colhead{range (d)} 
}
\startdata
IIb & $B     $ & $B\!-\!V       $ & $-1.171$ & $ 5.515$ & $ 3.747$ & $0.648$ & $ 87$ & & $ 0.782$ & $ 3.919$ & $ 3.511$ & $0.688$ & $ 44$ & $-20,117$ \\
    & $B     $ & $B\!-\!r       $ & $-0.905$ & $ 4.311$ & $ 3.725$ & $0.575$ & $ 74$ & & $-0.486$ & $ 4.077$ & $ 3.480$ & $0.537$ & $ 33$ & $-18,106$ \\
    & $B     $ & $B\!-\!R       $ & $-0.973$ & $ 3.542$ & $ 3.733$ & $0.424$ & $ 42$ & & $-0.650$ & $ 3.613$ & $ 3.417$ & $0.472$ & $ 27$ & $-18,106$ \\
\enddata
\tablecomments{The $t_c^\mathrm{max}$ values are listed in Column~11 of Table~\ref{table:BC_cal}.}
\tablenotetext{*}{This table is available in its entirety in machine-readable form.}
\end{deluxetable*}

\begin{deluxetable*}{lccccccccccccccc}
\tablenum{F15}
\tablecaption{Galactic broad-band extinction parameters for optical bands\tablenotemark{*}\label{table:AG_pars}}
\tablehead{
        &  &       \multicolumn{6}{c}{$t-t_{x_2}^\peak\leq t_c^\mathrm{max}$} & & \multicolumn{6}{c}{$t-t_{x_2}^\peak>t_c^\mathrm{max}$} & \\
                 \cline{3-8}  \cline{10-15} 
 \colhead{$x$} & \colhead{$c$}   & \colhead{$g_{x,c,0}$} & \colhead{$g_{x,c,1}$} & \colhead{$g_{x,c,2}$} & \colhead{$g_{x,c,3}$} & \colhead{$\ssd$} & \colhead{$N$} & & \colhead{$g_{x,c,0}$} & \colhead{$g_{x,c,1}$} & \colhead{$g_{x,c,2}$} & \colhead{$g_{x,c,3}$} & \colhead{$\ssd$} & \colhead{$N$} & \colhead{$t-t_{x_2}^\peak$} \\
  & \colhead{($x_1-x_2$)} &  &  &  &  &  &  &  & &  & & &  & & \colhead{range (d)}
}
\startdata
 & & \multicolumn{13}{c}{SNe~IIb} & \\
$B     $ & $B\!-\!V       $ & $ 4.101$ & $-0.130$ & $-0.103$ & $-0.445$ & $0.011$ & $ 91$ & & $ 4.035$ & $-0.078$ & $-0.104$ & $-0.643$ & $0.009$ & $ 44$ & $-20,117$ \\
$B     $ & $B\!-\!r       $ & $ 4.102$ & $-0.107$ & $-0.102$ & $-0.517$ & $0.009$ & $ 75$ & & $ 4.064$ & $-0.083$ & $-0.104$ & $-0.740$ & $0.007$ & $ 33$ & $-18,106$ \\
$B     $ & $B\!-\!R       $ & $ 4.103$ & $-0.089$ & $-0.102$ & $-0.450$ & $0.009$ & $ 44$ & & $ 4.060$ & $-0.069$ & $-0.102$ & $-0.730$ & $0.006$ & $ 27$ & $-18,106$ \\
\enddata
\tablecomments{The $t_c^\mathrm{max}$ values are listed in Column~11 of Table~\ref{table:BC_cal}.}
\tablenotetext{*}{This table is available in its entirety in machine-readable form.}
\end{deluxetable*}

\begin{deluxetable*}{lccccccccccccc}
\tablenum{F16}
\tablecaption{Parameters of the $A_x^\mathrm{h}$, $A_x^\mathrm{MW}$ and $K$-corrections for IR bands\label{table:IR_cor}}
\tablehead{
 \colhead{Type} & \colhead{$x$} & \colhead{$c$} & \colhead{$h_{x,c,0}$} & \colhead{$\ssd$} & \colhead{$N$} & \colhead{$k_{x,c,0}$} & \colhead{$k_{x,c,1}$} & \colhead{$k_{x,c,2}$} & \colhead{$\ssd$} & \colhead{$N$} & \colhead{$g_{x,c,0}$} & \colhead{$\ssd$} & \colhead{$N$}
}
\startdata
IIb & $Y$ & $V\!-\!Y$ & $ 0.849$ & $0.008$ & $27$ &$-1.547$ & $ 3.947$ & $ 1.672$ & $1.494$ & $22$ &$ 1.101$ & $0.013$ & $27$ \\
    & $J$ & $V\!-\!J$ & $ 0.666$ & $0.005$ & $31$ &$-3.809$ & $-0.950$ & $ 1.108$ & $2.271$ & $24$ &$ 0.827$ & $0.009$ & $31$ \\
    & $H$ & $V\!-\!H$ & $ 0.427$ & $0.001$ & $30$ &$-2.397$ & $ 0.501$ & $ 0.654$ & $0.764$ & $23$ &$ 0.517$ & $0.001$ & $31$ \\
    & $K$ & $V\!-\!K$ & $ 0.289$ & $0.001$ & $29$ &$-3.773$ & $ 0.870$ & $ 0.380$ & $1.153$ & $23$ &$ 0.350$ & $0.001$ & $31$ \\
Ib  & $Y$ & $V\!-\!Y$ & $ 0.931$ & $0.005$ & $10$ &$-0.662$ & $ 1.822$ & $ 2.086$ & $0.456$ & $ 7$ &$ 1.109$ & $0.013$ & $11$ \\
    & $J$ & $V\!-\!J$ & $ 0.721$ & $0.004$ & $13$ &$-3.991$ & $-2.180$ & $ 1.136$ & $0.276$ & $ 8$ &$ 0.824$ & $0.005$ & $13$ \\
    & $H$ & $V\!-\!H$ & $ 0.457$ & $0.001$ & $12$ &$-1.794$ & $-0.273$ & $ 0.718$ & $1.064$ & $ 7$ &$ 0.517$ & $0.001$ & $13$ \\
    & $K$ & $V\!-\!K$ & $ 0.308$ & $0.001$ & $13$ &$-1.961$ & $ 2.922$ & $ 0.408$ & $0.857$ & $ 6$ &$ 0.349$ & $0.001$ & $13$ \\
\enddata
\tablecomments{$h_{x,c,1}=g_{x,c,1}=g_{x,c,2}=g_{x,c,3}=0$.} 
\end{deluxetable*}

\bibliographystyle{aasjournal}
\bibliography{references} 



\end{document}